\UseRawInputEncoding

\documentclass[twocolumn]{aastex62}

\newcommand{\kms}{\,km\,s$^{-1}$}

\newcommand{\accunit}{$_{\odot}$ yr$^{-1}$}

\newcommand{\OI}{[O\,{\scriptsize I}]}
\newcommand{\OIa}{[O\,{\scriptsize I}]\,$\lambda$6300}

\def\arcsec{\hbox{$^{\hbox{\rlap{\hbox{\lower4pt\hbox{$\,\prime\prime$}}}\hbox{$\frown$}}}$}}

\shorttitle{}
\shortauthors{}

\begin{document}


\title{A High-resolution Optical Survey of Upper~Sco: Evidence for Coevolution of Accretion and Disk Winds}

\correspondingauthor{Min Fang}
\email{mfang@pmo.ac.cn}

\author{Min Fang}
\affiliation{Purple Mountain Observatory, Chinese Academy of Sciences, 10 Yuanhua Road, Nanjing 210023, China}
\affiliation{University of Science and Technology of China, Hefei 230026, People's Republic of China}
\affiliation{Department of Astronomy, University of Arizona, 933 North Cherry Avenue, Tucson, AZ 85721, USA}
\author{Ilaria Pascucci}
\affiliation{Department of Planetary Sciences, University of Arizona, 1629 East University Boulevard, Tucson, AZ 85721, USA}
\author{Suzan Edwards}
\affiliation{Five College Astronomy Department, Smith College, Northampton, MA 01063, USA}
\author{Uma Gorti}
\affiliation{SETI Institute/NASA Ames Research Center, Mail Stop 245-3, Moffett Field, CA 94035-1000, USA}
\author{Lynne A. Hillenbrand}
\affiliation{1Department of Astronomy, California Institute of Technology, Pasadena CA 91125}
\author{John M. Carpenter }
\affiliation{Joint ALMA Observatory, Alonso de Cordova 3107 Vitacura, Santiago, Chile}
\begin{abstract}
Magnetohydrodynamic (MHD) and photoevaporative winds are thought to play an important role in the evolution and dispersal of planet-forming disks. Here, we analyze high-resolution ($\Delta v \sim$\,7\,\kms) optical spectra from a sample of 115 T~Tauri stars in the  $\sim 5-10$\,Myr Upper~Sco association and focus on the \OIa\ and H$\alpha$ lines to trace disk winds and accretion, respectively. 
Our sample covers a large range in spectral type and we divide it into Warm (G0-M3) and Cool (later than M3) to facilitate comparison with younger regions. 
We detect the \OIa\ line in 45 out of 87 upper sco sources with protoplanetary disks and 32 out of 45 are accreting based on H$\alpha$ profiles and equivalent widths. All \OIa\ Upper~Sco profiles have a low-velocity (centroid $< -30$\,km/s, LVC) emission and most (36/45) can be fit by a single Gaussian (SC). The SC distribution of centroid velocities and FWHMs is consistent with MHD disk winds. We also find that  the Upper~Sco sample follows the same accretion luminosity$-$LVC \OIa\ luminosity relation and the same anti-correlation between SC FWHM and WISE W3-W4 spectral index as the younger samples. These results indicate that accretion and disk winds coevolve and that, as inner disks clear out, wind emission arises further away from the star. Finally, our large spectral range coverage reveals that Cool stars have larger FWHMs normalized by stellar mass than Warm stars indicating that \OIa\ emission arises closer in towards lower mass/lower luminosity stars. 
\end{abstract}



\section{Introduction} \label{sec:intr}

Circumstellar disks play an important role in both star and planet formation. While planet formation contributes to disk dispersal,  several other processes likely dominate, including  accretion of disk material onto the central star, fast ($\sim 100$\,km/s) jets, slower magnetohydrodynamic (MHD) winds, and photoevaporative winds driven by stellar high-energy photons \citep[e.g.,][]{2014prpl.conf..451F,2016SSRv..205..125G,2022arXiv220310068P}. With the realization that turbulence induced by magnetorotational instability is almost completely suppressed in the planet-forming region (e.g.,  \citealt{2014prpl.conf..411T}), theorists have turned to radially extended MHD winds to extract angular momentum from the disk and drive accretion. Numerical simulations show that MHD winds can be robustly launched over the planet-forming region ($\sim$\,1--30\,au) and enable accretion at the observed rates \citep[e.g.,][]{2015ApJ...801...84G,2016ApJ...821...80B,2017A&A...600A..75B}. Whether large-scale jets arise from the collimation of the inner MHD winds is still a matter of debate (e.g., \citealt{2019FrASS...6...54P}). Finally, photoevaporation has been invoked to explain the short ($\sim 100,000$\,yr) transition from disk-bearing to disk-less (e.g., \citealt{2014prpl.conf..475A}) and there is recent observational evidence that this process is significant in the later stages of disk evolution \citep{2020ApJ...903...78P}.

Low-excitation optical forbidden lines, especially from the strong \OIa\ transition, have long been established to trace jets/outflows from T~Tauri stars (e.g., \citealt{1987ApJ...321..473E,1995ApJ...452..736H}). 
Their line profiles typically consist of two distinct components: a high-velocity component (HVC), blueshifted by $\ge 100$\,km/s from the stellar velocity, and a low-velocity component (LVC), typically blueshifted by less than several tens km/s.  Spatially resolved observations demonstrated early on that  HVCs are produced in extended collimated jets  (e.g., \citealt{2000A&A...356L..41L,2000ApJ...537L..49B,2002ApJ...580..336W}). The origin of LVCs  was less well established but it was also proposed early on that LVCs  might trace a slow disk wind \citep{1995ApJ...454..382K}. 

Recently, with the advent of more sensitive high-resolution ($\Delta v \sim$7\,km/s) spectroscopy, it has been realized that  LVC profiles  sometimes are best fit by a single Gaussian while others require a two-component Gaussian fit, one of which is narrower and less blueshifted than the other, hence the naming NC and BC {\citep{2013ApJ...772...60R,2016ApJ...831..169S,2018AA...620A..87M,2018ApJ...868...28F}. Initially those LVC described by single Gaussians were flagged as either BC or NC, depending on their FWHM. However, a different approach was taken by \cite{2019ApJ...870...76B} who merged all single component LVC into the category SC and found they have some distinct behaviors, in particular correlations of FWHM with both inclination and infrared spectral index, not seen in other components. 

The trend for LVC component FWHMs to be larger in disks seen at higher inclination indicates that Keplerian broadening plays a role in setting the line widths, where BC widths  imply  Keplerian radii within 0.5\,au, thus excluding origin in a photoevaporative thermal wind, which must be launched beyond the gravitational radius.  \citep{2016ApJ...831..169S,2018ApJ...868...28F,2020MNRAS.496..223W}. While the Keplerian radii for NC are compatible with thermal winds, \cite{2019ApJ...870...76B} found that not only do the BC and NC kinematics correlate with each other, but they also depend on the strength of the HVC, linking both BC and NC to jets, and thus to an MHD origin.   \citep{2019ApJ...870...76B}. Additionally in one source found to date with a strong HVC, RU~Lup, spectro-astrometry  shows the \OIa\ LVC  with a velocity $\sim 30$\,km/s at just $\sim 2$\,au from the disk midplane, indicating it traces an MHD rather than a photoevaporative wind  \citep{2021ApJ...913...43W}.
 
 Previous studies on disk winds employing  optical forbidden lines mostly focused on  T~Tauri stars in young star-forming regions with ages less than 5\,Myr: Taurus, Lupus I and III, Cha~I and II, $\rho$~Oph, Corona
Australis, and NGC~2264 \citep{2016ApJ...831..169S,2018A&A...609A..87N,2018AA...620A..87M,2018ApJ...868...28F,2019ApJ...870...76B}.  Here, we expand upon these earlier studies by analyzing the \OIa\ line from a sample of 115 T~Tauri stars in Upper~Sco, an older $\sim 5-10$\,Myr association  \citep{2012ApJ...746..154P,2016A&A...593A..99F,2019ApJ...872..161D}. 
In addition, as the Upper~Sco protoplanetary disk sample includes a significant fraction of disks around M dwarfs (see e.g. Fig.~1 in \citealt{2016ApJ...827..142B}), we also investigate disk winds around lower mass stars more thoroughly than has been done before.

The paper is organized as follows. First, we describe the Upper~Sco sample, the observations and data reduction (Sect.~\ref{sect2}). 
Next, we present our homogeneous re-assessment of stellar properties, including stellar masses and mass accretion rates, and disk types (Sect.~\ref{Sect:spt+RV} and \ref{Sect:diskclass+accretion}). In Sect.~\ref{sect:results} we discuss the extraction and decomposition  of the Upper~Sco \OIa{} lines,  as well as detection rates and the relation between the \OIa\ and accretion. Sect.~\ref{Sect:discussion} further places our results in the broader context of disk evolution while Sect.~\ref{Sect:summary} summarizes our findings.

\begin{figure*}
\begin{center}
\includegraphics[width=1.6\columnwidth]{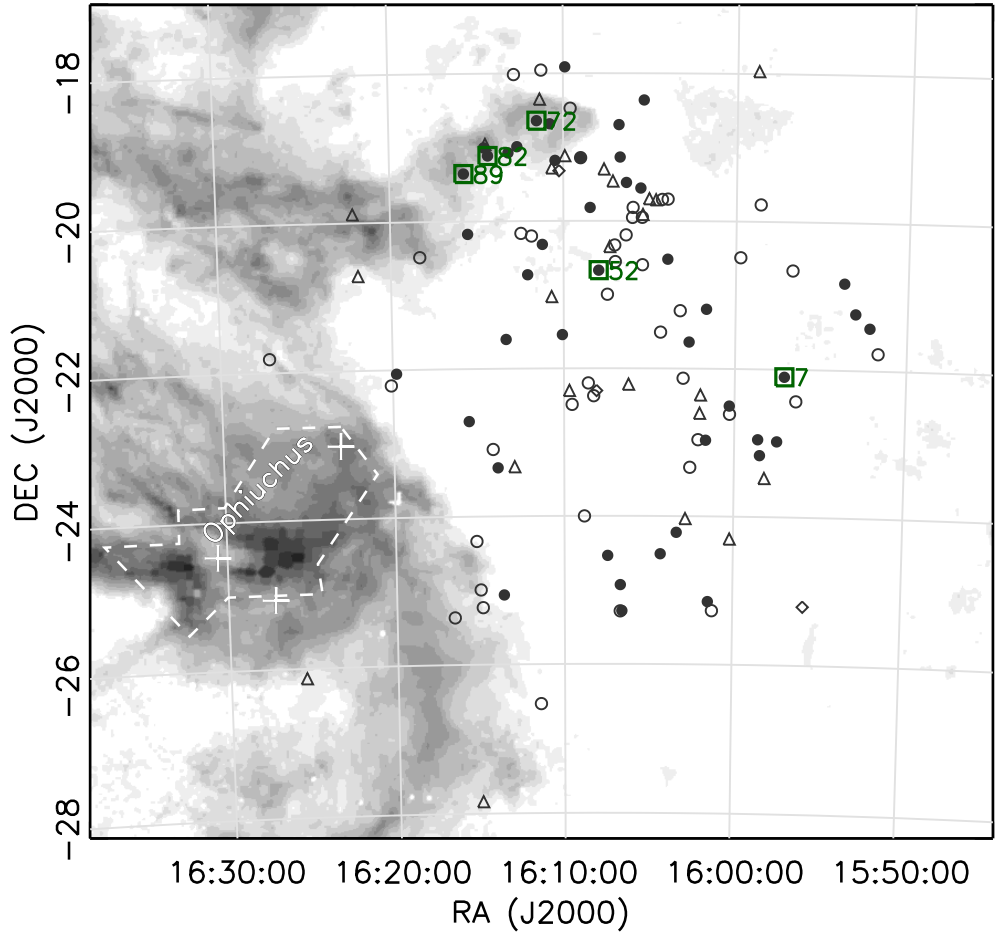}
\caption{
Distribution of our Upper Sco sources. 
We group full/evolved/transitional into protoplanetary disks (PDs: circles, filled for accreting CTTSs and open for WTTSs), debris/evolved transitional into debris disks (DBs: triangles), and separate sources with no disks (NDs: diamonds), see Sect.~\ref{sect:diskclass} and \ref{sect:CTTS} for details.    Green boxes mark sources with HVC emission in the \OIa\ line (Sect.~\ref{sect:results}). 
Sources are overplotted on the cumulative dust extinction map out to 300\,pc produced by \cite{2019ApJ...887...93G}.
The white dashed  polygon indicates the boundary between Ophiuchus and Upper~Sco from \cite{2018AJ....156...75E}. According to this boundary, three sources  (white plus symbols) could be members of Ophiuchus, hence are not considered in the analysis. }\label{Fig:YSODIS}
\end{center}
\end{figure*}

\begin{figure}
\begin{center}
\includegraphics[width=\columnwidth]{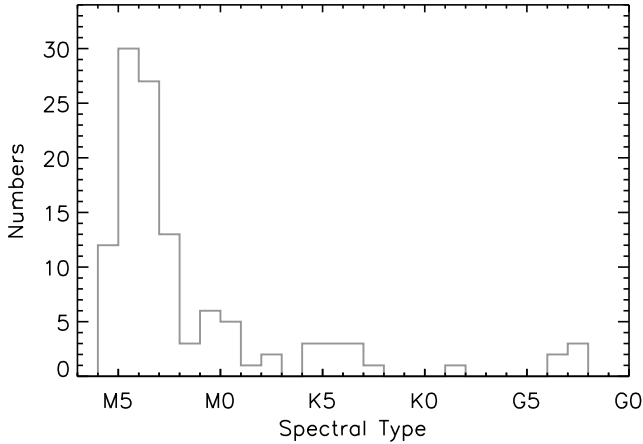}
\caption{Distribution of  spectral types for our Upper~Sco sample.  \label{Fig:SPTDIS}}
\end{center}
\end{figure}

\begin{figure}
\begin{center}
\includegraphics[width=\columnwidth]{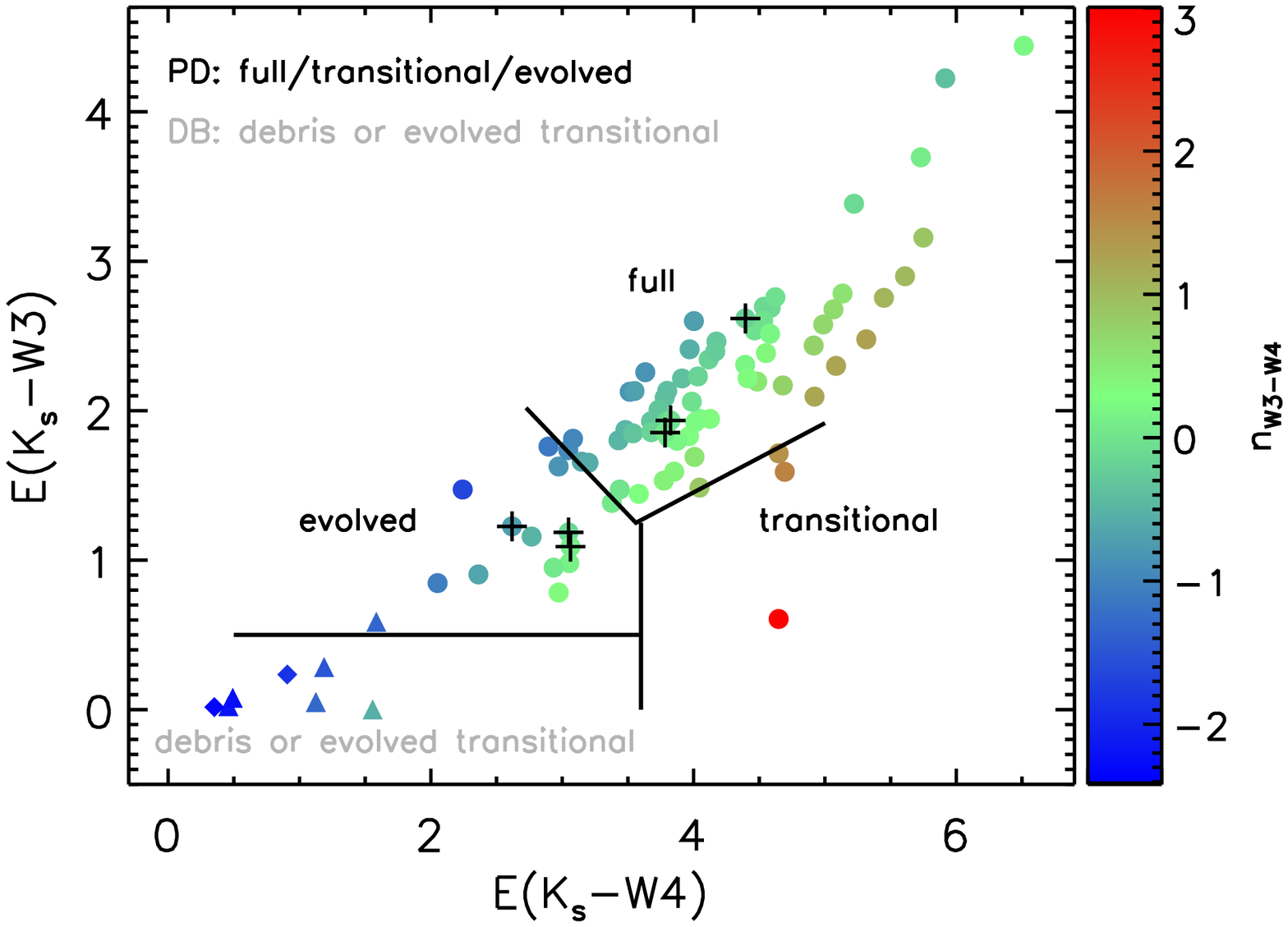}
\caption{Extinction-corrected $K_{\rm s}$-W3 and $K_{\rm s}$-W4 color excesses for  Upper~Sco sources with photometric uncertainty $\leq$0.2~mag in $K_{\rm s}$, W3, and W4 bands.
Symbols are color-coded by their $n_{\rm W3-W4}$ spectral index. Filled/empty symbols are for H$\alpha$ accreting/non-accreting stars (see Sect.~\ref{sect:CTTS}). The  black solid lines show the disk classification boundaries adopted in \cite{2022AJ....163...25L}.  The plus symbols mark the 6 sources classified in this work, other symbols as in Figure~\ref{Fig:YSODIS}.
}\label{Fig:twocolor}
\end{center}
\end{figure}

\section{Sample and Data Reduction}\label{sect2}
\subsection{Sample}\label{sect:sample}
Although this study is directed toward Upper Sco sources with disks, at an age of 5-10 My the majority of young stars in this region have already lost their disks. Indeed, \cite{2022AJ....163...25L} finds that only approximately 20\% of the Upper~Sco population  still retain protoplanetary disks. 

We start from a sample of 118 Upper~Sco candidates that have  Keck/HIRES high-resolution spectra covering the \OIa\ transition. Figure~\ref{Fig:YSODIS} shows their distribution over the sky. Based on the boundary between Ophiuchus and Upper~Sco identified in \cite{2018AJ....156...75E}, three sources\footnote{2MASS~J16230783-2300596, 2MASS~J16303390-2428062, and 2MASS~J162-71273-2504017} are likely Ophiuchus rather than Upper~Sco members (white crosses in Figure~\ref{Fig:YSODIS}). Thus, we remove them from our sample, leaving 115 Upper~Sco sources analyzed in this work (see Table~\ref{tabe_UpperSco}).  

Most of the observations were collected based on evidence of the sources having full or transitional disks following the spectral energy distribution (SED) convention as in \citet[][]{Espaillat2014prpl.conf..497E}, with the aim to complement ongoing ALMA observations of the disks. Of the 115 sources, 95 had been identified as disk candidates by \cite{2006ApJ...651L..49C} based on {\it Spitzer}, or \cite{2012ApJ...758...31L}  based on WISE excess emission, with 67 of them part of the ALMA disk survey by \cite{2016ApJ...827..142B}.
Two of the objects (HD~143006 and AS~205A, IDs~11 and 72, respectively in Table~\ref{tabe_UpperSco}) were already analyzed in \cite{2018ApJ...868...28F}. Our sample also includes another 20 Upper~Sco stars with Keck/HIRES spectra taken for other reasons (such as suspected binarity or interesting lightcurve morphology) that also have WISE photometry indicative of disks (see Sect.~\ref{sect:diskclass}). Fourteen of them are not present in  \cite{2012ApJ...758...31L} but collected in \cite{2022AJ....163...25L}.

Among our sample of 115 disk candidates, 104  have {\it Gaia} Early Data Release 3 (EDR3) parallactic distances from the geometric-distance table provided by \citet{2021AJ....161..147B}. With these new distances, we noted that Source ID~19 has a much larger EDR3 distance than the mean distance of $\sim$142~pc to Upper~Sco but also a large renormalized unit weight error (RUWE$=$14.338). This suggests  that Source ID~19 is a multiple system or may otherwise be problematic in terms of the astrometric measurement. Thus, for  ID~19, as well as the other 11 sources without {\it Gaia} parallaxes, we take the mean distance of 142~pc to Upper~Sco estimated from the other members considered in this work.

\subsection{Observations and data reduction}
The spectral data for our sample are collected from the Keck/HIRES archive and cover observations carried out from 2006 to 2015. An observation log is provided in Table~\ref{Tab:obs_log} of Appendix~\ref{Appen:Obs_log}. The table includes the dates of the observations, nominal spectral resolution, Program ID, and principle investigator. The nominal resolution for our sample is almost always 34,000 ($\Delta v \sim 8.8$\,km/s). 
We estimate the  spectral resolution near the \OIa\ line by fitting 4 telluric lines around this transition toward 6 sources observed at high signal to noise  (larger than 30). From the FWHM of the telluric lines we  obtain a slightly better resolution of 7\,\kms, which is used hereafter.  This is essentially the same resolution achieved in our previous \OI\ studies of young star-forming regions as measured in  \cite{2015ApJ...814...14P}.

We reduced the raw data using the Mauna Kea Echelle Extraction (MAKEE) pipeline written by Tom~Barlow\footnote{http://www2.keck.hawaii.edu/inst/common/makeewww/index.html}. MAKEE is designed to run non-interactively using a set of default parameters, and carry out bias-subtraction, flat-fielding, and spectral extraction,  including sky subtraction, and wavelength calibration with ThAr calibration lamps. The wavelength calibration is performed in air by setting the keyword ``novac" in the pipeline inputs. Heliocentric correction is also applied to the extracted spectra. 

 We corrected for the telluric contamination near the \OIa\ line using the same method as in \cite{2018ApJ...868...28F}. In short, we first remove any telluric absorption near the \OIa\ line using O/B telluric standards; a detailed description of telluric removal can be found in  the HIRES Data Reduction manual\footnote{see the description in https://www2.keck.hawaii.edu\\/inst/common/makeewww/Atmosphere/index.html}.

\section{Stellar Parameters}\label{Sect:spt+RV}
\subsection{Spectral classification and stellar properties}\label{Sect:spt}
We classify the Keck spectra of each Upper~Sco source using the grid of pre-main sequence spectral templates collected from the VLT/X-Shooter instrument \citep{2013A&A...551A.107M,2017A&A...605A..86M,2018A&A...609A..70R}. The spectral templates have been re-classified using the scheme for M types from \cite{1999ApJ...525..466L,2003ApJ...593.1093L,2004ApJ...617.1216L,2006ApJ...645..676L}, and for K and earlier types from \cite{2013ApJS..208....9P}, see a detailed description in \cite{2021ApJ...908...49F}. We normalized the Keck spectra and the spectral templates in the same way, and find the template that best matches each Keck spectrum.
We  also derive the extinction toward each target using the best-fit template that reproduces the the {\it Gaia} photometry.  

In our approach we fit 2 free parameters:  the extinction in the $V$ band ($A_{V}$) and the absolute flux density at 7500~\AA\ ($I_{\rm 7500}$). First, we redden the template with  $A_{V}$ using the extinction law from  \cite{1989ApJ...345..245C}, adopting a total to selective extinction value typical for the interstellar medium  ($R_{V}$=3.1). The reddened template is normalized to be one at 7500~\AA\ and then  multiplied by $I_{\rm 7500}$. Next, the ``flux-calibrated" template is used to produce the synthetic {\it Gaia} photometry. The best-fit parameters are derived by minimizing the difference between the synthetic {\it Gaia} photometry and the observed {\it Gaia} photometry using the $\chi^2$ method. In this way, we can obtain both the extinction and continuum flux, estimated from the best-fit template, near the accretion-related lines and oxygen forbidden line, from which we can also estimate the line fluxes.  A detailed description of the spectral classification and flux calibration is presented in Appendix~\ref{Appen:SPT} while a discussion of the flux calibration uncertainties is presented in Appendix~\ref{Appen:UNC}.      

We obtain the effective temperature of each star from the spectral type to effective temperature relation in \cite{2017AJ....153..188F}. The stellar luminosity is estimated from $I_{\rm 7500}$ using the bolometric corrections from \cite{2014ApJ...786...97H}. We derive  stellar masses from the luminosity and effective temperature using the non-magnetic pre-main sequence evolutionary tracks of \citet{2016A&A...593A..99F}.  A detailed description is presented in Appendix~\ref{Appen:mass}. The resulting mass ($M_{\star}$), as well as  the stellar spectral type, visual extinction ($A_{\rm V}$), luminosity ($L_{\star}$),  and radius ($R_{\star}$) are listed in Table~\ref{tabe_UpperSco}. Figure~\ref{Fig:SPTDIS} shows the distributions of spectral types for our Upper~Sco sample.
The majority (69/115) of the sources are later than M3.

\subsection{Stellar Radial velocity}\label{Sect:RV}
We derive the stellar radial velocity (RV)  by cross-correlating each spectrum with a stellar synthetic spectrum of the same effective temperature as our source. The model spectra are from \cite{2013A&A...553A...6H} for a solar abundance and surface gravity log\,$g$=4.0. We degrade the model spectrum to match the spectral resolution of HIRES, and rotationally broaden it to match the source photospheric features. We did the cross-correlation separately for each echelle order  without strong emission lines and excluding the wavelength ranges with strong telluric absorption.  The total number of available orders varies from source to source but it is at least 4 and up to 17. The resulting RVs for individual sources are listed in Table~\ref{tabe_UpperSco}. The typical RV uncertainty is $\sim$1.4\,\kms.

Since arc calibration frames for the absolute wavelength solution were taken only at the beginning of each night, systematic shifts of the wavelength solution of individual spectra are possible. In \cite{2018ApJ...868...28F}, we assessed the shifts  by cross-correlating the telluric lines with the model atmospheric transmission curve for the Keck Observatory, and found that the wavelength calibration tended to be drifted throughout the night by $\sim$0.5--2\kms. In the same way, we estimate the shift for each observation and list the RV correction in Table~\ref{tabe_UpperSco}.  The RV of  each source need to be corrected by adding the value in the ``Correction" column to its RV. The shift only changes the heliocentric RV derived from the spectra, and does not affect the relative shift between the stellar photospheric absorption features and the \OIa\ line.

\section{Disk Types and Accretion}\label{Sect:diskclass+accretion}

\subsection{Disk classification}\label{sect:diskclass}
As mentioned in Sect.~2.1 our sample is drawn from disk candidates from \cite{2016ApJ...827..142B} and our own SED classification. Because the majority (109 out of 115) of our Upper~Sco sources have been recently assigned a disk classification by \cite{2022AJ....163...25L} we use their classification for the 109 sources and follow his approach to classify the remaining 6 sources for a homogeneous re-classification. \cite{2022AJ....163...25L} uses colors, particularly in the WISE bands, to distinguish 5 groups:  Class~III or  diskless;  debris/evolved transitional disks; evolved disks; transitional disks; and full disks. Figure~\ref{Fig:twocolor} shows the boundaries of this classification in an extinction-corrected color excess \footnote{The intrinsic colors corresponding to individual spectral types used to derive the color excess are from \cite{2013ApJS..208....9P}} plot where our sources are color-coded by their W3-W4 spectral index, calculated as $n=d log(\lambda F_{\lambda})/d log\lambda$. The used WISE photometry are collected from AllWISE Source Catalog \citep{2014yCat.2328....0C}.

To simplify the discussion and more readily compare our results with those in younger star-forming regions, we  adopt only 3 categories in this work: No disk (ND), Debris Disk (DB), and Protoplanetary Disk (PD). These 3 categories correspond to the Class~III, the debris/evolved transitional disks, and the evolved/transitional/full disks in \cite{2022AJ....163...25L}, respectively.  
For the remaining 6 sources (IDs~99, 102, 105, 107, 108, 112; the plus symbols in Figure~\ref{Fig:twocolor}) without classification in \cite{2022AJ....163...25L}, we adopt the same approach and find that all of them are PDs. Based on this classification, our sample includes  4 ND sources\footnote{These sources (IDs~5, 53, 64, and 94) were initially classified as debris/evolved transitional disks in \cite{2012ApJ...758...31L} but reclassified as Class~III (diskless) in \cite{2022AJ....163...25L} either due to  insignificant excess emission or  unreliable WISE detection.}, 24 DBs, and 87 PDs.  The spectral energy distributions (SEDs) of these sources are shown in Appendix~\ref{Appen:SED}.

We also include in Figure~\ref{Fig:twocolor} a color bar indicating the magnitude of the WISE W3-W4 spectral index, as we will compare this index with the kinematic properties of the \OIa\ line (see  Sect.~\ref{sect:IRindex}). Note that although there is only one bona-fide transition disk in our sample, there is a clear gradient in the W3-W4 color as the  $K_s-W4$ excess increases in full and evolved disks.

\begin{figure}
\begin{center}
\includegraphics[width=\columnwidth]{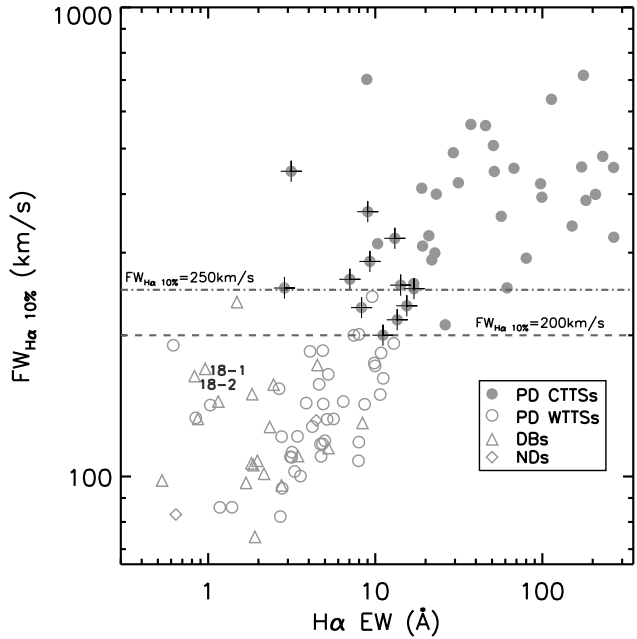}
\caption{H$\alpha$ $FW_{\rm H\alpha 10\%}$ vs. $EW$ for our Upper~Sco sample:  filled circles for PDs classified as CTTSs;  open circles for PDs classified as WTTSs;  open triangles for the DBs;  and  open diamonds for NDs. The black pluses mark the 12 additional sources which were classified as WTTSs based on their H$\alpha$ $EW$, but re-classified as CTTSs  based on their  $FW_{\rm H\alpha 10\%}$. }\label{Fig:EW_FW10}
\end{center}
\end{figure}

\begin{figure}
\begin{center}
\includegraphics[width=\columnwidth]{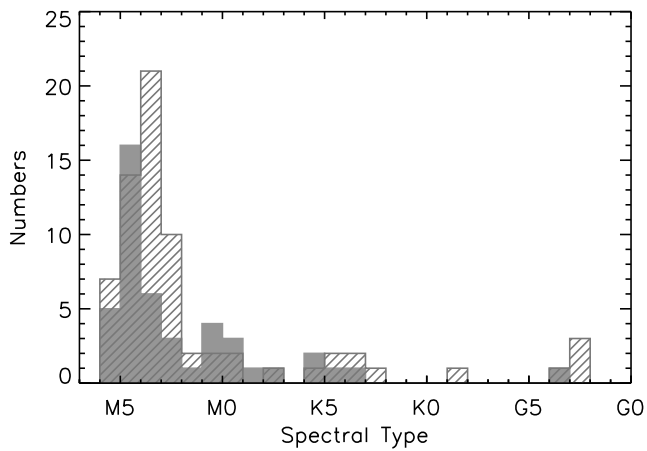}
\caption{Distributions of  spectral types for the CTTSs (grey color filled histograms) and WTTSs (grey line-filled histograms).  \label{Fig:ACCSPTDIS}}
\end{center}
\end{figure}

\subsection{CTTS/WTTS classification}\label{sect:CTTS}
To identify which sources are truly accreting, we use two quantities: the H$\alpha$ equivalent width ($EW$) and  the full width of the H$\alpha$ line at 10\% of the peak intensity ($FW_{\rm H\alpha,10\%}$), which are commonly employed as accretion indicators \citep{2003AJ....126.2997B,2003ApJ...592..282J,2003ApJ...582.1109W,2004A&A...424..603N,2009A&A...504..461F,2013ApJS..207....5F}. First, we classify young stars into accreting or non-accreting using the spectral type dependent H$\alpha$ $EW$ criteria described in \cite{2009A&A...504..461F}. Next, we check their $FW_{\rm H\alpha,10\%}$ and assign accreting (CTTS) status to those sources using the criteria defined in Appendix~\ref{Appen:FW10}, even if their H$\alpha$ $EW$ is below the accreting thresholds. This way sources with low H$\alpha$ $EW$s but broad  profiles typical to accreting stars (12 in our sample) are properly classified as CTTS. The resulting classification is summarized in Table~\ref{tabe_UpperSco}.
 
 In Figure~\ref{Fig:EW_FW10}, we show the H$\alpha$ $FW_{\rm H\alpha,10\%}$ vs. $EW$ for our sample and mark the CTTSs we have identified.  Source~18 is a known eclipsing binary (see USco~48 in \citealt{David2019ApJ...872..161D} for details) and its disk is classified  as a DB. Its HIRES spectrum shows two well-separated H$\alpha$  emission lines whose  $EWs$ and $FW_{\rm H\alpha,10\%}$ are consistent with WTTSs (see 18-1 and 18-2 in Figure~\ref{Fig:EW_FW10}).

Our study covers 32 sources with X-Shooter spectra whose accretion status has been recently determined by \citet{2020A&A...639A..58M}. Taking advantage of the broad spectral coverage of X-Shooter, \citet{2020A&A...639A..58M} estimate the accretion luminosity by simultaneously fitting the stellar photosphere and Balmer continuum emission and classify a source as CTTS if its accretion luminosity is above the typical chromospheric value corresponding to its spectral type (see also \citealt{2017A&A...605A..86M}). Among the 32 sources in common, 26 have our same classification and are not discussed further while  6, mostly low accretors, have a different  classification. Recently, \cite{2022AJ....163...74T} proposed a new scheme to classify low accretors using the He~{\scriptsize I}~$\lambda$10830 line profile. According to this scheme, IDs~1, 20, 41, and 58 would be classified as in \cite{2020A&A...639A..58M} while IDs~81 and 91 would follow our classification, see a detail discussion in Appendix~\ref{Appen:ACC_com}. 

 Given that there is no clear preference between the two classifications among the 6 sources discussed above and that only 43\% (50/115) of the Upper~Sco sample have He~{\scriptsize I}~$\lambda$10830 data, we will adopt here the CTTS/WTTS classification based only on the H$\alpha$ line from the HIRES data. Based on this classification, our Upper~Sco sample comprises  45 CTTSs  (45 PDs) and 70 WTTSs (42 PDs, 24 DBs, and 4 NDs).  Figure~\ref{Fig:ACCSPTDIS} shows the distributions of spectral types for the CTTSs and WTTSs. The K-S test returns a high probability that these spectral types are drawn from the same parent population (p=0.2).
As noted above,  low accretors and/or accretors with significant accretion variability might have HIRES H${\alpha}$ profiles similar to those of DBs, where H${\alpha}$ originates from chromospheric emission, and thus might be classified as WTTSs here. For those WTTSs, a PD classification and an \OIa\ detection  most mean that they are accreting at low levels.

\begin{figure}
\begin{center}
\includegraphics[width=\columnwidth]{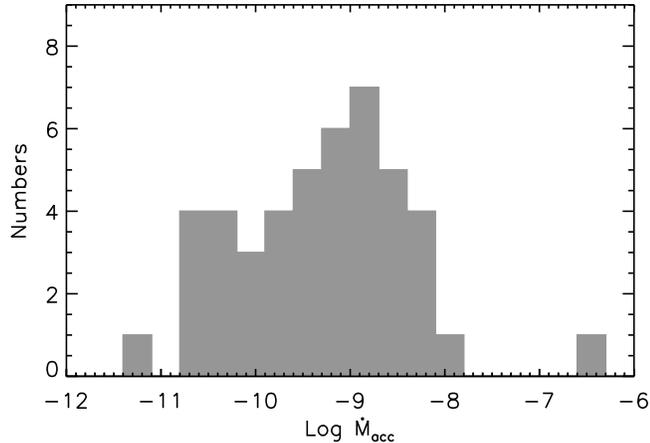}
\caption{Distribution of Upper~Sco CTTSs accretion rates.   \label{Fig:ACCDIS}}
\end{center}
\end{figure}

\subsection{Mass accretion rates}\label{sect:Macc}
We estimate the accretion luminosity using four permitted emission lines covered by the HIRES spectra,
two Balmer lines (H$\beta$, and H$\alpha$) and two  He~{\scriptsize I} lines (5876 and 6678\AA), as their  luminosities are known to be tightly correlated  with the accretion luminosity (e.g., \citealt{2008ApJ...681..594H,2009A&A...504..461F,2012A&A...548A..56R,2014A&A...561A...2A,2017A&A...600A..20A}). Line luminosities are derived from the measured equivalent widths of individual lines times the adjacent continuum flux, the latter is obtained from the template that best matches the Keck spectra and GAIA photometry of each target (see Sect.~\ref{Sect:spt}). We then convert line luminosities to accretion luminosities via the empirical relations  in \cite{2018ApJ...868...28F}. For two objects in common with \cite{2018ApJ...868...28F} (HD~143006 and AS~205A, IDs~11 and 72, respectively in Table~\ref{tabe_UpperSco}) we collect mean accretion luminosities directly from that work. 
For an additional two sources accretion luminosities are only from the H$\alpha$ line. For the remaining 41 CTTSs in Upper~Sco: 31 have mean accretion luminosities from the four permitted lines mentioned above; 8 from the H$\alpha$, H$\beta$, and He~I$\lambda$5876 lines; and  2 only from the H$\alpha$ and H$\beta$ lines.
 
Finally, accretion luminosities are converted into mass accretion rates using the following relation \citep{1998ApJ...492..323G}:
\begin{equation}
\dot{M}_{\rm acc}=\frac{L_{{\rm acc}}R_{\star}}{{\rm G}M_{\star} (1-\frac{R_{\star}}{R_{\rm in}})},
\end{equation}
\noindent where $R_{\rm in}$ denotes the truncation radius of the disk, which is taken to be 5\,$R_{\star}$ \citep{1998ApJ...492..323G}, G is the gravitational constant, $M_{\star}$ is the stellar mass, and $R_{\star}$ is the stellar radius.  The resulting mass accretion rates are also listed in Table~\ref{tabe_UpperSco}  and their distribution is illustrated in Figure~\ref{Fig:ACCDIS}. The $Log~\dot{M}_{acc}$ in $M$\accunit\ range from $-$11.3 to  $-$6.6 and the medians for the Upper~Sco sample is $-$9.3.

\setcounter{table}{2}
\begin{table}
\scriptsize
\renewcommand{\tabcolsep}{0.04cm}
\caption{\OIa\ samples compared in this work. The SpTy range for the Warm sample is  G0-M3  while for the Cool sample it is M3-M5.2.  }\label{Table:csamples}
\begin{center}
\begin{tabular}{lcccccc}
\hline
\hline
Population & Age & Sample   & $\Delta v$ & Ref. & Median $L_{\rm acc}$ & Median $\dot{M}_{\rm acc}$ \\
     & [Myr] &  & [km/s] & & [L$_\odot$] & [M$_\odot$/yr] \\
\hline 
SFB & 1-3 & Warm & 6.6 & 1, 2, 3 &0.04 &3.6$\times$10$^{-9}$\\
Cha+Lupus & 1-3 & Warm & 16--40 & 4&0.02 &2.3$\times$10$^{-9}$\\
          &     & Cool &16--40  &4 &0.0014&3.9$\times$10$^{-10}$ \\   
NGC~2264 & 3-5 & Warm & 11.3 & 5&0.05 &1.0$\times$10$^{-8}$ \\
Upper~Sco & 5-10 & Warm & 7 & 6 &0.01 &1.1$\times$10$^{-9}$ \\
          &     & Cool &7  &6 &0.0009 &2.3$\times$10$^{-10}$ \\    
\hline
\end{tabular}
\end{center}
\tablerefs{1. \citet{2016ApJ...831..169S}; 2. \citet{2018ApJ...868...28F};
3. \citet{2019ApJ...870...76B}; 4. \cite{2018A&A...609A..87N}; 
5. \cite{2018AA...620A..87M}; 6. this work}
\end{table}

\begin{figure}
\begin{center}
\includegraphics[width=\columnwidth]{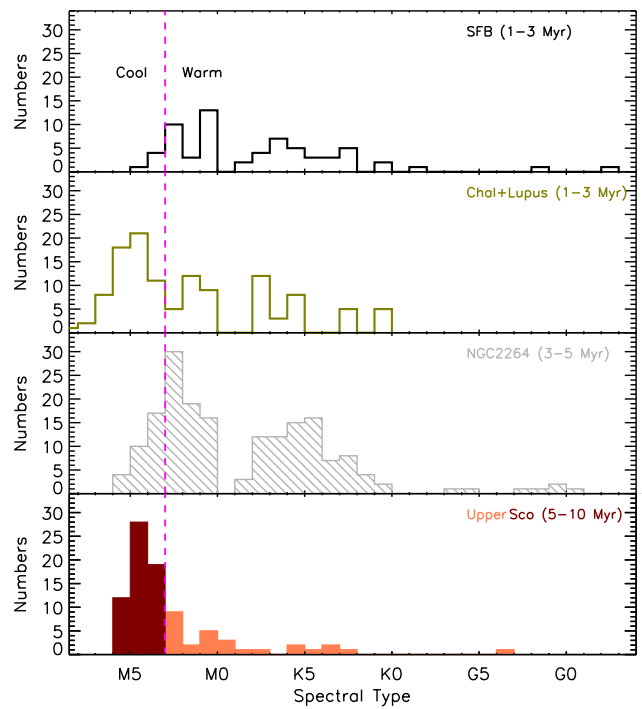}
\caption{Distribution of spectral types for the SFB (black open  histogram), Cha+Lupus (olive open  histogram), NGC~2264 (gray line-filled histogram), and Upper~Sco sample including only PDs.). 
\label{Fig:allSPTDIS}}
\end{center}
\end{figure} 

\section{Mass Outflow Diagnosed via the \OIa\ line} \label{sect:results}
In the following subsections, we focus on the analysis of the \OIa\  lines, their detection rates, line profiles, and relations between the \OIa\ luminosity and stellar/disk properties. We focus on trends with spectral type and age, and throughout the analysis we compare our Upper~Sco population with several other regions of recent star formation.

Specifically, the comparison samples are:
the $\sim 1-3$\,Myr old population of selected sources from Taurus, Lupus, $\rho$~Oph, and Corona Australis observed at  high resolution ($R \sim 48,000$:  \citealt{2016ApJ...831..169S,2018ApJ...868...28F,2019ApJ...870...76B}， hereafter SFB); the similarly young population of all disks in  Chamaeleon~I/II and Lupus observed at a much lower resolution  with X-Shooter ($R \sim 7,400-18,200$: \citealt{2018A&A...609A..87N}, hereafter Cha+Lupus); and the slightly older and further away ($\sim 3-5$\,Myr, $d=760$\,pc) population of NGC~2264  observed at intermediate resolution ($R \sim 26,500$: \citealt{2018AA...620A..87M}). Figure~\ref{Fig:allSPTDIS} shows the distributions of the spectral type of these surveys that can be used for comparison with Upper~Sco. As clearly shown in Figure~\ref{Fig:allSPTDIS},  
 the $1-3$\,Myr-old SFB and the $3-5$\,Myr-old NGC~2264 populations have few sources later than M3. Thus, we split the Upper~Sco sample into two subgroups: the ones later than M3 will be grouped into the Cool star sample while earlier spectral types (G0-M3) will constitute the Warm sample. Cool stars range in stellar mass from $\sim 0.08-0.28$\,$M_\odot$ while Warm stars from $\sim 0.28-1.79$\,$M_\odot$. Comparisons of the wind properties between the Upper~Sco samples, the $1-3$\,Myr-old SFB and the $3-5$\,Myr-old NGC~2264 populations will be restricted to the G0-M3 spectral type range (Warm sample). A summary of these surveys used for the comparisons is provided in Table~\ref{Table:csamples}. The same table also gives the coverage in spectral type that can be used for comparison with Upper~Sco.

We have verified that, within this restricted spectral type range (warm sample), the K-S test gives a high probability ($0.2-0.5$) that the stars in Upper~Sco, SFB, and NGC~2264 are drawn from the same parent population. The Cha+Lupus survey covers a significant number of late-type stars and, as Upper~Sco, two samples of Cool and Warm stars can be used for the comparison. The last two columns of Table~\ref{Table:csamples} provide the median accretion luminosity and median mass accretion rate from each sample calculated in this work.

\begin{figure*}
\begin{center}
\includegraphics[width=2\columnwidth]{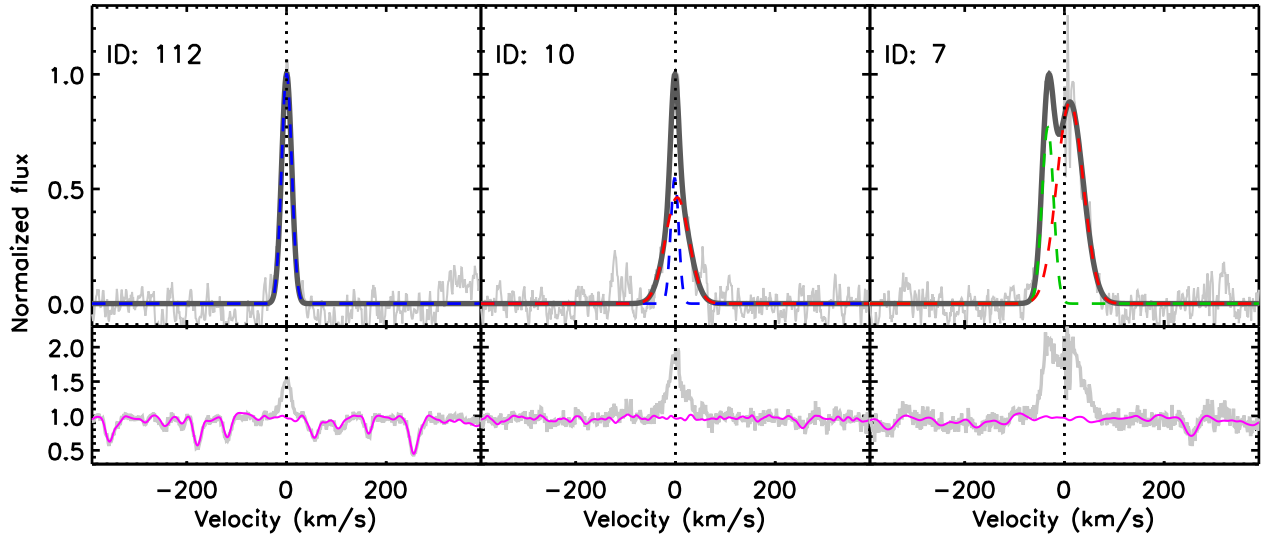}
\caption{ Examples of  corrected \OIa\ profiles as well their fits (top panels) and their uncorrected \OIa\ profiles (bottom panels) with their photospheric corrections (shown in magenta), for three Upper~Sco sources: SC (ID~109), NC+BC (ID~10, blue dashed line for NC and red dashed line for BC), and SCJ (ID~7, which shows both one SC and one HVC). SC dominates the Upper~Sco sample (36/45) while there are only 7 BC+NC and just 2 SCJ, see Table~\ref{Table:statistics}. The gray solid line is the sum of the  individual components. Line profile fits for all sources are shown in Fig.~\ref{Fig:SED_LR1}. }\label{Fig:example_line}
\end{center}
\end{figure*}

\subsection{Line profiles and decomposition}\label{Sect:corrected_profiles}

We subtract photospheric features  following the procedure outlined in \citet{1989ApJS...70..899H}:  a photospheric standard with spectral type similar to the target is rotationally broadened, veiled  (parameter $r_{\lambda}$), and shifted in velocity to best match the photospheric lines of the target star.  
Veiling is used to mimic the filling effect in photospheric lines from the excess emission produced by the accretion shocks, and it is defined as the ratio of the excess continuum emission ($F_{\rm excess}$) to the photospheric flux ($F_{\rm phot}$): $r_{\lambda}=F_{\rm excess}/F_{\rm phot}$.  
The best-fit photospheric spectrum is achieved by  minimizing the $\chi^{2}$, defined as  $\chi^{2}=\sum (F_{\rm target}-F_{\rm Standard})^{2}$. Corrected line profiles are obtained by subtracting the best-fit photospheric spectra from the target spectra.  Examples of  corrected line profiles for three sources are shown in Figure~\ref{Fig:example_line} while all other line profiles  are available in Figure~\ref{Fig:SED_LR1} of Appendix~\ref{Appen:SED}. 

Following the same procedure used in our series of previous papers  \citep{2013ApJ...772...60R,2016ApJ...831..169S,2018ApJ...868...28F,2019ApJ...870...76B},  we decompose the \OIa\ corrected line profiles into Gaussian components. To find the minimum number of Gaussians that describe the observed profile we use an iterative approach based on the IDL procedure {\it mpfitfun}. The minimum number is achieved when the root mean square (rms) of the residual (Gaussians minus original spectrum) is within 2$\sigma$ of the rms of the original spectrum  next to the \OIa\ line. 
 
Figure~\ref{Fig:example_line} shows examples of the best-fit Gaussian components for profiles of three sources in Upper Sco. The other sources are shown in Appendix~\ref{Appen:SED}, Figure~\ref{Fig:SED_LR1}. The best fit parameters resulting from the decomposition
of each individual components, which are the line luminosities, the velocity centroid ($v_{\rm c}$), the full width at half maximum (FWHM), and the equivalent width (EW), are listed in Table~\ref{tabe_lineprofile} . The majority of LVC in Upper Sco are fit by single 
Gaussian components, with only a few BC+NC. For this reason we adopt the terminology for the \OIa{} components as in \cite{2019ApJ...870...76B}, so that LVC are either SC or BC+NC, as further discussed in Sect.~\ref{sect:IndividualKinematicComponents}. 

 The overall uncertainty of \OIa{} line fluxes is about 
0.08~dex including a flux calibration uncertainty of 15\% in our Keck spectra and a median uncertainty of 12\% in the line integration. For the line centroids and FWHMs, their median uncertainties of individual measurements are 1.3 and 1.5\,\kms, respectively. The dominant uncertainty sources are the noise in the  spectra of the Upper~Sco sources and contamination from the residual of telluric \OIa{} line, and the contribution in the uncertainty from the photospheric templates is negligible since these templates are always high quality and reproduce the photospheric features of the  spectra of the Upper~Sco sources very well (see Figs.~\ref{Fig:example_line} and~\ref{Fig:SED_LR1}.).

\subsection{Detection rates}\label{sect:DetectionRates}
All of our 45 \OIa\  detections in Upper~Sco are from PD sources with a detection rate of 52\% (45/87) among the PD; no  emission is seen  toward  DB or ND sources. The majority of the detections (32/45, $\sim$71\%) are from CTTSs but a considerable fraction (13/45, $\sim$29\%) are associated with stars that appear not to be accreting based on their H$\alpha$ profiles, hence are classified as WTTSs in this work (see Appendix~\ref{App:outliers}). As we discuss  in Sect.~\ref{sect:SCcentroids} the distribution of LVC centroids and FWHMs for CTTS and nominal WTTS with \OIa\  is indistinguishable suggesting that the WTTS with \OI\ detections are likely accreting CTTS but at levels below what can be discerned using H$\alpha$, see also Sect.~\ref{sect:CTTS}. 

Table~\ref{Table:sdet} compares  \OIa\ detection statistics among CTTS and WTTS in Upper~Sco with the younger regions in this comparative study, broken out by spectral type range, i.e. warm or cool samples and whether detections are in CTTS or sources classified as WTTS depending on  H$\alpha$ properties. In order to compare  \OIa\ detection rates among the different regions, we use the Fisher exact test on a $2\times2$ contingency table and test the null hypothesis that the number of \OIa\ detections/non-detections in two populations is equally likely. The test returns the probability (p-value) of obtaining by chance a contingency table at least as extreme as the one that was actually observed. If the p-value is below 5\% we conclude that the observed imbalance is statistically significant. The two contrasted populations are spectral ranges and evolutionary state and results are summarized in Table~\ref{Table:contingency}.

We start by exploring any spectral type (stellar mass) dependence. Here, the columns of the $2\times2$ contingency table are the Warm and Cool samples, hence this test can be carried out only for Cha+Lupus and Upper~Sco. For the 1-3\,Myr-old Cha+Lupus sample \citep{2018A&A...609A..87N} the \OI\ detection rates among CTTS are  statistically indistinguishable: $\sim 75$\%  (50/67) for the Warm and $\sim 85$\% (29/34) for the Cool CTTSs, Fisher p-value of 31\%. The same is true for the  5-10\,Myr-old Upper~Sco region. Among the Warm sample \OI\ detection rates are 50\% (14/28) for CTTS and 64\% (18/28) for CTTS+WTTS while among the Cool samples rates are  31\% (18/59) and 46\% (27/59), respectively. The Fisher  p-values are 10\% and 12\% when considering CTTS or the combined CTTS+WTTS. Therefore, we conclude that the \OIa\ line has a similar detection rate toward Warm (G0$-$M3) and Cool (M3$-$M5.2) stars.

\setcounter{table}{4}
\begin{table}
\scriptsize
\renewcommand{\tabcolsep}{0.04cm}
\caption{Summary of CTTS and WTTS in different samples with \OIa\ detections. }\label{Table:sdet}
\begin{center}
\begin{tabular}{lc|c|cc|cc|c}
\hline
\hline
Population & Sample & Total & \multicolumn{2}{c|}{CTTS} & \multicolumn{2}{c|}{WTTS}  & CTTS+WTTS     \\
 &        &   &   & w\OI  &  & w\OI &  w\OI \\
\hline 
SFB & Warm & 60 & 54 & 52  & 6 & 3 & 55  \\
Cha+Lupus & Warm & 67 & 62 &  50  & 5 & 0 & 50\\
          & Cool &34 & 34 & 29  & 0 & 0 & 29 \\    
NGC~2264 & Warm & 165 & 165 & 104  &0 & 0 & 104  \\
Upper~Sco & Warm & 28 & 18 & 14  & 10 & 4 & 18 \\
          & Cool & 59 & 27 & 18 & 32 & 9 & 27 \\    
\hline
\end{tabular}
\end{center}
\tablecomments{The NGC~2264 sample was selected to include only CTTS \citep{2018AA...620A..87M}}
\end{table}

\begin{table}
\scriptsize
\renewcommand{\tabcolsep}{0.04cm}
\caption{Results of the Fisher exact test. For each $2 \times 2$ contingency table we report the percentage probability that the number of \OIa\ detections/non-detections in two samples is equally likely.  }\label{Table:contingency}
\begin{center}
\begin{tabular}{cc|c|c}
\hline
\hline
Pop 1 & Pop 2 & CTTS &  CTTS+WTTS     \\
\hline 
\multicolumn{4}{c}{Warm Samples} \\
\hline
Upper~Sco & SFB &  0.04 & 0.4 \\
Upper~Sco & Cha+Lupus &  3 & 33 \\
Upper~Sco & NGC~2264 &  21 & 100 \\
\hline 
\multicolumn{4}{c}{Cool Samples} \\
\hline
Upper~Sco & Cha+Lupus &  3$\times 10^{-5}$ & 0.02 \\
\hline 
\multicolumn{4}{c}{Cool vs. Warm Samples} \\
\hline
\multicolumn{2}{c}{Cha+Lupus} &  31 & 31 \\
\multicolumn{2}{c}{Upper~Sco} &  10 & 12 \\
\hline
\end{tabular}
\end{center}
\end{table}

In relation to evolutionary trends, we note that Table~\ref{Table:sdet} indicates a decreasing \OI\ detection rate with increasing age and decreasing accretion luminosity. To test if this trend is statistically significant, we construct $2\times2$ contingency tables where one of the columns reports the number of Upper~Sco \OIa\ detections/non-detections  while the other column gives the values for a younger region. Restricting ourselves to the Warm  CTTS samples, we find that the imbalance between the Upper~Sco \OI\ detection rate (50\%) and that of the 1-3 Myr-old SFB (87\%) and Cha+Lupus (75\%) regions is statistically significant, p-values of 0.04\% and 3\%. When including WTTS (CTTS+WTTS column), p-values increase to 0.43\% and 33\%, i.e. only the difference between SFB and Upper~Sco remains statistically significant. However, among the Cool samples the \OI\ detection rates for Upper~Sco are statistically lower than those for Cha+Lupus for CTTS alone as well as CTTS+WTTS, p-values of $3\times 10^{-5}$ and 0.02\%, respectively. The Fisher exact test reports high p-values of 21\% and 100\% when comparing the \OI\ detection rates in the Warm samples of Upper~Sco and NGC~2264. It is possible that the \OI\ detection rate does not decrease from 3-5\,Myr, the age of NGC~2264, to 5-10\,Myr or, as pointed out in \cite{2018AA...620A..87M},
the lower rate in NGC~2264 is caused by the cluster larger distance in combination with strong nebular emission. In conclusion, we find strong evidence that  \OIa\ detection rate decreases going from 1-3\,Myr-old regions to the 5-10\,Myr-old sources in Upper~Sco. Along with the \OI\ detection rate the accretion luminosity decreases too and the K-S test returns a low probability (0.4\%) for the  Warm Upper~Sco sample to be drawn from the same $L_{acc}$ population as the younger SFB and Cha+Lupus sources. The relation between the \OIa\ and accretion will be further explored in the next subsection.

 \begin{figure}
\begin{center}
\includegraphics[width=0.9\columnwidth]{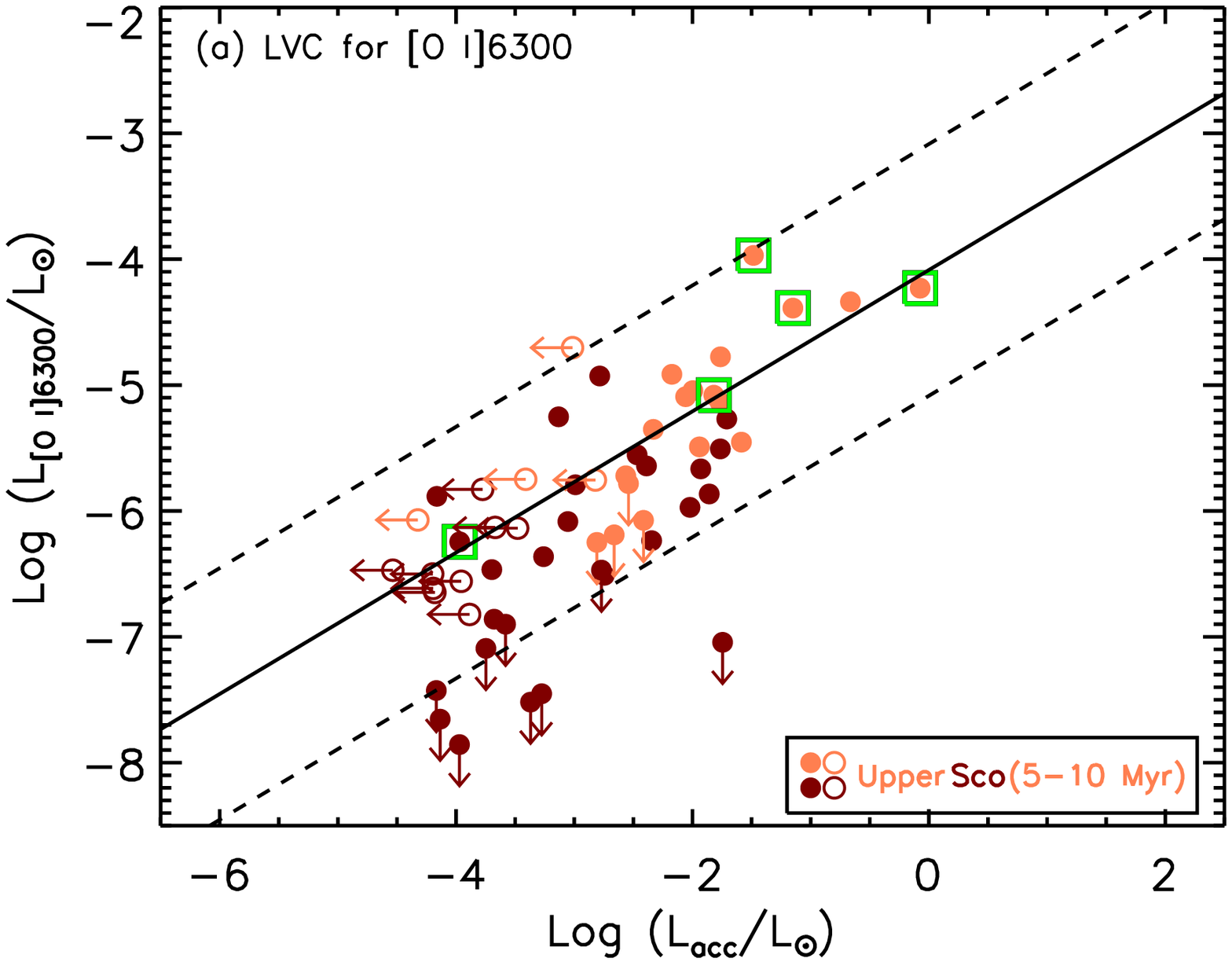}
\includegraphics[width=0.9\columnwidth]{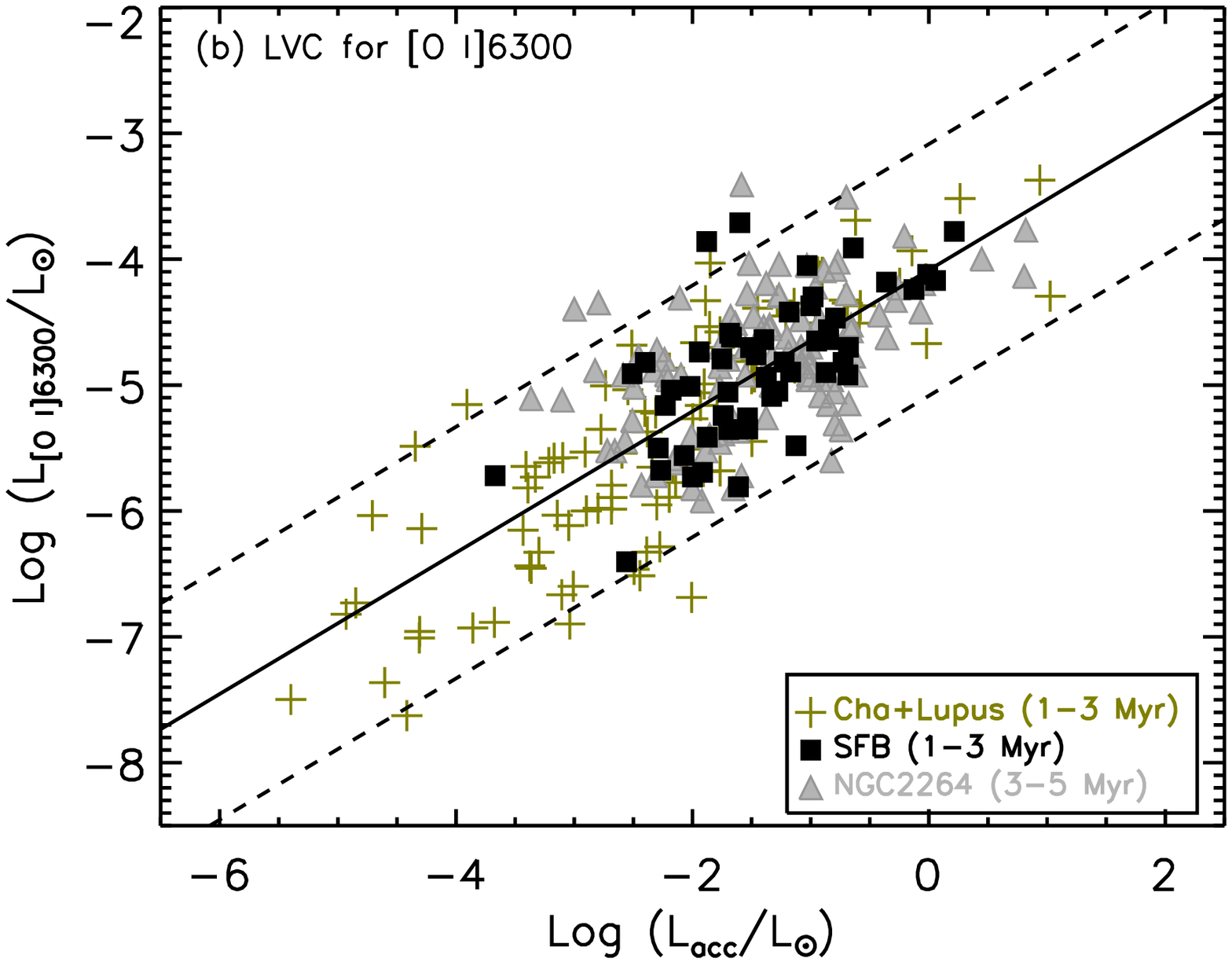}
\caption{(a): LVC \OIa\ luminosities   vs accretion luminosities for the Upper~Sco sample: filled(open) circles for CTTS(WTTS), orange(maroon) for Warm(Cool) sources. Green squares mark sources with HVCs in their \OIa\ profiles. (b): LVC \OIa\ luminosities  vs accretion luminosities for other  samples discussed in this work: black filled squares for the 1-3\,Myr SFB, gray filled triangles  for the 3-5\,Myr NGC~2264, and  olive pluses  for the 1-3\,Myr Cha+Lupus population. In each panel,  the solid line is the best linear fit to all LVC detections for all the sample shown in  both the panels. The dashed lines mark a $\pm$1~dex difference from the solid line.  
}\label{Fig:LVC_acc}
\end{center}
\end{figure}

\subsection{Relation between the [O~{\scriptsize I}] luminosity and accretion luminosity}\label{sect:OIlumAcc}

Studies of young $\sim 1-3$\,Myr old stars have shown that there exists a decent correlation between the  accretion luminosity and the total \OIa\  line luminosity, as well its  individual HVC and LVC components \citep{2013ApJ...772...60R,2014A&A...569A...5N,2018A&A...609A..87N,2018ApJ...868...28F}. Our older Upper~Sco \OIa{} sample is dominated by LVC, hence we test whether a  correlation  between the \OIa\ LVC luminosity and the accretion luminosity still persists at $\sim 5-10$\,Myr. 

Figure~\ref{Fig:LVC_acc} (a) shows the \OIa{} LVC luminosity vs accretion luminosity for our Upper~Sco sample. Warm and Cool samples are identified by color and CTTS and WTTS by open and closed symbols. Upper limits for 13 CTTS without \OIa{} and for 13 WTTS with \OIa{} are also shown. In general the Upper Sco relation follows the same relation as the younger samples, which are shown in Figure~\ref{Fig:LVC_acc} (b). The WTTS with \OIa{} also follow the relation if the accretion upper limits are equated with a luminosity. An important question is whether these 13 nominal WTTS are misidentified by H$\alpha$ criteria and in fact are weakly accreting, as the CTTS/WTTs boundary is based on the width of this line, which can be ambiguous for diagnosing very low accretion rates (see Fig.~\ref{Fig:WTTS_Halpha_OI}).  In contrast, in the Cha+Lupus sample accretion is measured via the Balmer jump rather than H$\alpha$ and thus includes sources with lower accretion luminosities than Upper Sco.
In contrast, the CTTS without \OIa{} lie somewhat below the LVC vs accretion luminosity relation and may indicate that accretion and \OI are not uniquely coupled. Further discussion of the 13 PD WTTS with \OI\ detection and the 13 CTTS with no \OI\ detection is provided in Appendix~\ref{App:outliers}.

Figure~\ref{Fig:LVC_acc} (b) summarizes results from the 1-3\,Myr-old SFB, the similarly old Cha+Lupus, and the 3-5\,Myr-old NGC~2264 populations. It is apparent that  Upper~Sco  follows the same relation as the younger samples.  The Pearson correlation coefficient is 0.78 for the \OIa{} LVC luminosity vs accretion luminosity of the combined sample including only the CTTSs with \OIa{} detection, with a low probability ($<10^{-16}$) that the two quantities are uncorrelated. We perform a two-variable linear regression including only the detections and obtain the following relation:

\begin{equation}
\label{Equ2:LVC_acc}
Log~L_{\rm OI63, LVC}=(0.56\pm0.03)Log~L_{\rm acc} -(4.09\pm0.06)
\end{equation}

with a slope  similar to those reported in the literature for younger samples, e.g., 0.52$\pm$0.07, 0.59$\pm$0.04, and 0.60$\pm$0.03 in  \cite{2013ApJ...772...60R}, \cite{2018A&A...609A..87N}, and \cite{2018ApJ...868...28F}, respectively.

\subsection{Individual kinematic components}\label{sect:IndividualKinematicComponents}
 The identification of kinematic components within the LVC requires high spectral resolution ($R \ge 25,000$), hence the Cha+Lupus sample will not be included in the following comparison, however the 3-5\,Myr-old NGC~2264 sample can be used.  As we focus on \OIa\ detections, we also include the nominal WTTS with \OIa\ detections. In the discussion that follows we adopt the characterization of the LVC used by  \cite{2019ApJ...870...76B}, designating an LVC either as a BC+NC or SC. This necessitated re-assigning LVC from the literature  previously designated as either BC or NC depending on their FWHM, into the single category SC \citep[e.g.,][]{2016ApJ...831..169S,2018AA...620A..87M}. 
 
  \subsubsection{HVC and LVC detection frequencies}
 High-velocity(HV) \OIa\ emission at v$> 30$\,km/s is typically attributed to jets and low velocity (LVC) to slow disk winds (e.g., \citealt{1995ApJ...452..736H,2000A&A...356L..41L,2000ApJ...537L..49B,2002ApJ...580..336W,2016ApJ...831..169S,2018ApJ...868...28F,2019ApJ...870...76B}). In our 45 \OIa\ detections in  Upper~Sco  we find only 5 sources with HVC, all of which also have LVC. Four of the HVC are among the highest accretion rate CTTS in the sample, as can be seen in Figure~\ref{Fig:LVC_acc} green squares.  Of the total 45 LVC profiles only 7 are BC+NC, 3 of which also have HVC, and the rest are fit with a single Gaussian, 2 of which also have HVC. Again following \cite{2019ApJ...870...76B}, these are separately designated as SCJ  or SC, with 2 of the former and 36 of the latter in Upper Sco. A comparison of profile types in the different regions is summarized in Table~\ref{Table:statistics}.

\begin{figure*}
\begin{center}
\includegraphics[width=0.68\columnwidth]{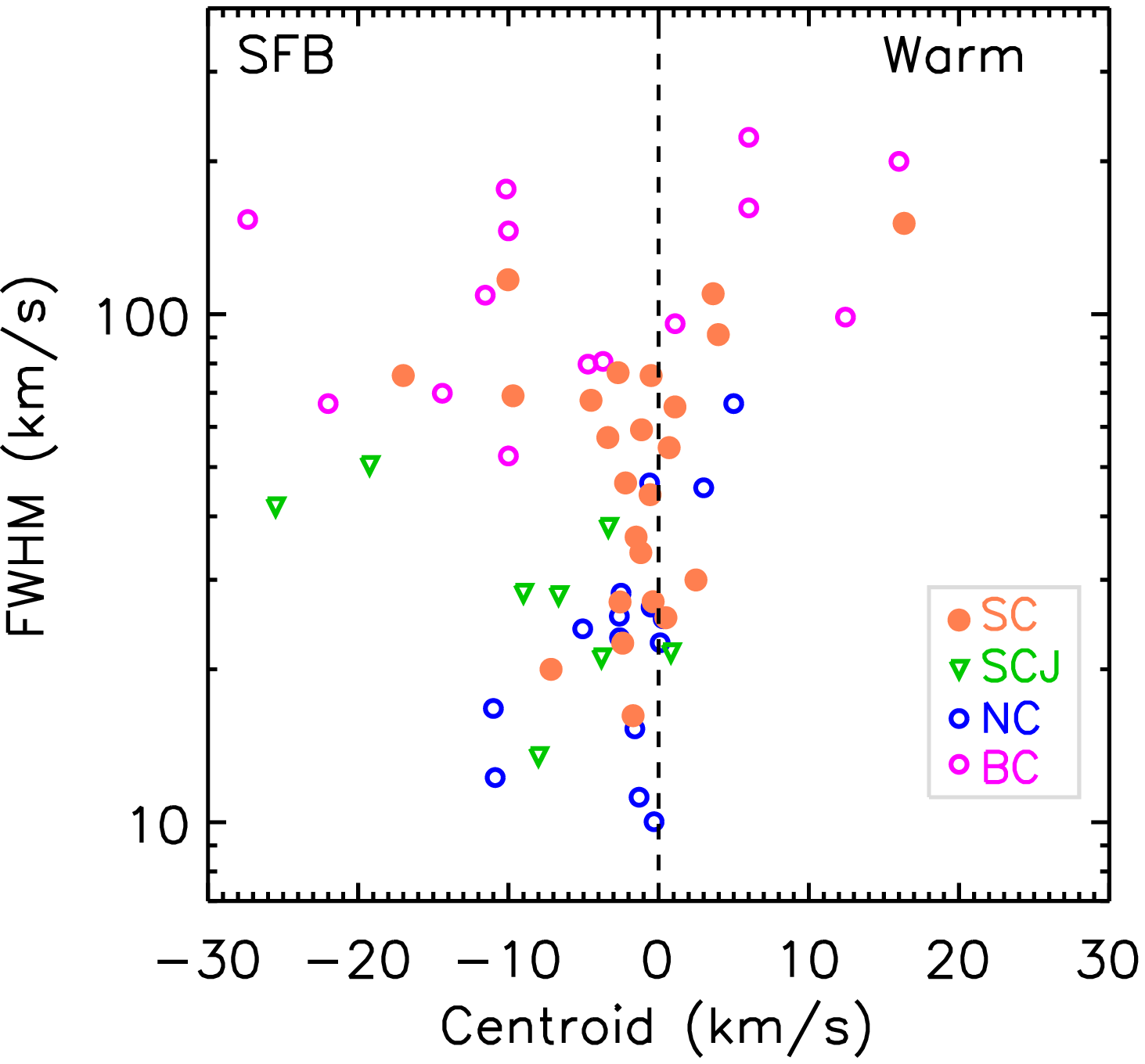}
\includegraphics[width=0.68\columnwidth]{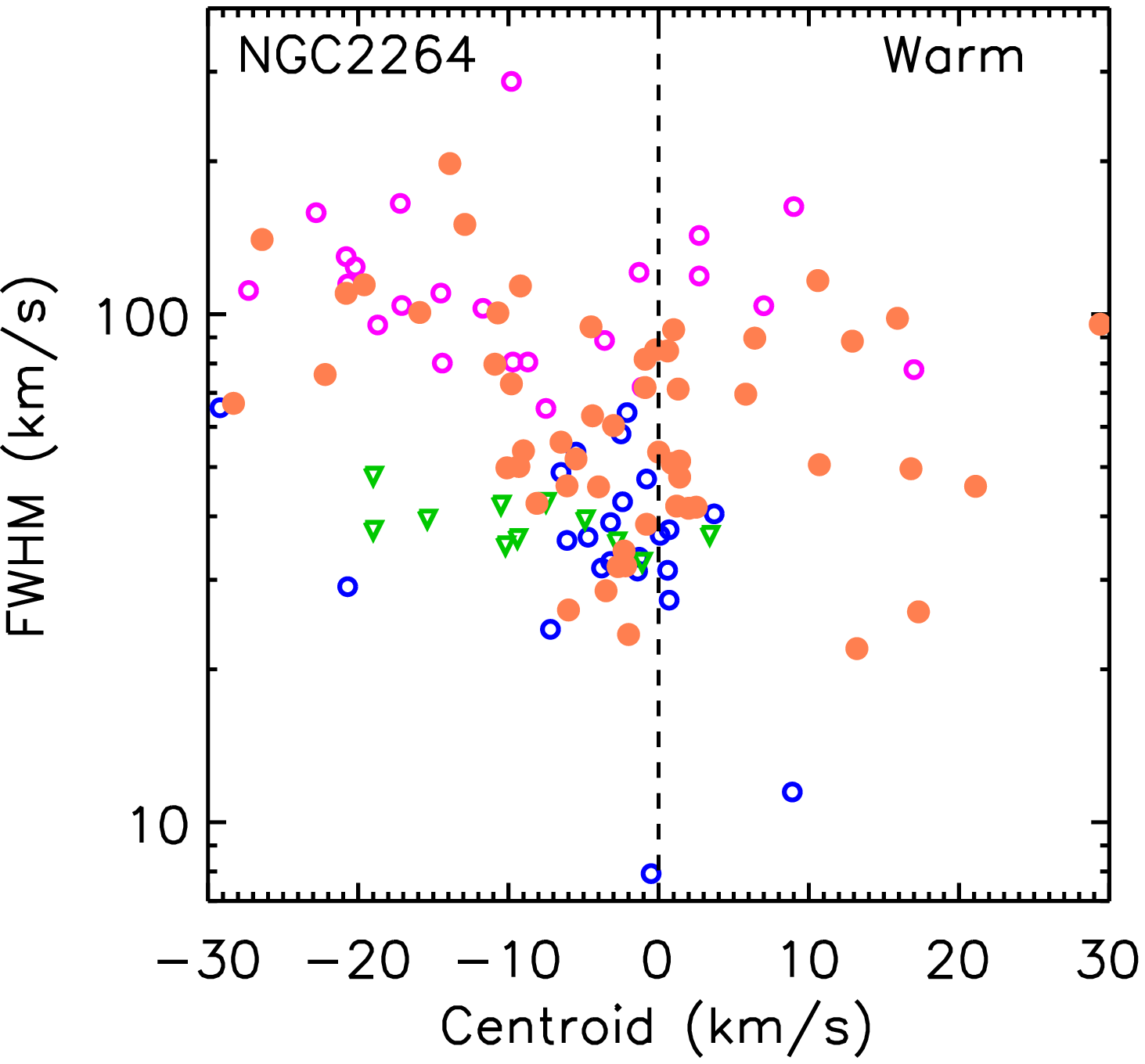}
\includegraphics[width=0.68\columnwidth]{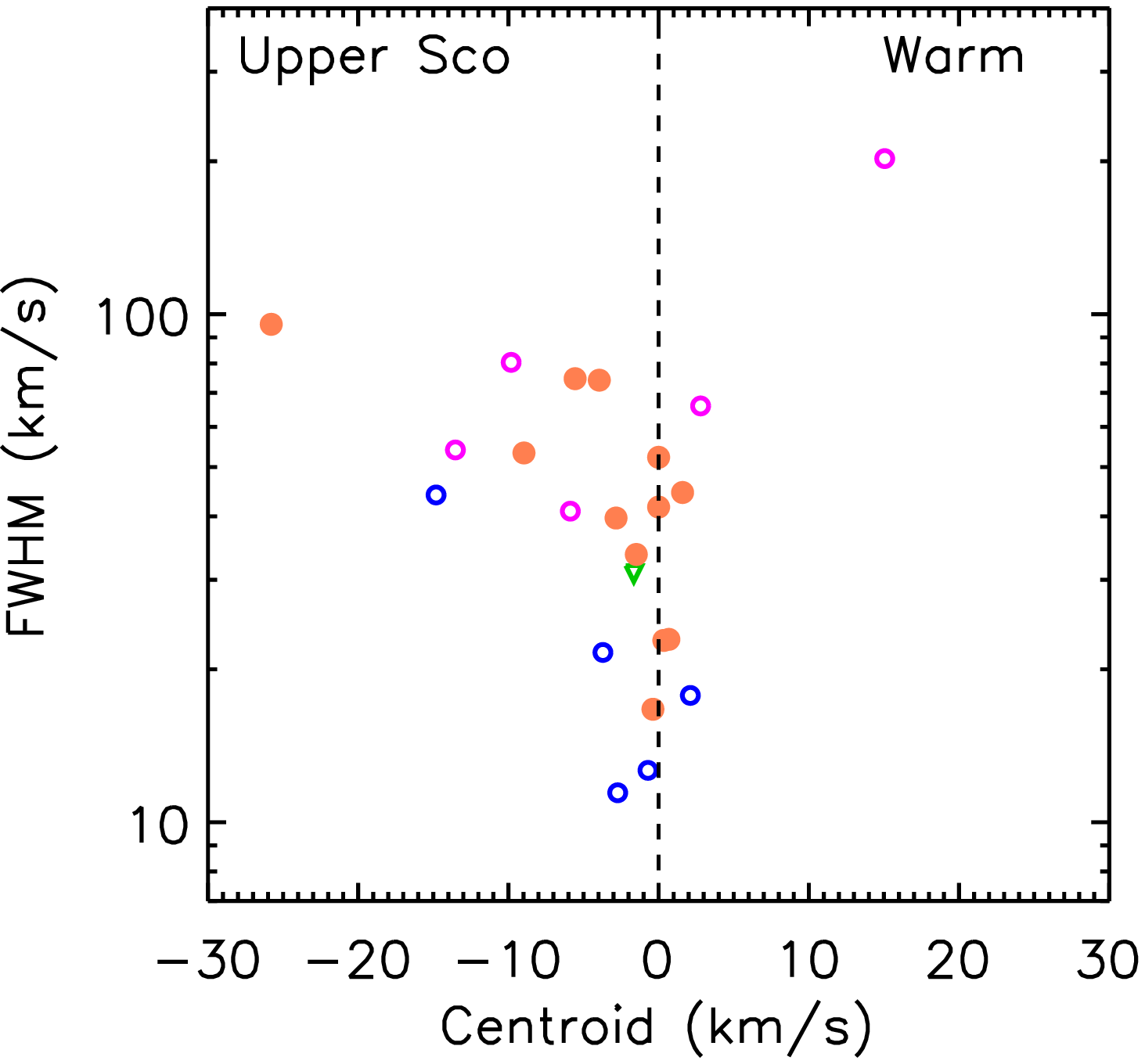}
\caption{FWHM vs. centroids of various LVC components (SC, NC, BC, and SCJ) in the \OIa\ line profiles for warm samples of three populations. In each panel, the orange filled circles are for the SC, blue open circles for the NC, magenta open circles for the BC, and the green upside down triangles for SCJ.} \label{Fig:allvc_FWHM_dis}
\end{center}
\end{figure*}

\setcounter{table}{6}
\begin{table}
\scriptsize
\renewcommand{\tabcolsep}{0.04cm}
\caption{Statistics on the line profiles  including both the CTTS and WTTS}\label{Table:statistics}
\begin{center}
\begin{tabular}{lcccc}
\hline
\hline
 Population    &SC\  & SCJ &BC+NC  &HVC \\
\hline
\multicolumn{5}{c}{Warm Samples (G0-M3)}\\
\hline
SFB&44\%~(24/55) &16\%~~(9/55)  &29\%~(16/55)&47\%~(26/55)\\
NGC~2264&52\%(54/104) &21\%(22/104)  &22\%(23/104)&31\%(32/104)\\
Upper~Sco&67\%~~(12/18) &6\%~~(1/18) &28\%~~(5/18)&22\%~~(4/18)\\
\hline\multicolumn{5}{c}{Cool Sample (M3-M5.2)}\\
\hline
Upper~Sco &89\%(24/27) &4\% (1/27) & 7\% (2/27)  &4\% (1/27)\\
\hline
\end{tabular}
\end{center}
\tablecomments{The only Cool sample that can be compared with Upper~Sco is Cha+Lupus but the spectral resolution of this latter dataset is too low to distinguish different line profiles (see also Sect.~\ref{sect:IndividualKinematicComponents}), hence it is excluded here} 
\end{table}

 The comparison in Table~\ref{Table:statistics} shows that the frequency of HVCs decreases while that of SCs increases with age. We argue that this trend is actually due to the mass accretion rate which, overall, decreases with age \citep[e.g., Table~\ref{Table:csamples} and][]{Hartmann2016ARA&A..54..135H}.  For example among the Warm sample, in Upper~Sco (5-10\,Myr-old) the median $\dot{M}_{acc}=1.1\times10^{-9}~M_{\odot}~yr^{-1}$ while in SFB (1-3\,Myr-old) and NGC~2264 (3-5\,Myr-old)  the medians are $\dot{M}_{acc}=3.6\times10^{-9}~M_{\odot}~yr^{-1}$ and $\dot{M}_{acc}=1.0\times10^{-8}~M_{\odot}~yr^{-1}$. In fact, within the three samples we also see that the median $L_{acc}$ for SC sources is always  lower than that of BC+NC.  For the  Upper~Sco Warm sample  median $\dot{M}_{acc}$ are $3.5\times10^{-9}$, $5.9\times10^{-9}$, and $1.1\times10^{-9}$~$M_{\odot}~yr^{-1}$ respectively for sources with HVC, BC+NC, and SC profiles. Thus, Upper~Sco sources also follow the general trend in the \OIa\ profile which simplifies and loses first the HVC and then the BC+NC as the mass accretion rate declines (e.g., Sect.~2.3 in \citealt{2022arXiv220310068P}).

\subsubsection{Distribution of LVC centroids and line widths}\label{sect:SCcentroids}

To further understand the evolution of the character of the LVC with age/accretion rate we show in   Figure~\ref{Fig:allvc_FWHM_dis}  the relation between the FWHM and centroid velocities of the BC, NC, SC, and SCJ for the warm samples in SFB, NGC~2264, and Upper~Sco, where BC and NC are only from profiles with BC+NC. Because observations were carried out at slightly different spectral resolutions, we deconvolve all observed FWHMs for instrumental broadening taking a Gaussian  FWHM appropriate for each sample (see  Table~\ref{Table:csamples})\footnote{$FWHM=\sqrt{FWHM_{obs}^2-FWHM_{ins}^2}$, where  $FWHM_{obs}$ is the observed width and $FWHM_{ins}$ is the instrumental broadening}. A summary of the corresponding median FWHMs and centroid velocities  
 are listed in Table~\ref{Table:allFWHM_vc}. 
 
Several trends are apparent from  Figure~\ref{Fig:allvc_FWHM_dis}. In each region, the largest blueshifts are found among the components with the largest FWHM. These are predominantly the BC and SCJ components, but also some of the broader SC (which in previous works would also have been identified as BC). While the older Upper Sco follows the same trend, both the FWHM of the broadest lines and the $v_{\rm c}$  of the most blueshifted lines are less extreme. There are some redshifted components among the BC and broader SC, especially in NGC~2264, which are not well explained in a disk wind scenario, although they could arise from inclination effects \citep{2016ApJ...831..169S}. As seen in Table~\ref{Table:allFWHM_vc}, for each population the {\it median} $v_{\rm c}$ of the SC is comparable to that of the corresponding NCs, i.e. less blue-shifted than BCs and SCJs.

\setcounter{table}{7}
\begin{table}
\scriptsize
\renewcommand{\tabcolsep}{0.04cm}
\caption{Median FWHMs (\kms) and centroids (\kms) of various LVC components}\label{Table:allFWHM_vc}
\begin{center}
\begin{tabular}{lccccc}
\hline
\hline
 Population    &&SC\   &NC & BC & SCJ \\
\hline
\multicolumn{5}{c}{Warm Samples (G0-M3)}\\
\hline
SFB~~~~~~~~  $v_{\rm c}$&    &$-$1.3  &$-$1.3  &$-$7.4 &$-$5.2\\
~~~~~~~~~~~~~~  FWHM &  &55.9  &24.0 &98.7  &28.2 \\
\hline
NGC~2264 $v_{\rm c}$ & &$-$2.3 &$-$2.4 &$-$9.8 &$-$9.8\\
~~~~~~~~~~~~~~  FWHM&    &58.2  &36.4 &110.0 &38.5 \\
\hline
Upper~Sco $v_{\rm c}$&& $-$0.9  &$-$2.7   & $-$5.9 &$-$1.7 \\
~~~~~~~~~~~~~~  FWHM&&43.1  &17.8   &66.0  &30.9 \\
\hline
\hline\multicolumn{5}{c}{Cool Sample (M3-M5.2)}\\
\hline
Upper~Sco $v_{\rm c}$&&$-$0.5  &\nodata  &\nodata   &\nodata \\
~~~~~~~~~~~~~~  FWHM& &67.4   &\nodata   &\nodata   &\nodata \\
\hline
\end{tabular}
\end{center}
\tablecomments{The median values are calculated when there are more than three sources with such types, except for the SCJ (only 1 case) in Upper~Sco. 
}
\end{table}

\begin{figure}
\begin{center}
\includegraphics[width=\columnwidth]{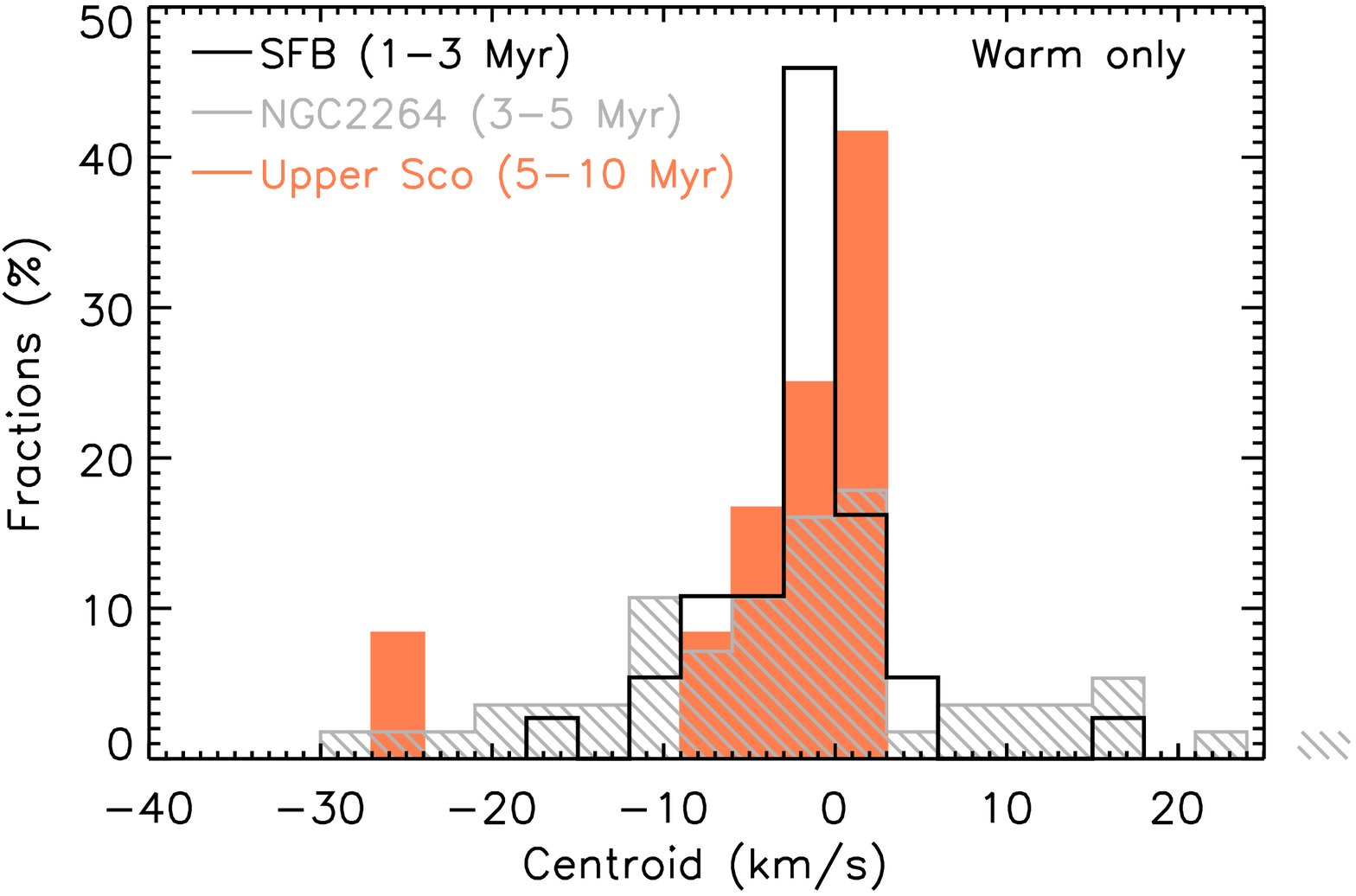}
\includegraphics[width=\columnwidth]{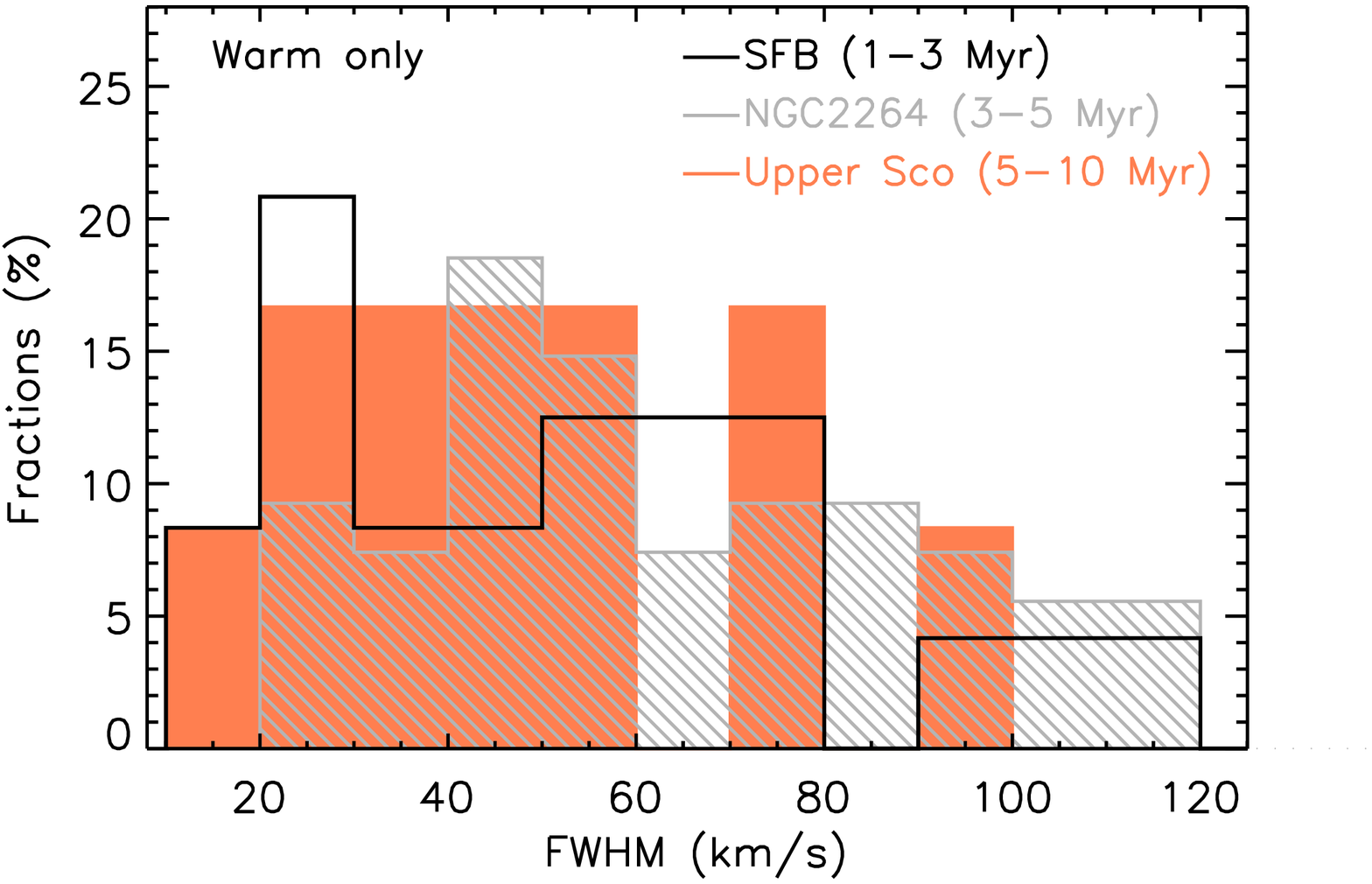}
\caption{Distribution of centroids (Upper panel) and FWHMs   corrected for instrumental broadening (Lower panel) for the \OIa\ SCs around Warm (G0-M3) stars: orange color  filled histogram for Upper~Sco;  black open histogram for the 1-3\,Myr-old SFB; and gray line filled histogram for the 3-5\,Myr-old NGC~2264 population. Centroids and FWHMs are statistically indistinguishable for the three samples, see text for details. } \label{Fig:vc_FWHM_dis}
\end{center}
\end{figure}

\begin{figure}
\begin{center}
\includegraphics[width=\columnwidth]{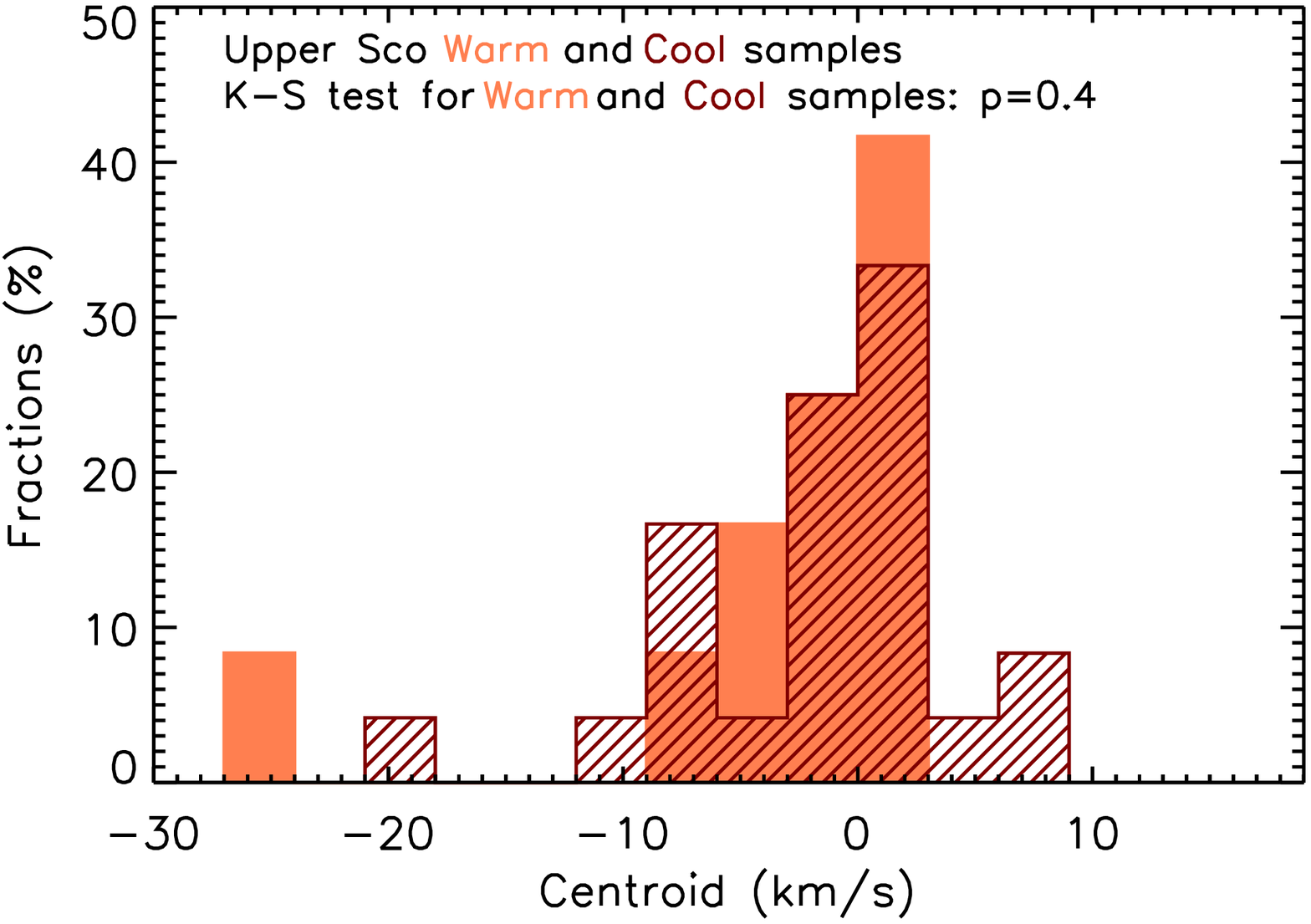}
\includegraphics[width=\columnwidth]{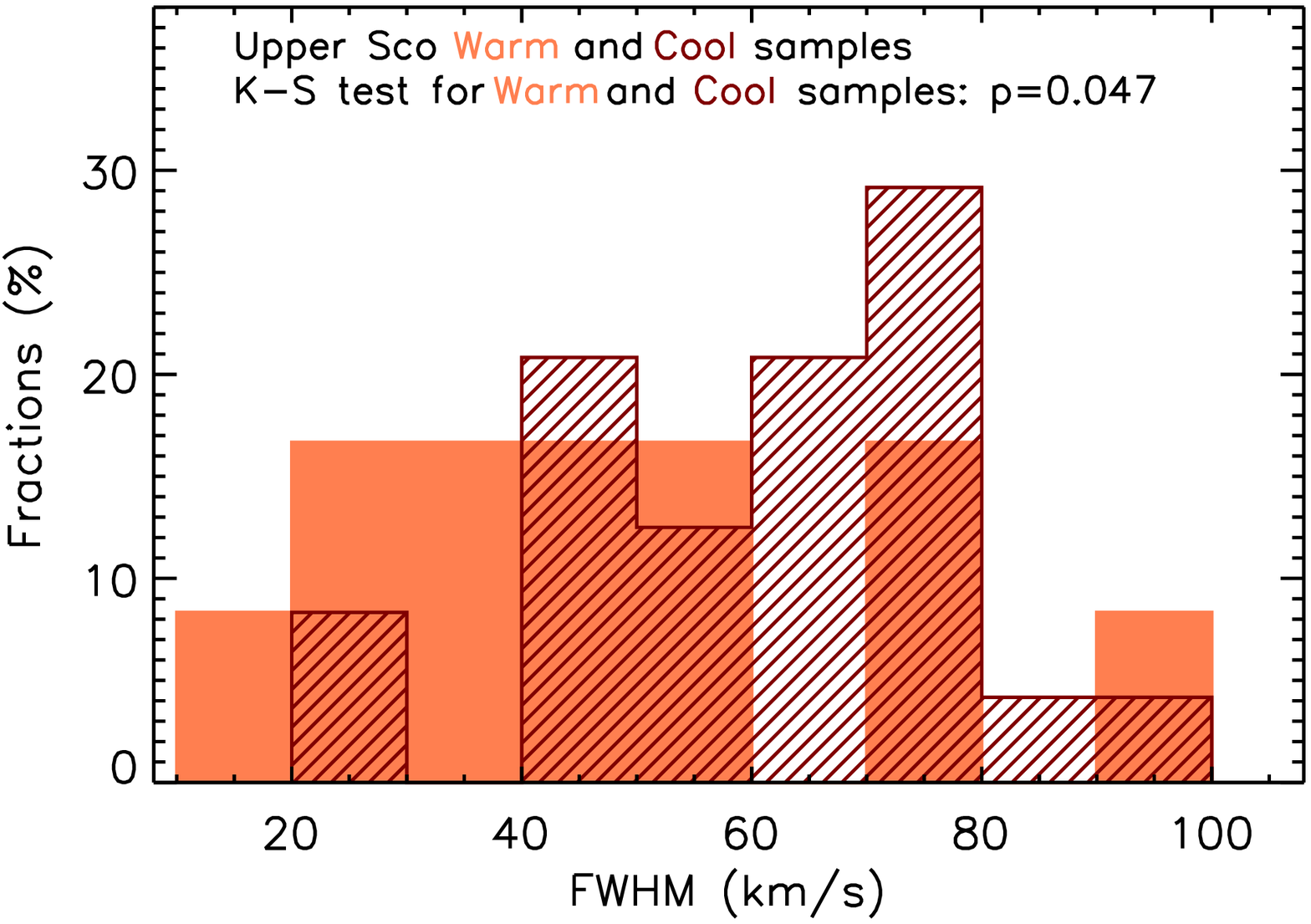}
\caption{Distribution of centroids (Upper panel) and FWHMs  corrected for instrumental broadening (Bottom panel) for the \OIa\ SCs in Upper~Sco (orange for the Warm and maroon line filled histogram for the Cool sample). Only FWHMs for the Warm and Cool samples are statistically different.  \label{Fig:vc_FWHM_dis_us} }
\end{center}
\end{figure}

 Since the majority of \OIa{} in Upper~Sco are SC profiles, especially among the Cool Sample, we confine the rest of the comparison to the SC parameters among the different populations. First we explore whether SC centroids  ($\rm{v_c}$) and FWHMs change with spectral type and age. Figure~\ref{Fig:vc_FWHM_dis}  compares distributions of the $\rm{v_c}$ and FWHMs corrected for instrumental broadening for the Warm sources in the  1-3\,Myr-old SFB, the  3-5\,Myr-old  NGC~2264, and  our 5-10\,Myr-old Upper Sco sample.  As summarized in Table~\ref{Table:allFWHM_vc},  the median $\rm{v_c}$ and FWHM for the three populations are: $-$1.3\kms\ and 55.9\,\kms\ for SFB,  $-$2.3\kms\ and 58.2\,\kms\ for NGC~2264, and   $-$0.9\kms\ and 43.1\kms\ for Upper~Sco. Although the median values for Upper~Sco are lower than in the younger regions, the K-S test returns high probabilities that the $\rm{v_c}$ and FWHMs are drawn from the same parent population (p$=0.27-0.97$ and 0.16$-$0.43, respectively).

Figure~\ref{Fig:vc_FWHM_dis_us} focuses on comparing the Warm and Cool samples in Upper Sco, showing the distribution of SC $\rm{v_c}$ and FWHMs corrected for instrumental broadening,  including both the CTTS and nominal WTTS with \OIa{} detections. The median $\rm{v_c}$ and FWHM are  $-$0.9\,\kms\ and 43.1\,\kms\ for the Warm sample while they are somewhat less blueshifted, $-$0.5\,\kms\ , and somewhat broader, 67.4\,\kms\ in the Cool sample.
 The K-S test returns a high probability that the SC centroids are drawn from the same parent population (p$=0.4$). In contrast, the K-S probability for their FWHMs is low (p$=0.047$) suggesting that Cool sources do have broader \OIa\ lines than Warm ones.

These results suggest that the \OIa{} in Upper~Sco, which predominantly have a SC profile, is also tracing a slow disk wind as proposed for the young populations. We also note that, within each Upper~Sco  sample (Warm or Cool), CTTSs and WTTSs have indistinguishable distributions of centroids and FWHMs, suggesting that their \OIa\ lines trace the same wind.

Next, we explore relations between the SC FWHM and disk inclination. \cite{2019ApJ...870...76B} reported a positive linear correlation between these two quantities for their mostly young sample of T~Tauri stars with SpTy from M5 to G8. We repeat the test for all the sources with disk inclination in Upper~Sco, SFB, and NCG~2264 (see Table~\ref{tabe_Mwind}) but correct their FWHMs for instrumental broadening given the different spectral resolutions of the surveys.  For SFB, which is a subset of \cite{2019ApJ...870...76B}, we find a Pearson's correlation coefficient of 0.43 and a probability of 6\%, slightly above the 5\% cut that is typically adopted to indicate a correlation. The difference with the \citet{2019ApJ...870...76B} result can be attributed to the sample, here restricted to stars with SpTy no later than M3 and ages $\sim 1-3$\,Myr. The Warm NGC~2264 
sample  does not show any hint for a positive  correlation (even when we exclude the cluster of disks with inclinations of more than 80$^\circ$, see Figure~\ref{Fig:newFWHMvsdiskincli}): the Pearson's correlation coefficient and probability are $-$0.40 and 50\% (0.26 and 16\% when removing the disks with inclination larger than 90$^\circ$). While the Warm Upper~Sco sample is too small for such a test, the SC FWHM and disk inclination for the larger Upper~Sco Cool sample are also not correlated (Pearson's correlation coefficient and probability of 0.25 and 31\%). 
We note however that Upper~Sco disk inclinations  have significant uncertainties, due to the relatively low sensitivity and resolution of the ALMA survey, while the NGC~2264 system inclinations were deduced from the star's rotation properties. Interestingly, when combining all samples in Figure~\ref{Fig:newFWHMvsdiskincli} and removing the suspicious clusters of disks at high inclinations ($> 80^\circ$), the Pearson's correlation coefficient and probability are 0.34 and 0.6\%, hinting at a positive relation between the SC FWHM and disk inclination as reported in \citet{2019ApJ...870...76B}.

When normalizing the FWHM  by stellar mass, the relation between disk inclination and the SC FWHMs becomes more obvious (Figure~\ref{Fig:FWHMvsdiskincli}). As for the young  populations \citep{2016ApJ...831..169S,2018ApJ...868...28F},  the SC FWHM in Upper~Sco increases with  disk inclination,  suggesting that Keplerian
broadening contributes significantly to the line widths.
Compared to Warm sources, Cool ones in Upper~Sco tend to have larger normalized FWHMs, hinting at smaller emitting radii. 
Therefore, we estimate emitting radii and look for any difference with stellar mass.  Following  \cite{2018ApJ...868...28F}, we calculate the emitting radius at the base of the wind ($r_{\rm base}$) from half of the deprojected \OIa\ FWHM assuming Keplerian rotation. Figure~\ref{Fig:FWHM_line} provides the distribution of $r_{\rm base}$ for these samples and highlights the smaller values for Cool stars: the median $r_{\rm base}$ for the Warm USco sample is 1.2\,au while that for the Cool sample is 0.2\,au. 
These radii are well within the so-called gravitational radius, the radius at which the sound speed equals the Keplerian orbital speed. Hence, these winds are not thermal/photoevaporative in origin (see also \citealt{2016ApJ...831..169S}).

\begin{figure}
\begin{center}
\includegraphics[width=\columnwidth]{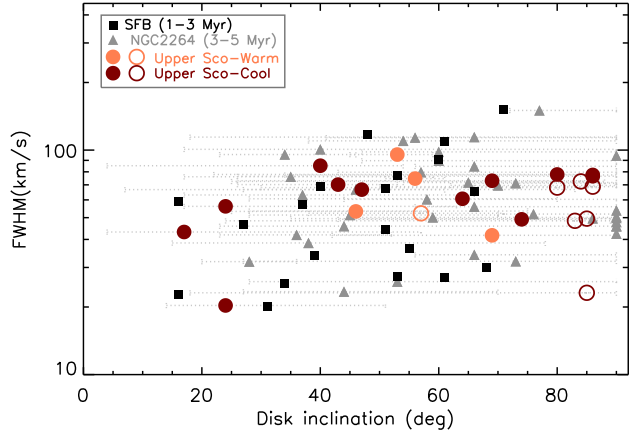}
\caption{SC \OIa\  FWHM vs.  disk inclination for our Upper~Sco sample (circles, orange for Warm and maroon for Cool sources, filled for CTTS and open for WTTS), the SFB 1-3\,Myr sample (black filled squares), and the 3-5\,Myr NGC~2264 population (gray filled triangles). FWHMs are corrected for instrumental broadening.} 
\label{Fig:newFWHMvsdiskincli}
\end{center}
\end{figure}

\begin{figure}
\begin{center}
\includegraphics[width=\columnwidth]{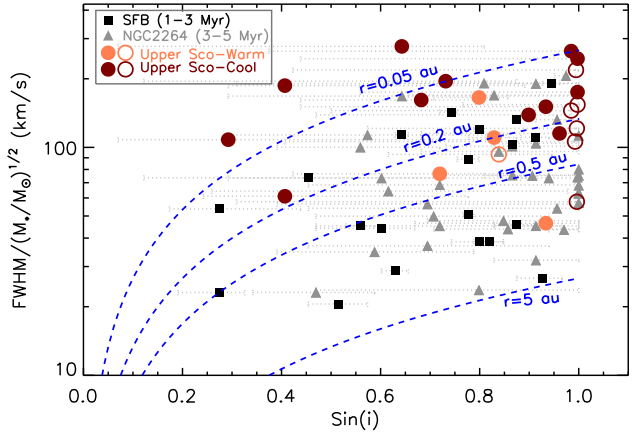}
\caption{Same as Fig.~\ref{Fig:newFWHMvsdiskincli} but that the  FWHMs are corrected for instrumental broadening and normalized by stellar masses, and disk inclinations is shown in sin function. 
 Blue dashed lines show Keplerian FWHM as a function of disk inclination at radii of 0.05, 0.2, 0.5, and 5\,au.} 
\label{Fig:FWHMvsdiskincli}
\end{center}
\end{figure}

\begin{figure}
\begin{center}
\includegraphics[width=\columnwidth]{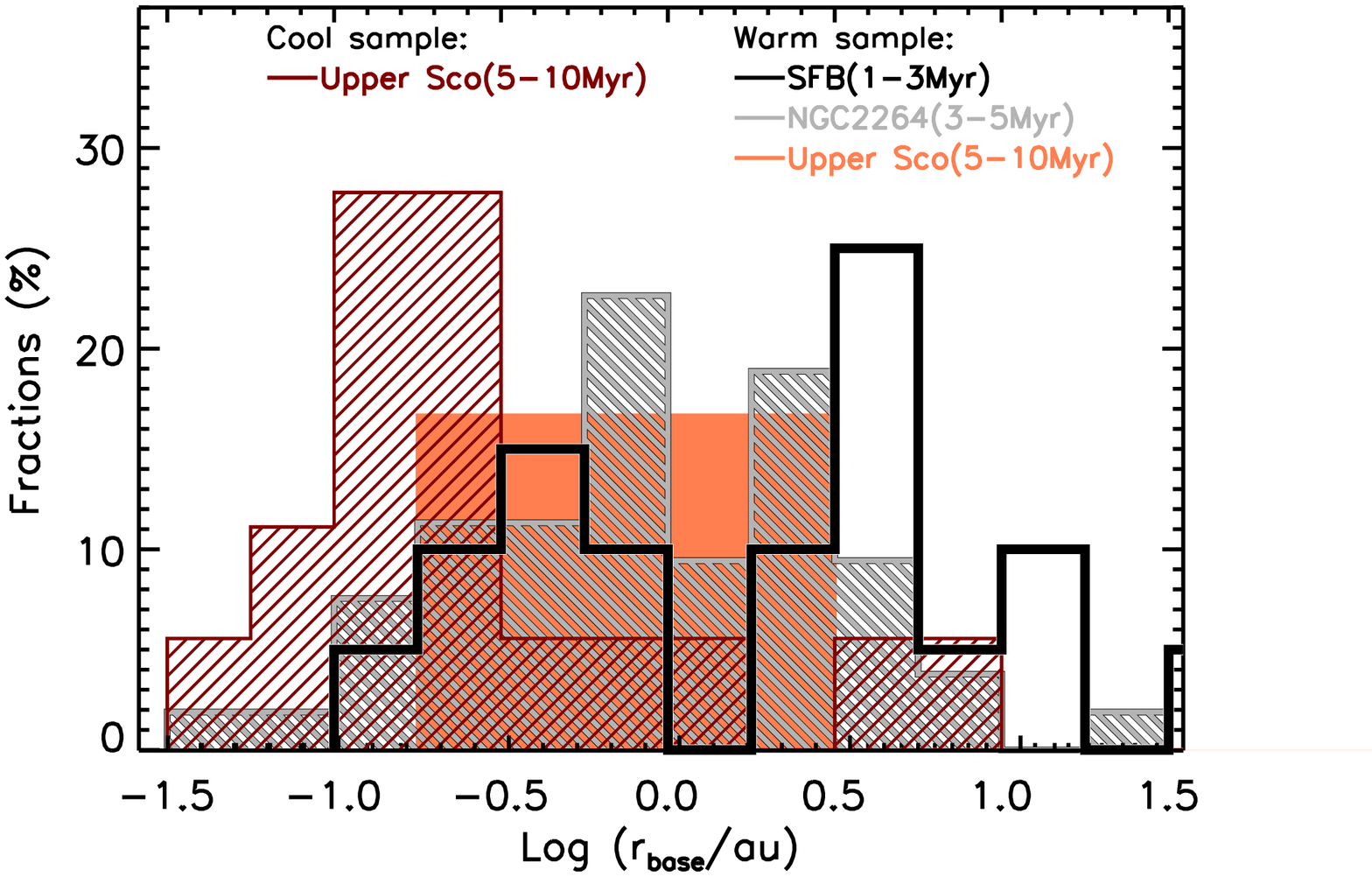}
\caption{Distribution of emitting radii at the base of the wind for SFB (black open histogram), NGC~2264 (gray line-filled histogram), and Upper~Sco (orange color/maroon line-filled histogram for Warm/Cool sources) disks with known inclinations, see Table~\ref{tabe_Mwind}. Cool sources have smaller emitting radii than Warm sources.
 }\label{Fig:FWHM_line}
\end{center}
\end{figure}

\subsection{Relations with infrared spectral index}\label{sect:IRindex} 
Previous work on $\sim 1-3$\,Myr-old stars identified relations between the \OIa\ EW/line luminosity as well as its FWHM and the infrared spectral index. In particular, \cite{2019ApJ...870...76B} showed that the SC \OIa\ EW is anti-correlated with the $n_{13-31}$ index and  \cite{2020ApJ...903...78P} reported a likely anti-correlation with the luminosity for a subset of the sample. Furthermore, \cite{2019ApJ...870...76B} found that the \OIa\ FWHM is  anti-correlated with $n_{13-31}$.  These results have been interpreted as co-evolution of inner disks (larger $n_{13-31}$ indicate more depleted inner dust disks) and winds (lower \OIa\ luminosities and smaller FWHM indicate less dense and further out winds). As 80\% of the \OIa\ detections in Upper~Sco are SCs, we explore whether such relations persist at 5-10\,Myr. However, as most Upper~Sco sources lack low-resolution {\it Spitzer} spectra to compute the $n_{13-31}$ spectral index, we use instead the WISE W3 (12\,\micron) and W4 (22\,\micron) bands to generate an index as close as possible to the  $n_{13-31}$.

\begin{figure}
\begin{center}
\includegraphics[width=\columnwidth]{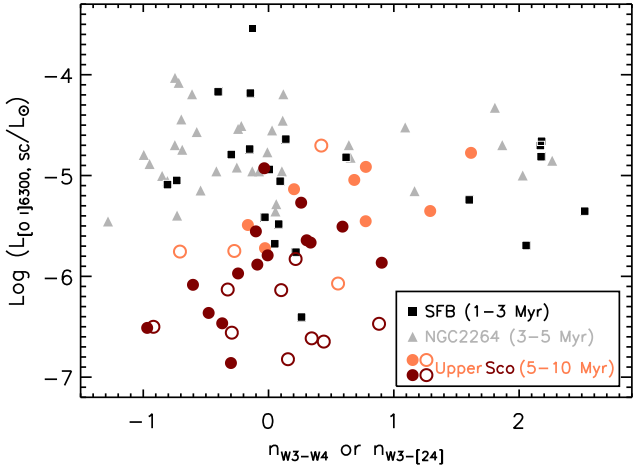}
\caption{SC \OIa\  luminosity vs. infrared spectral index for our Upper~Sco sample (circles, orange for Warm and maroon for Cool sources, filled for CTTS and open for WTTS), the SFB 1-3\,Myr sample (black filled squares), and the 3-5\,Myr NGC~2264 population (gray filled triangles).The SC \OIa\  luminosity is not correlated with these spectral indices.}\label{Fig:SCLineLumvsSpectralIndex} 
\end{center}
\end{figure} 

Figure~\ref{Fig:SCLineLumvsSpectralIndex} shows the \OIa\ line luminosity vs.  $n_{\rm W3-W4}$  for our SC Upper~Sco sample. As noted above, the 
Upper~Sco sample is dominated by Cool stars and it is apparent from this figure that their \OIa\ luminosity is lower than that of Warm stars and their disks cover a narrower 
$n_{\rm W3-W4}$ range. The Pearson's r-test lends high probabilities that the two quantities are uncorrelated, 0.07 and 0.21 for the Warm and Cool samples, respectively. 

In the same figure we also overplot the 1-3\,Myr-old SFB (black squares) and the 3-5\,Myr-old NGC~2264 (gray triangles) samples for which we have computed their spectral indices. 
Because of the large distance of  NGC~2264 and possible contamination in the longest wavelength W4 band, we prefer the {\it Spitzer} 24\,$\mu$m photometry \citep{2009AJ....138.1116S} for this sample. The spatial resolution of {\it Spitzer}  at 24\,$\mu$m is two time higher than that of WISE/W4. Therefore for NCG~2264 we  calculate the $n_{\rm W3-[24]}$ spectral index. The figure shows that the younger samples tend to have less dust-depleted disks and  overall brighter \OIa\ luminosities than the Upper~Sco Warm sample. The Pearson's r-test on the young samples also gives a high probability (0.12) that the two quantities are uncorrelated suggesting that a spectral index covering longer wavelengths, like $n_{13-31}$, can better link the depletion of inner disks with the \OIa\ luminosity.

\begin{figure}
\begin{center}
\includegraphics[width=\columnwidth]{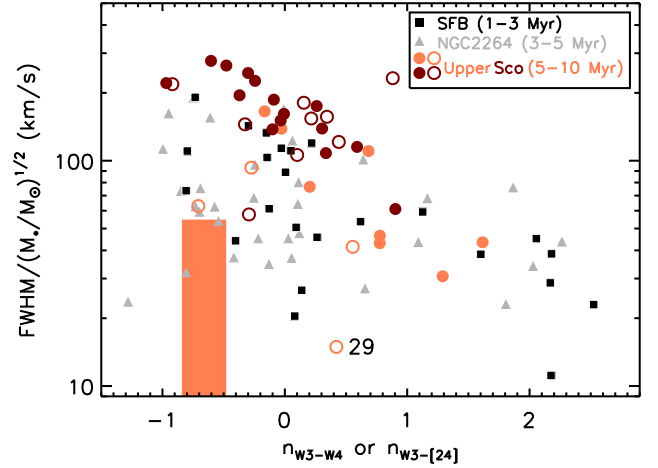}
\caption{SC \OIa\  FWHM vs. infrared spectral index. FWHMs are corrected for instrumental broadening and normalized by stellar masses. Symbols are the same as in Figure~\ref{Fig:SCLineLumvsSpectralIndex}.
The two quantities are anti-correlated.
} 
\label{Fig:FWHMvsSpectralIndex}
\end{center}
\end{figure}

Figure~\ref{Fig:FWHMvsSpectralIndex} shows the \OIa\ FWHM normalized by stellar mass (see Sect.~\ref{Sect:spt}) vs. the $n_{\rm W3-W4}$ spectral index for our SC Upper~Sco sample. The 1-3\,Myr-old SFB  and the 3-5\,Myr-old NGC~2264 populations are also overplotted. Stellar masses for these samples are from the references in Table~\ref{Table:csamples}.
Two trends are apparent: i) the Cool sample has larger FWHM normalized by stellar mass than the Warm sample and ii) the Cool and Warm samples follow the same anti-correlation between the logarithm of the FWHM normalized by stellar mass and spectral index reported for the younger samples. The Pearson's correlation coefficients and probabilities for Upper~Sco are -0.49 (0.11) and -0.48 (0.02) for the Warm and Cool samples, respectively. The  insignificant correlation for the Warm sample is due to  source ID~29. This source has a disk with a large cavity (radius$\sim$70\,au) and a disk inclination  close to face-on (6$^{\circ}\pm 1.5^{\circ}$, \citealt{2017ApJ...836..201D}), and thus shows a narrow \OIa\  profile. Excluding this source,  the Pearson's correlation coefficient and probability is  $-$0.63 and 0.036 for the Warm Upper~Sco sample. The same test on the younger samples combined lends a Pearson's coefficient and probability of $-$0.50 and  3$\times10^{-5}$. Hence, there is a statistically significant anti-correlation between the FWHM and the infrared spectral index.

\section{\OI\ results in broader context}\label{Sect:discussion}

\subsection{The coevolution of accretion and disk winds}
The past ten years have seen magnetized disk winds re-emerging as the prime driver of accretion, hence disk evolution, in what appears to be mostly low turbulence disks \citep[e.g.,][]{2014prpl.conf..411T,Lesur2022arXiv220309821L,Pinte2022arXiv220309528P}. In this paradigm, whenever MHD winds are present, stars accrete disk gas and the two phenomena, disk winds and accretion, are expected to coevolve. Upper~Sco is a particularly interesting region to test this paradigm given its relatively old age and significant evolution of its disk population.

The analysis carried out in  previous sections shows that the \OIa\ line is a good MHD disk wind tracer in Upper~Sco sources. First,  the distribution of velocity centroids for the \OIa\ SC profiles is asymmetric with a tail on the blueshifted side  and a median of -0.9\,km/s indicative of slow outflowing gas. Second, most SC FWHMs are broad enough (median of $\sim 43$\,km/s) to be incompatible with a thermal wind (e.g., \citealt{2016ApJ...831..169S} and Sect.~\ref{sect:SCcentroids}). Furthermore, the distribution of SC velocity centroids and FWHMs from Upper~Sco are indistinguishable from the full SC Warm sample, which covers an age range from $\sim 1-3$\,Myr through to $\sim 5-10$\,Myr.

As expected in the wind-driven accretion paradigm, we find that the well established correlation between the \OIa\ luminosity and  $L_{\rm acc}$ seen in younger regions is maintained by the older stars in Upper Sco, which have overall lower accretion rates. In the  Warm Upper~Sco sample the median $L_{\rm acc}$  is a factor of $\sim 3$ lower than Warm stars in SFB and Cha+Lupus (Sects.~\ref{sect:DetectionRates} and~\ref{sect:OIlumAcc}) and the Cool Upper Sco sample further extends this relation to the lowest $L_{\rm acc}$  (e.g., Figures~\ref{Fig:ACCDIS} and \ref{Fig:LVC_acc}). Thus, the connection between disk winds and accretion  persists at $\sim 5-10$\,Myr for stars that are still  surrounded by accreting disks.

In addition to the \OIa\ luminosity declining with $L_{\rm acc}$, the character of the \OIa\ profiles evolves from a high frequency of HVC (jets) and LVC structure with both BC+NC at high $L_{\rm acc}$ to predominantly SC profiles, as seen here in Upper Sco and previously described in \cite{2022arXiv220310068P}. The transition of the LVC from BC+NC to SC is not yet understood and likely holds clues to the character of the disk wind as the accretion rate drops.

\subsection{Wind mass loss versus mass accretion rates}
Given that the \OIa\ emission appears to trace a wind throughout disk evolution (e.g., Figure~\ref{Fig:vc_FWHM_dis}), here we use its properties to estimate wind mass loss rates  from the modestly blueshifted SC components and explore  whether there is any evidence of evolution in the wind mass loss vs. stellar accretion rate in these slow winds.

Following \cite{2018ApJ...868...28F} we assume that the \OIa\ line is optically thin, traces gas at 5,000\,K, and take an oxygen abundance of $3.2 \times 10^{-4}$. With these assumptions the wind mass loss rate can be written as:
\begin{equation} 
    \dot{M}_{\rm loss} = 2.4\times10^{-4} \left( \frac{V_{\rm wind}}{10\,km/s} \right) \left( \frac{1\,au}{l_{\rm wind}} \right)  \left( \frac{L_{6300}}{L_\odot} \right) \, M_\odot /yr
\label{eq:Mloss}    
\end{equation}

\noindent where $L_{6300}$ is the \OIa\ luminosity while $V_{\rm wind}$ and $l_{\rm wind}$ are the wind velocity and  height, respectively. As in \cite{2018ApJ...868...28F}, the $V_{\rm wind}$ of each sub-sample is taken as the median value of the \OIa\ centroids. In particular here we focus on the SCs and obtain $-$0.5\kms\ and  $-$0.9\kms\ for the Upper~Sco Warm and Cool samples, $-$1.5\kms\ for the Warm SFB sample\footnote{Note that the median \OIa\ NC and BC SFB centroids are different, $-$2.6\kms\ and  $-$9.7\kms\, respectively  \citep{2018ApJ...868...28F}}, and $-$2.7\kms\ for the Warm NGC~2264 sample. 
The wind height is not constrained via high-resolution spectroscopy alone. Hence, as in \cite{2018ApJ...868...28F}, we assume that it is a factor $f$ of the wind base ($r_{\rm base}$). This also means that the only sources for which we can estimate wind mass loss rates are those with a known disk inclination, i.e. those summarized in Table~\ref{tabe_Mwind}. 

\begin{figure}
\begin{center}
\includegraphics[width=\columnwidth]{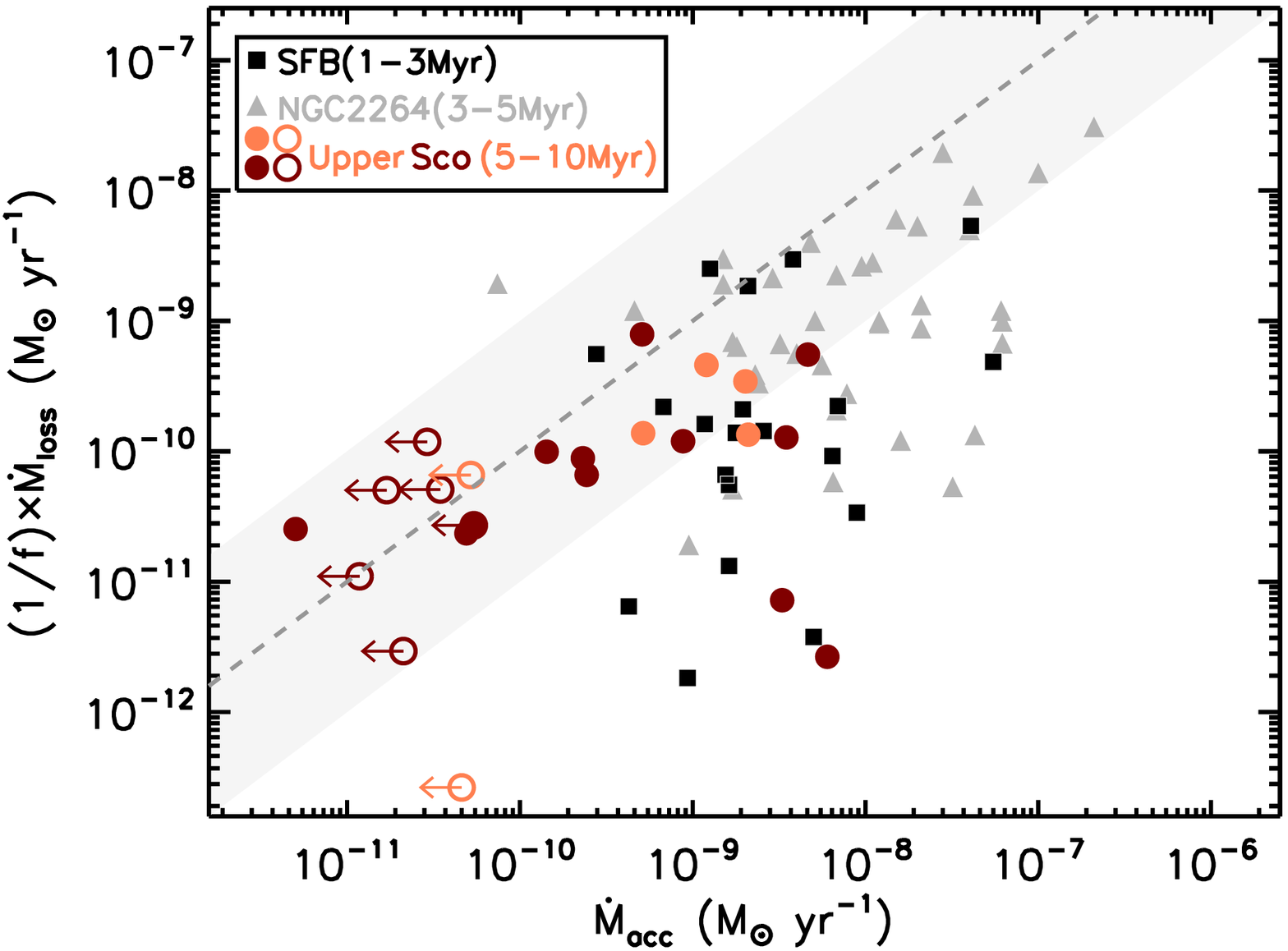}
\includegraphics[width=\columnwidth]{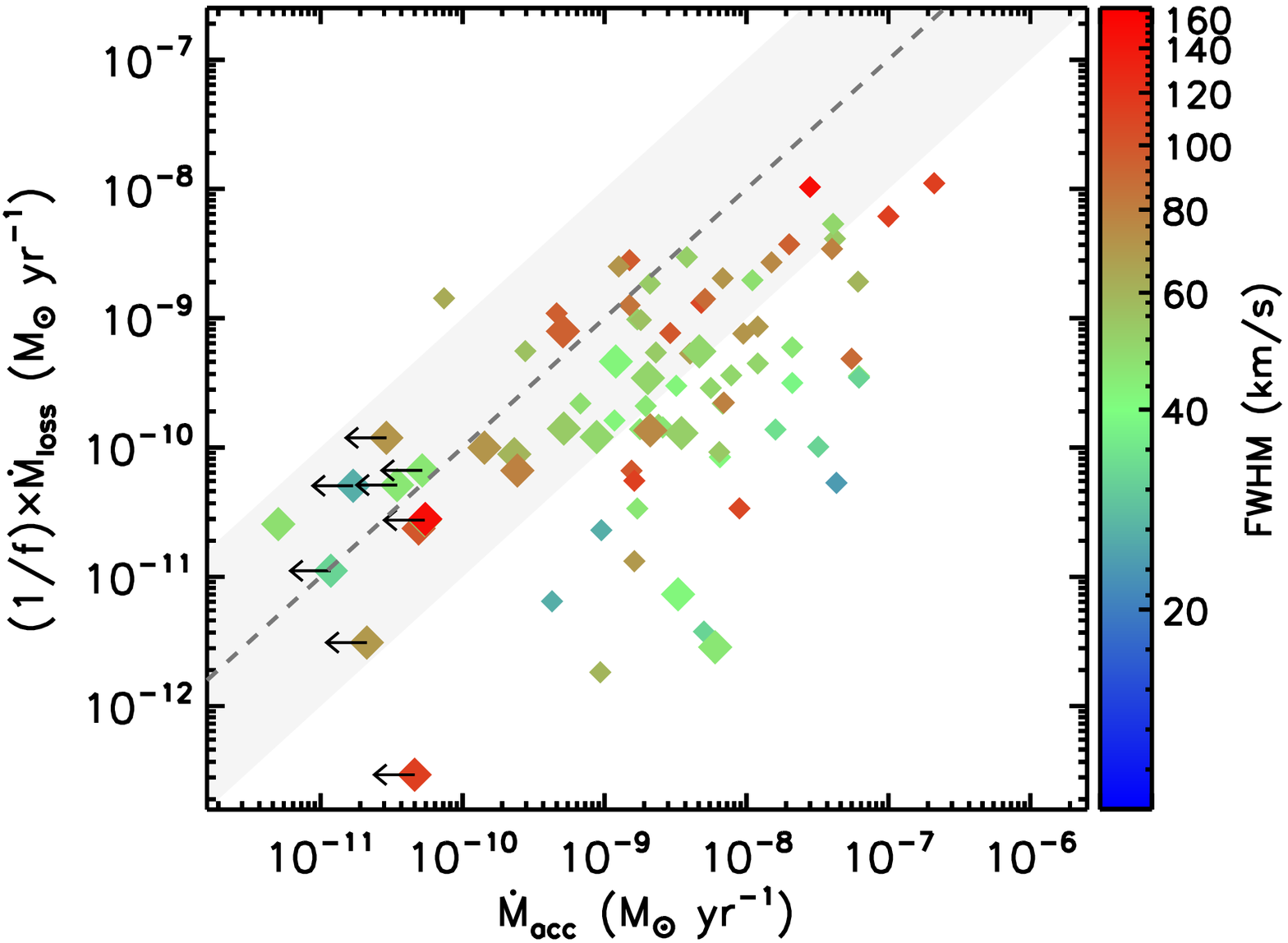}
\caption{Top: $\dot{M}_{\rm loss}$ vs. $\dot{M}_{\rm acc}$ for the three populations. The symbols are same as in Figure~\ref{Fig:SCLineLumvsSpectralIndex}. The dash line mark the y=x/10 relation, and the gray shaded region covers one order of mag difference on both sides around the y=x/10 relation. Bottom: Same as the top panel, but the symbols are coded with the \OIa\ FWHMs.}\label{Fig:Lacc_Mloss}
\end{center}
\end{figure}

Figure~\ref{Fig:Lacc_Mloss} shows the  wind mass loss rate assuming $f=1$ as a function of the stellar mass accretion rate. For a wind vertical extent that is 10 (0.1) times $r_{\rm base}$, each point would shift down (up) by an order of magnitude. The dash line marks the one-to-one relation and the gray shaded region covers one order of magnitude difference on both sides. About 54\% (43/84)  of the sources fall within the gray shaded region  where  wind mass loss rates roughly match accretion rates. 
As reported in the Protostars and Planets VII chapter on disk winds, a similar ratio between ejection and accretion is found for  the Class 0/I and II sources with spatially resolved outflows  \citep[][for a recent review]{2022arXiv220310068P}.  We note that the older Upper~Sco sources tend to have higher  ratios between  wind mass loss rates and accretion rates than younger samples. Among the Upper~Sco sample, 79\% (19/24) of the source fall within the gray shaded region in Figure~\ref{Fig:Lacc_Mloss}, while among the younger sources there are only 51\% of sources located within the shaded region and 47\% of them have lower ratios (less than 0.01).
We also note that the CTTS in Figure~\ref{Fig:Lacc_Mloss} with much lower $\dot{M}_{\rm loss}/\dot{M}_{acc}$ tend to have narrower SC profiles ($< 40$\,km/s), hence larger $r_{\rm base}$ and $l_{\rm wind}$ which lower $\dot{M}_{\rm loss}$ (Equation~\ref{eq:Mloss}). This ratio could be higher if  in these sources the bulk of the wind is molecular, hence any atomic tracer would only provide a lower limit to the mass loss rate \citep[e.g.,][]{2020ApJ...903...78P}. 
More work needs to be carried out to understand these high accreting stars with low $\dot{M}_{\rm loss}$ based on the \OIa\ line.

\subsection{\OIa\ emitting radii and dust disk evolution}
While covering a narrower $n_{\rm W3-W4}$ spectral index than Warm stars, Cool stars display the same anti-correlation between this index and the \OIa\ FWHM corrected for stellar mass (Figure~\ref{Fig:FWHMvsSpectralIndex}). In the context of the Warm sample, \cite{2009ApJ...703.1964F} find that $n_{\rm 13-31} \ge 1$ indicate disks with a large dust inner cavity, also called transition disks. Therefore, the anti-correlation has been interpreted as launching radii moving outward as the  dust inner disk is depleted \citep{2019ApJ...870...76B}. 

Although the Upper~Sco sample includes only one bona-fide transition disk and that disk does not have an \OIa\ detection, an increase in the W3-W4 spectral index still points to inner disk clearing (see Figure~\ref{Fig:twocolor} and Sect.~\ref{sect:diskclass}). Therefore, as in the younger samples, the Upper~Sco data indicate that the \OIa\ emitting radius, and possibly the wind launching radius, moves further out as the inner dust disk of both Warm and Cool stars is depleted.

\section{Summary}\label{Sect:summary}
We have analyzed high-resolution ($\Delta v \sim$\,7\,km/s) optical spectra  for a sample of 115 stars in the $\sim 5-10$\,Myr-old Upper~Sco association to search for potential signatures of disk winds via the  \OIa\ profiles. Of our sample, 87 are surrounded by a protoplanetary disk (here including full, evolved, and transition), 24 by a debris disk, and 4 have no disk. A total of 45 stars, all surrounded by a protoplanetry disk, are determined to be accreting (CTTS) based on H$\alpha$ profiles, with accretion rates ranging from $5 \times 10^{-12}$ to $2.5 \times 10^{-7}$\,M$_\odot$/yr.

Our spectra cover a wide range of spectral types, from G0 to M5.2, with the majority later than M3. In order to facilitate comparison of \OIa\ emission among stars of comparable mass in younger ($< 5$\,Myr-old) star-forming regions, we separate our sample into Warm (G0 to M3) and Cool (later than M3), corresponding to a boundary in stellar mass around  0.28\,M$_\odot$. Two of the studies from younger regions which we use for comparison primarily consist of Warm stars at similar spectral resolution as our sample (NGC 2264 and a selection of stars from  Taurus, Lupus, $\rho$ Oph and CrA).  A third study includes both Warm and Cool stars but at lower spectral resolution (Cha + Lup). The Upper~Sco dataset enables us to characterize \OIa\ emission both as a function of age and as a function of stellar mass. Our main results can be summarized as follows:

\begin{enumerate}

\item The \OIa\ line is detected in 45 out of 115 Upper~Sco sources, all with protoplanetary disks. In general, The Upper~Sco sample follows the same accretion luminosity$-$LVC \OIa\ luminosity relation as the younger samples。 The decline of accretion luminosity with age is indicated by the fact that the median accretion luminosity of the warm Upper~Sco sample is a factor of $\sim3$ lower than that of younger stars when matched to the same spectral type range. The link between accretion and \OIa\ may not be uniquely coupled however, as there are some CTTS with no \OIa\ and some WTTS with \OIa. The possibility that these outliers may be due to sensitivity limits is discussed in the text.

\item All 45 \OIa\ detections in Upper~Sco show low velocity emission (LVC) and only 5 also show high velocity emission (HVC) attributed to a jet. Most  LVC profiles are well fit by a single Gaussian (SC) but 7 show a composite form with both broad (BC) and narrow (NC) components.

\item As in other regions, the HVC and BC+NC are associated with higher accretion luminosities and their fraction in Upper Sco is lower than  in younger regions with higher accretion luminosities. In contrast  \OIa\  in Upper Sco has the highest proportion of SC profiles.

\item In Upper~Sco both the Warm and Cool samples have a distribution of LVC SC centroid velocities consistent with those in younger regions and as such are attributed to slow disk winds. While these components in all regions have smaller blueshifts than the BC or SCJ components, and are more comparable to the NC, their centroid distributions have modest median blueshifts and are asymmetric with a clear tail on the negative side.

\item The Upper~Sco distribution of FWHM for the LVC SC differs between the Warm and Cool samples, with the Cool sample showing a larger FWHM normalized by stellar mass than the Warm one. If the lines are rotationally broadened this implies closer emitting radii (and perhaps wind launching)  in lower mass stars. This comparison cannot yet be made for younger sources as the only study including low mass stars has a resolution too low for profile decomposition.

\item The Upper~Sco \OIa\ profiles show an anti-correlation between LVC SC FWHM and the WISE W3-W4  spectral index, following a similar trend reported for the younger samples. The correlation is in the sense that \OIa\  profiles become narrower as the infrared spectral index increases. This index indicates inner disk clearing, although the only bona fide transition disk in the Upper~Sco  sample does not have an \OIa\   detection.
\end{enumerate}

Our main findings expand upon the emerging view of disk winds and their role in disk evolution in several ways. First, they demonstrate that disk winds persist at $\sim 5-10$\,Myr, as long as stars are still accreting disk material. Second, as stars age and their accretion luminosity declines so does their \OIa\ luminosity, in the same proportion as seen for younger stars. The character of the LVC tracing a disk wind evolves with decreasing accretion luminosity, where simple SC profiles with low wind velocities $\sim 1-2$\kms\ become the dominant mode in older regions like Upper Sco.  The winds indicated by the SC components show a large spread in the ratio of mass loss rate to mass accretion rate, in some cases with ratios near unity but others with ratios less than 0.1. It is possible that in sources with low ratios (large accretion rates) the wind is mostly molecular and the \OIa\  provides only a lower limit to the mass loss rate. Finally, all accreting stars, down to the late M types, present disk winds but the lowest mass stars differ from their Warm counterparts in the radii where \OIa\ emission arises, where larger FWHM suggest smaller wind launching radii radii in the lower mass stars.

\acknowledgments
Many thanks to the anonymous referee for comments that helped to improve this paper. We thank the additional non-coauthor Keck/HIRES PIs and observers who provided the data products recorded in Table~\ref{Tab:obs_log}. This material is based upon work supported by the National Aeronautics and Space Administration (NASA) under Agreement No. NNX15AD94G for the program ''Earths in Other Solar Systems" and under Agreement No. 80NSSC21K0593 for the program "Alien Earths".

%

\vspace{5mm}
\facilities{Keck:I (HIRES)}


\newpage
\setcounter{table}{0}

\begin{longrotatetable}
\begin{deluxetable*}{rcccrccccccrcccccccc}
\renewcommand{\tabcolsep}{0.02cm}
\tablecaption{A List of the Sources in This Work, as well as Their Stellar and Disk Properties \label{tabe_UpperSco}}
\tablewidth{700pt}
\tabletypesize{\scriptsize}
\tablehead{
\colhead{ID}&\colhead{Name} & \colhead{Dis} & \colhead{SpT}& \colhead{Log$L_{\star}$} &\colhead{$A_{\rm V}$} &\colhead{$R_{\star}$} & \colhead{Mass} &\colhead{Disk} & \colhead{CTTS} &\colhead{Log~$\dot{M}_{\rm acc}$} &\colhead{$RV_{\rm Helio}$} &\colhead{Correction}  & \colhead{Log$L_{\rm NC}$}& \colhead{Log$L_{\rm BC}$}& \colhead{Log$L_{\rm HVC}$} & \colhead{[O {\tiny I] }} & \colhead{LVC} & \colhead{HVC} \\
\colhead{}&\colhead{} & \colhead{(pc)} & \colhead{}& \colhead{($L_{\odot}$)}& \colhead{(mag)} &\colhead{($R_{\odot}$)} & \colhead{($M_{\odot}$)} &\colhead{Type} &\colhead{} &\colhead{($M$\accunit)} &\colhead{(\kms)}&\colhead{(\kms)} &\colhead{($L_{\odot}$)}&\colhead{($L_{\odot}$)}&\colhead{($L_{\odot}$)} & \colhead{Det} & \colhead{Comp.} & \colhead{}}
\startdata
       1& 2MASSJ15514032-2146103& 139.7&M4.8&$ -1.39$&   0.2&0.74&0.11&PD &N &$<-10.77$&$-4.9$&$-0.6$& \nodata& $ -6.50$& \nodata &Y &SC &  \\
      2& 2MASSJ15521088-2125372& 154.2&M4.9&$ -2.70$&   0.0&0.17&0.09&PD &Y &$-11.30$&$-4.7$&$-1.9$& \nodata& $ -5.88$& \nodata &Y &SC &  \\
      3& 2MASSJ15530132-2114135& 141.8&M4.7&$ -1.41$&   0.2&0.71&0.12&PD &Y &$-10.31$&$-5.3$&$-1.2$& \nodata& $ -6.47$& \nodata &Y &SC &  \\
      4& 2MASSJ15534211-2049282& 134.8&M3.8&$ -1.27$&   0.4&0.75&0.19&PD &Y &$ -9.94$&$-6.0$&$-1.8$& $ -6.24$& $ -5.30$& \nodata &Y &BC+NC &  \\
      5& 2MASSJ15554883-2512240& 143.0&G3.0&$  0.37$&   0.7&1.55&1.35&ND &N &\nodata&$-0.9$&$ 0.1$& \nodata& \nodata& \nodata &N &   &  \\
      6& 2MASSJ15562477-2225552& 141.1&M4.2&$ -1.30$&   0.4&0.77&0.16&PD &N &$<-10.67$&$-5.0$&$-1.2$& $ -6.56$& \nodata& \nodata &Y &SC &  \\
      7& 2MASSJ15570641-2206060& 143.9&M4.8&$ -1.46$&   0.3&0.68&0.11&PD &Y &$-10.56$&$-5.2$&$-1.2$& \nodata& $ -6.25$& $ -6.66$ &Y &SCJ &Y\\
      8& 2MASSJ15572986-2258438& 142.0&M4.7&$ -1.42$&   0.2&0.71&0.12&PD &Y &$-10.59$&$-0.6$&$-1.1$& \nodata& \nodata& \nodata &N &   &  \\
      9& 2MASSJ15581270-2328364& 144.9&G2.0&$  0.51$&   0.9&1.73&1.44&DB &N &\nodata&$-3.7$&$-1.0$& \nodata& \nodata& \nodata &N &   &  \\
     10& 2MASSJ15582981-2310077& 140.0&M4.8&$ -1.39$&   0.7&0.74&0.11&PD &Y &$ -8.92$&$-4.5$&$-1.5$& $ -6.78$& $ -6.38$& \nodata &Y &BC+NC &  \\
     11& 2MASSJ15583692-2257153& 166.5&G3.0&$  0.50$&   0.4&1.80&1.52&PD &Y &$ -7.99$&$-0.2$&$-1.3$& $ -5.01$& $ -4.44$& \nodata &Y &BC+NC &  \\
     12& 2MASSJ15584772-1757595& 137.7&K3.0&$  0.11$&   1.5&1.83&1.25&DB &N &$< -9.95$&$-6.3$&$-0.8$& \nodata& \nodata& \nodata &N &   &  \\
     13& 2MASSJ16001330-2418106& 146.2&K9.9&$ -0.56$&   0.5&1.22&0.56&DB &N &$<-10.23$&$-4.3$&$-0.5$& \nodata& \nodata& \nodata &N &   &  \\
     14& 2MASSJ16001730-2236504& 144.0&M3.7&$ -0.95$&   0.0&1.07&0.21&PD &N &$<-10.00$&$-5.1$&$-1.9$& \nodata& \nodata& \nodata &N &   &  \\
     15& 2MASSJ16001844-2230114& 168.1&M5.2&$ -1.00$&   0.4&1.20&0.11&PD &Y &$ -8.22$&$-2.6$&$-1.3$& $ -5.86$& \nodata& \nodata &Y &SC &  \\
     16& 2MASSJ16014086-2258103& 135.0&M4.0&$ -1.12$&   0.0&0.92&0.18&PD &Y &$ -8.46$&$-5.2$&$-1.0$& \nodata& $ -5.51$& \nodata &Y &SC &  \\
     17& 2MASSJ16014157-2111380& 140.3&M5.0&$ -1.65$&   1.0&0.56&0.09&PD &Y &$ -9.85$&$-7.3$&$-1.0$& \nodata& $ -6.36$& \nodata &Y &SC &  \\
     18& 2MASSJ16020039-2221237& 141.4&M2.0&$ -0.45$&   0.2&1.62&0.33&DB &N &$< -9.38$&$-6.7$&$-0.9$& \nodata& \nodata& \nodata &N &   &  \\
     19& 2MASSJ16020287-2236139& 142.0&M0.7&$ -1.61$&   0.4&0.39&0.53&DB &N &\nodata&$ 4.6$&$-0.3$& \nodata& \nodata& \nodata &N &   &  \\
     20& 2MASSJ16020757-2257467& 139.6&M2.5&$ -0.86$&   0.3&1.05&0.31&PD &N &$<-10.28$&$-4.5$&$-0.9$& \nodata& $ -5.75$& \nodata &Y &SC &  \\
     21& 2MASSJ16024152-2138245& 139.4&M4.9&$ -1.55$&   0.0&0.63&0.09&PD &Y &$ -9.94$&$-6.8$&$-0.8$& \nodata& \nodata& \nodata &N &   &  \\
     22& 2MASSJ16025123-2401574& 144.5&K5.0&$ -0.34$&   0.1&1.32&0.85&DB &N &$<-10.26$&$-5.0$&$-1.3$& \nodata& \nodata& \nodata &N &   &  \\
     23& 2MASSJ16030161-2207523& 139.1&M5.1&$ -1.69$&   0.5&0.54&0.08&PD &N &$<-10.82$&$-4.9$&$-2.6$& \nodata& \nodata& \nodata &N &   &  \\
     24& 2MASSJ16031329-2112569& 136.1&M4.9&$ -1.45$&   0.4&0.70&0.09&PD &N &$<-10.87$&$-5.5$&$-1.6$& \nodata& \nodata& \nodata &N &   &  \\
     25& 2MASSJ16032225-2413111& 139.0&M3.9&$ -1.01$&   0.6&1.03&0.19&PD &Y &$ -9.06$&$-5.7$&$-1.0$& \nodata& $ -5.64$& \nodata &Y &SC &  \\
     26& 2MASSJ16035767-2031055& 142.6&K5.1&$ -0.20$&   0.9&1.56&0.81&PD &Y &$ -9.29$&$-5.9$&$-0.5$& \nodata& $ -4.91$& \nodata &Y &SC &  \\
     27& 2MASSJ16035793-1942108& 152.9&M2.7&$ -0.94$&   0.6&0.97&0.31&PD &N &$<-10.43$&$-1.9$&$29.5$& \nodata& \nodata& \nodata &N &   &  \\
     28& 2MASSJ16041740-1942287& 152.8&M3.6&$ -1.11$&   0.4&0.89&0.22&PD &N &$<-10.46$&$-2.7$&$-0.9$& \nodata& $ -6.13$& \nodata &Y &SC &  \\
     29& 2MASSJ16042165-2130284& 144.6&K2.3&$ -0.01$&   1.4&1.49&1.25&PD &N &$<-10.34$&$-4.4$&$-0.9$& $ -4.70$& \nodata& \nodata &Y &SC &  \\
     30& 2MASSJ16043916-1942459& 149.6&M3.7&$ -1.21$&   0.2&0.80&0.20&DB &N &$<-11.13$&$-3.1$&$-0.4$& \nodata& \nodata& \nodata &N &   &  \\
     31& 2MASSJ16050231-1941554& 152.6&M4.2&$ -1.46$&   0.6&0.64&0.15&DB &N &$<-11.50$&$-2.8$&$-1.6$& \nodata& \nodata& \nodata &N &   &  \\
     32& 2MASSJ16052459-1954419& 151.2&M3.4&$ -1.05$&   0.3&0.93&0.24&DB &N &$<-10.74$&$-2.8$&$-0.5$& \nodata& \nodata& \nodata &N &   &  \\
     33& 2MASSJ16052556-2035397& 143.5&M5.1&$ -1.42$&   0.4&0.74&0.10&PD &N &$<-10.63$&$-2.7$&$-0.7$& \nodata& \nodata& \nodata &N &   &  \\
     34& 2MASSJ16052661-1957050& 142.0&M5.1&$ -1.28$&   0.0&0.87&0.09&PD &N &$<-10.32$&$-2.6$&$-1.1$& \nodata& $ -6.82$& \nodata &Y &SC &  \\
     35& 2MASSJ16053215-1933159& 151.3&M4.8&$ -1.61$&   0.4&0.57&0.10&PD &Y &$ -9.38$&$-2.3$&$-1.7$& \nodata& $ -6.51$& \nodata &Y &SC &  \\
     36& 2MASSJ16055863-1949029& 151.1&M3.8&$ -1.27$&   0.1&0.76&0.19&PD &N &$<-10.66$&$-2.9$&$-0.8$& \nodata& \nodata& \nodata &N &   &  \\
\enddata
\end{deluxetable*}
\end{longrotatetable}

\setcounter{table}{0}
\begin{longrotatetable}
\begin{deluxetable*}{rcccrccccccrcccccccc}
\renewcommand{\tabcolsep}{0.02cm}
\tablecaption{{\it (continuued)}}
\tablewidth{700pt}
\tabletypesize{\scriptsize}
\tablehead{
\colhead{ID}&\colhead{Name} & \colhead{Dis} & \colhead{SpT}& \colhead{Log$L_{\star}$} &\colhead{$A_{\rm V}$} &\colhead{$R_{\star}$} & \colhead{Mass} &\colhead{Disk} & \colhead{CTTS} &\colhead{Log~$\dot{M}_{\rm acc}$} &\colhead{$RV_{\rm Helio}$} &\colhead{Correction}  & \colhead{Log$L_{\rm NC}$}& \colhead{Log$L_{\rm BC}$}& \colhead{Log$L_{\rm HVC}$} & \colhead{[O {\tiny I] }} & \colhead{LVC} & \colhead{HVC} \\
\colhead{}&\colhead{} & \colhead{(pc)} & \colhead{}& \colhead{($L_{\odot}$)}& \colhead{(mag)} &\colhead{($R_{\odot}$)} & \colhead{($M_{\odot}$)} &\colhead{Type} &\colhead{} &\colhead{($M$\accunit)} &\colhead{(\kms)}&\colhead{(\kms)} &\colhead{($L_{\odot}$)}&\colhead{($L_{\odot}$)}&\colhead{($L_{\odot}$)} & \colhead{Det} & \colhead{Comp.} & \colhead{}}
\startdata
     37& 2MASSJ16060061-1957114& 155.0&M4.2&$ -1.16$&   0.0&0.90&0.17&PD &N &$<-10.53$&$-3.4$&$-0.2$& \nodata& \nodata& \nodata &N &   &  \\
     38& 2MASSJ16061330-2212537& 139.0&M3.6&$ -0.72$&   0.4&1.39&0.21&DB &N &$<-10.15$&$-4.9$&$-1.0$& \nodata& \nodata& \nodata &N &   &  \\
     39& 2MASSJ16062196-1928445& 142.0&M0.9&$ -0.45$&   0.5&1.50&0.42&PD &Y &$ -8.90$&$-2.8$&$-0.6$& $ -5.47$& $ -5.33$& \nodata &Y &BC+NC &  \\
     40& 2MASSJ16062277-2011243& 152.1&M4.1&$ -1.31$&   0.3&0.75&0.17&PD &N &$<-10.93$&$-3.6$&$-1.0$& \nodata& $ -6.65$& \nodata &Y &SC &  \\
     41& 2MASSJ16063539-2516510& 137.4&M5.1&$ -1.64$&   0.0&0.58&0.08&PD &Y &$-10.70$&$-4.1$&$-1.5$& \nodata& \nodata& \nodata &N &   &  \\
     42& 2MASSJ16064102-2455489& 151.8&M4.8&$ -1.55$&   1.0&0.62&0.09&PD &Y &$ -9.64$&$-3.1$&$-0.9$& \nodata& $ -6.08$& \nodata &Y &SC &  \\
     43& 2MASSJ16064115-2517044& 150.8&M3.5&$ -1.26$&   0.5&0.74&0.22&PD &N &$<-11.06$&$-3.7$&$-0.5$& \nodata& \nodata& \nodata &N &   &  \\
     44& 2MASSJ16064385-1908056& 145.3&K7.9&$ -0.47$&   0.5&1.24&0.69&PD &Y &$ -9.95$&$-3.6$&$-2.1$& \nodata& \nodata& \nodata &N &   &  \\
     45& 2MASSJ16070014-2033092& 138.3&M2.7&$ -0.93$&   0.6&0.99&0.31&PD &N &$<-11.21$&$-5.5$&$-1.5$& $ -6.07$& \nodata& \nodata &Y &SC &  \\
     46& 2MASSJ16070211-2019387& 145.6&M5.0&$ -1.66$&   0.5&0.55&0.09&PD &N &$<-10.86$&$-3.0$&$-0.3$& \nodata& \nodata& \nodata &N &   &  \\
     47& 2MASSJ16070873-1927341& 147.7&M4.0&$ -1.35$&   0.5&0.71&0.17&DB &N &$<-11.05$&$-3.3$&$ 0.0$& \nodata& \nodata& \nodata &N &   &  \\
     48& 2MASSJ16071971-2020555& 154.6&M3.8&$ -1.16$&   0.5&0.85&0.20&DB &N &$<-10.89$&$-2.4$&$-1.1$& \nodata& \nodata& \nodata &N &   &  \\
     49& 2MASSJ16072625-2432079& 143.4&M3.9&$ -0.93$&   0.0&1.13&0.19&PD &Y &$ -9.61$&$-2.7$&$-1.4$& \nodata& $ -5.79$& \nodata &Y &SC &  \\
     50& 2MASSJ16072747-2059442& 179.3&M4.9&$ -0.89$&   0.5&1.33&0.13&PD &N &$< -9.90$&$-5.6$&$-1.1$& \nodata& \nodata& \nodata &N &   &  \\
     51& 2MASSJ16073939-1917472& 138.5&M2.7&$ -0.82$&   0.5&1.13&0.29&DB &N &$<-10.91$&$-4.8$&$-1.7$& \nodata& \nodata& \nodata &N &   &  \\
     52& 2MASSJ16075796-2040087& 135.9&K4.0&$ -0.96$&   1.4&0.59&0.71&PD &Y &$ -8.96$&$ 2.5$&$-1.9$& \nodata& $ -3.97$& $ -3.87$ &Y &BC+NC &Y\\
     53& 2MASSJ16080555-2218070& 143.5&M2.5&$ -0.82$&   0.4&1.10&0.31&ND &N &$<-10.16$&$-5.7$&$-0.8$& \nodata& \nodata& \nodata &N &   &  \\
     54& 2MASSJ16081566-2222199& 138.6&M2.7&$ -0.87$&   0.3&1.06&0.30&PD &N &$<-10.52$&$-4.6$&$-2.2$& \nodata& \nodata& \nodata &N &   &  \\
     55& 2MASSJ16082751-1949047& 142.0&M4.9&$ -1.20$&   0.6&0.93&0.11&PD &Y &$ -9.76$&$-5.7$&$-1.0$& \nodata& \nodata& \nodata &N &   &  \\
     56& 2MASSJ16083455-2211559& 137.9&M4.9&$ -1.60$&   0.4&0.59&0.09&PD &N &$<-11.21$&$-7.0$&$-0.8$& \nodata& \nodata& \nodata &N &   &  \\
     57& 2MASSJ16084894-2400045& 143.5&M3.9&$ -1.34$&   0.1&0.71&0.18&PD &N &$<-11.11$&$-1.5$&$-1.1$& \nodata& \nodata& \nodata &N &   &  \\
     58& 2MASSJ16090002-1908368& 135.7&M5.0&$ -1.41$&   0.2&0.74&0.10&PD &N &$<-10.40$&$-7.3$&$-1.4$& \nodata& \nodata& \nodata &N &   &  \\
     59& 2MASSJ16090075-1908526& 136.9&M0.6&$ -0.51$&   0.9&1.36&0.46&PD &Y &$ -8.92$&$-5.8$&$-0.9$& \nodata& $ -5.04$& \nodata &Y &SC &  \\
     60& 2MASSJ16093558-1828232& 157.4&M3.7&$ -1.16$&   1.0&0.85&0.21&PD &N &$<-10.27$&$-3.0$&$-1.0$& \nodata& $ -6.14$& \nodata &Y &SC &  \\
     61& 2MASSJ16094098-2217594& 144.8&K9.0&$ -0.15$&   0.5&1.86&0.56&DB &N &$< -9.77$&$-5.0$&$-0.8$& \nodata& \nodata& \nodata &N &   &  \\
     62& 2MASSJ16095361-1754474& 157.1&M5.0&$ -1.54$&   0.4&0.63&0.10&PD &Y &$-10.26$&$-3.3$&$-1.0$& \nodata& $ -6.86$& \nodata &Y &SC &  \\
     63& 2MASSJ16095441-1906551& 137.6&M1.8&$ -0.66$&   0.9&1.26&0.35&DB &N &$<-10.25$&$-5.9$&$-1.0$& \nodata& \nodata& \nodata &N &   &  \\
     64& 2MASSJ16101473-1919095& 137.9&M3.4&$ -0.93$&   0.6&1.06&0.24&ND &N &$<-11.26$&$-5.7$&$-0.7$& \nodata& \nodata& \nodata &N &   &  \\
     65& 2MASSJ16102819-1910444& 150.4&M5.0&$ -1.78$&   0.2&0.48&0.08&PD &Y &$-10.80$&$-3.4$&$-0.8$& \nodata& \nodata& \nodata &N &   &  \\
     66& 2MASSJ16103956-1916524& 155.0&M2.8&$ -1.00$&   0.6&0.92&0.30&DB &N &$<-10.35$&$-3.8$&$-2.3$& \nodata& \nodata& \nodata &N &   &  \\
     67& 2MASSJ16104202-2101319& 140.1&K4.9&$ -0.00$&   1.3&1.93&0.79&DB &N &$< -9.54$&$-4.8$&$ 0.5$& \nodata& \nodata& \nodata &N &   &  \\
     68& 2MASSJ16104636-1840598& 140.1&M4.9&$ -1.52$&   0.9&0.64&0.09&PD &Y &$-10.30$&$-5.6$&$-1.4$& \nodata& \nodata& \nodata &N &   &  \\
     69& 2MASSJ16111330-2019029& 152.9&M4.3&$ -0.98$&   0.0&1.12&0.16&PD &Y &$ -8.48$&$-2.4$&$-1.8$& \nodata& $ -5.67$& \nodata &Y &SC &  \\
     70& 2MASSJ16111534-1757214& 135.3&M1.2&$ -0.55$&   0.7&1.36&0.40&PD &N &$< -9.69$&$-5.7$&$-2.5$& \nodata& $ -5.75$& \nodata &Y &SC &  \\
     71& 2MASSJ16112057-1820549& 135.8&K4.8&$ -0.09$&   1.1&1.72&0.83&DB &N &\nodata&$-6.2$&$-1.1$& \nodata& \nodata& \nodata &N &   &  \\
     72& 2MASSJ16113134-1838259& 131.7&K5.0&$  0.89$&   1.8&5.44&0.73&PD &Y &$ -6.60$&$-5.3$&$-1.8$& $ -4.62$& $ -4.46$& $ -4.09$ &Y &BC+NC &Y\\
     73& 2MASSJ16115091-2012098& 139.8&M3.8&$ -1.15$&   0.3&0.86&0.20&PD &N &$<-10.54$&$-6.8$&$ NaN$& \nodata& $ -5.83$& \nodata &Y &SC &  \\
     74& 2MASSJ16122737-2009596& 142.2&M5.2&$ -1.54$&   1.0&0.65&0.08&PD &N &$<-10.74$&$-3.1$&$-1.4$& \nodata& \nodata& \nodata &N &   &  \\
     75& 2MASSJ16123916-1859284& 134.7&M2.0&$ -0.54$&   0.9&1.47&0.33&PD &Y &$ -8.69$&$-2.1$&$-2.7$& \nodata& $ -5.49$& \nodata &Y &SC &  \\
     76& 2MASSJ16124893-1800525& 152.0&M3.8&$ -1.07$&   0.2&0.95&0.20&PD &N &$<-10.82$&$-4.2$&$-3.4$& \nodata& \nodata& \nodata &N &   &  \\
      77& 2MASSJ16125533-2319456& 152.5&G2.0&$  0.78$&   0.7&2.38&1.79&DB &N &\nodata&$-3.3$&$-0.4$& \nodata& \nodata& \nodata &N &   &  \\
     78& 2MASSJ16130996-1904269& 135.1&M4.9&$ -1.11$&   1.2&1.04&0.12&PD &Y &$ -8.93$&$-6.0$&$-0.4$& \nodata& $ -5.55$& \nodata &Y &SC &  \\
     79& 2MASSJ16133650-2503473& 142.0&M3.8&$ -0.89$&   1.3&1.17&0.20&PD &Y &$ -8.33$&$-1.2$&$-0.6$& \nodata& $ -5.27$& \nodata &Y &SC &  \\
     80& 2MASSJ16135434-2320342& 142.0&M4.8&$ -1.02$&   0.0&1.13&0.13&PD &Y &$ -8.20$&$-5.0$&$-3.2$& \nodata& \nodata& \nodata &N &   &  \\
     81& 2MASSJ16141107-2305362& 142.0&K3.0&$  0.50$&   1.0&2.86&1.23&PD &N &\nodata&$-8.4$&$-1.2$& \nodata& \nodata& \nodata &N &   &  \\
     82& 2MASSJ16142029-1906481& 138.8&K9.0&$ -0.68$&   2.2&1.02&0.67&PD &Y &$ -9.04$&$-5.4$&$-1.2$& $ -5.08$& \nodata& $ -4.19$ &Y &SCJ &Y\\
     83& 2MASSJ16142893-1857224& 135.1&M3.2&$ -0.73$&   0.4&1.30&0.25&DB &N &$< -9.48$&$-4.6$&$-0.7$& \nodata& \nodata& \nodata &N &   &  \\
     84& 2MASSJ16143367-1900133& 135.8&M3.3&$ -0.45$&   1.7&1.82&0.23&PD &Y &$ -9.29$&$-7.0$&$-1.3$& \nodata& $ -4.93$& \nodata &Y &SC &  \\
     85& 2MASSJ16145918-2750230& 149.2&G8.0&$  0.07$&   0.7&1.33&1.24&DB &N &\nodata&$ 1.1$&$-0.6$& \nodata& \nodata& \nodata &N &   &  \\
     86& 2MASSJ16145928-2459308& 157.7&M4.7&$ -1.31$&   0.2&0.80&0.12&PD &N &$<-10.78$&$-1.3$&$-2.3$& \nodata& $ -6.62$& \nodata &Y &SC &  \\
     87& 2MASSJ16151239-2420091& 145.2&M4.7&$ -1.70$&   0.8&0.51&0.10&PD &N &$<-11.23$&$-2.7$&$-1.4$& \nodata& $ -6.47$& \nodata &Y &SC &  \\
     88& 2MASSJ16153456-2242421& 136.9&M0.2&$ -0.41$&   0.3&1.49&0.48&PD &Y &$ -8.68$&$-2.4$&$-0.8$& \nodata& $ -5.14$& \nodata &Y &SC &  \\
     89& 2MASSJ16154416-1921171& 125.9&K8.0&$ -0.38$&   2.0&1.38&0.66&PD &Y &$ -8.23$&$-3.8$&$-4.0$& $ -4.85$& $ -4.57$& $ -4.58$ &Y &BC+NC &Y\\
     90& 2MASSJ16163345-2521505& 158.4&M0.1&$ -0.73$&   1.2&1.02&0.56&PD &N &$<-10.91$&$ 0.0$&$-2.8$& \nodata& \nodata& \nodata &N &   &  \\
     91& 2MASSJ16181904-2028479& 138.3&M5.0&$ -1.47$&   0.8&0.69&0.10&PD &N &$<-10.63$&$-5.2$&$-1.1$& \nodata& \nodata& \nodata &N &   &  \\
     92& 2MASSJ16215466-2043091& 108.5&K7.0&$ -0.35$&   0.7&1.41&0.68&DB &N &\nodata&$-4.1$&$-2.0$& \nodata& \nodata& \nodata &N &   &  \\
     93& 2MASSJ16220961-1953005& 133.3&M3.7&$ -0.65$&   0.8&1.53&0.20&DB &N &$<-10.12$&$-7.4$&$-0.7$& \nodata& \nodata& \nodata &N &   &  \\
     94& 2MASSJ16235385-2946401& 134.6&G2.5&$  0.66$&   0.7&2.12&1.69&ND &N &\nodata&$ 0.5$&$-0.9$& \nodata& \nodata& \nodata &N &   &  \\
     95& 2MASSJ16270942-2148457& 137.0&M4.9&$ -1.70$&   1.3&0.52&0.09&PD &N &$<-10.68$&$-5.4$&$-1.8$& \nodata& \nodata& \nodata &N &   &  \\
     96& 2MASSJ15564244-2039339& 140.2&M3.4&$ -1.05$&   0.5&0.92&0.24&PD &N &$<-10.84$&$-4.9$&$-0.6$& \nodata& \nodata& \nodata &N &   &  \\
     97& 2MASSJ15583620-1946135& 155.6&M3.9&$ -1.16$&   0.2&0.87&0.19&PD &N &$<-10.84$&$-2.3$&$-0.9$& \nodata& \nodata& \nodata &N &   &  \\
     98& 2MASSJ15594426-2029232& 138.0&M4.2&$ -1.18$&   0.5&0.88&0.17&PD &N &$<-10.38$&$-6.3$&$-1.1$& \nodata& \nodata& \nodata &N &   &  \\
     99& 2MASSJ16011398-2516281& 142.0&M4.1&$ -1.13$&   0.0&0.92&0.17&PD &N &$<-10.40$&$-1.8$&$-1.3$& \nodata& \nodata& \nodata &N &   &  \\
    100& 2MASSJ16012902-2509069& 134.5&M4.2&$ -1.21$&   0.1&0.84&0.16&PD &Y &$ -8.71$&$-5.1$&$-1.4$& \nodata& $ -5.97$& \nodata &Y &SC &  \\
    101& 2MASSJ16023587-2320170& 139.1&M3.7&$ -1.07$&   0.1&0.93&0.21&PD &N &$<-10.12$&$-5.6$&$-0.5$& \nodata& \nodata& \nodata &N &   &  \\
    102& 2MASSJ16041893-2430392& 142.0&M2.7&$ -0.52$&   0.1&1.58&0.29&PD &Y &$ -9.22$&$-4.9$&$-1.6$& \nodata& $ -5.72$& \nodata &Y &SC &  \\
    103& 2MASSJ16052157-1821412& 148.9&K3.8&$ -0.04$&   0.9&1.66&1.05&PD &Y &$ -8.96$&$-3.5$&$-1.6$& \nodata& $ -4.78$& \nodata &Y &SC &  \\
    104& 2MASSJ16064794-1841437& 155.8&K9.0&$ -0.17$&   0.8&1.83&0.56&PD &Y &$ -9.42$&$-2.1$&$-1.2$& \nodata& \nodata& \nodata &N &   &  \\
    105& 2MASSJ16093164-2229224& 153.9&M2.5&$ -0.48$&   0.5&1.63&0.30&PD &N &$< -9.63$&$-2.7$&$-1.3$& \nodata& \nodata& \nodata &N &   &  \\
    106& 2MASSJ16100501-2132318& 145.4&K9.0&$ -0.42$&   0.6&1.37&0.61&PD &Y &$ -8.63$&$-4.3$&$-1.3$& $ -5.45$& \nodata& \nodata &Y &SC &  \\
    107& 2MASSJ16112601-2631558& 142.0&M2.6&$ -0.72$&   0.4&1.25&0.29&PD &N &$<-10.18$&$-4.1$&$-1.6$& \nodata& \nodata& \nodata &N &   &  \\
    108& 2MASSJ16120505-2043404& 122.5&M1.5&$ -0.57$&   1.0&1.37&0.37&PD &Y &$ -9.50$&$-5.6$&$-1.3$& \nodata& \nodata& \nodata &N &   &  \\
    109& 2MASSJ16120668-3010270& 131.9&M0.0&$ -0.59$&   0.3&1.19&0.55&PD &Y &$ -9.40$&$-1.3$&$-1.5$& $ -5.35$& \nodata& \nodata &Y &SC &  \\
    110& 2MASSJ16132190-2136136& 144.8&M3.3&$ -0.87$&   0.5&1.12&0.25&PD &Y &$ -9.51$&$-3.4$&$-0.6$& \nodata& \nodata& \nodata &N &   &  \\
    111& 2MASSJ16145244-2513523& 160.0&M3.5&$ -1.09$&   0.5&0.90&0.23&PD &N &$<-10.33$&$-2.6$&$-1.5$& \nodata& \nodata& \nodata &N &   &  \\
    112& 2MASSJ16153220-2010236& 142.0&M2.4&$ -0.57$&   1.0&1.46&0.31&PD &Y &$ -9.13$&$-6.7$&$-1.8$& \nodata& \nodata& \nodata &N &   &  \\
    113& 2MASSJ16194711-2203112& 126.8&M4.2&$ -1.30$&   0.1&0.77&0.16&PD &Y &$-10.30$&$-3.3$&$-1.7$& \nodata& \nodata& \nodata &N &   &  \\
    114& 2MASSJ16200616-2212385& 137.3&M3.6&$ -1.22$&   0.2&0.78&0.21&PD &N &$<-11.23$&$-2.2$&$-1.2$& \nodata& \nodata& \nodata &N &   &  \\
    115& 2MASSJ16252883-2607538& 139.5&M3.1&$ -0.87$&   0.4&1.10&0.27&DB &N &$<-10.44$&$-3.1$&$-1.9$& \nodata& \nodata& \nodata &N &   & \\ 
\enddata
\end{deluxetable*}
\end{longrotatetable}

\clearpage

\setcounter{table}{1}
\clearpage
\startlongtable
\begin{deluxetable*}{rccccccccccccccccc}
\tablewidth{700pt}
\tablecaption{Observation log summary \label{Tab:obs_log}}
\tabletypesize{\scriptsize}
\tablehead{
\colhead{ID}&\colhead{Name} & \colhead{RA} & \colhead{DEC} & \colhead{Data-Obs}& \colhead{Nominal}& \colhead{Program} &\colhead{PI} & \colhead{S/N} \\
\colhead{}&\colhead{} & \colhead{(J2000)} &  \colhead{(J2000)} &\colhead{} & \colhead{Resolution}& \colhead{ID}  &\colhead{} & \colhead{}  }
\startdata
      1& 2MASSJ15514032-2146103 &  15 51 40.32&$-$21 46 10.3&2013-06-04&34000&C189Hr  &Carpenter&  9.1\\
      2& 2MASSJ15521088-2125372 &  15 52 10.88&$-$21 25 37.2&2015-06-02&34000&C247Hr  &Carpenter&  1.0\\
      3& 2MASSJ15530132-2114135 &  15 53 01.32&$-$21 14 13.5&2013-06-04&34000&C189Hr  &Carpenter&  6.8\\
      4& 2MASSJ15534211-2049282 &  15 53 42.11&$-$20 49 28.2&2006-08-12&45000&H212Hr  &Shkolnik&  3.2\\
      5& 2MASSJ15554883-2512240 &  15 55 48.83&$-$25 12 24.0&2015-06-01&34000&C247Hr  &Carpenter& 40.3\\
      6& 2MASSJ15562477-2225552 &  15 56 24.77&$-$22 25 55.2&2007-05-24&34000&C269Hr  &Dahm    & 16.4\\
      7& 2MASSJ15570641-2206060 &  15 57 06.41&$-$22 06 06.0&2007-05-25&34000&C269Hr  &Dahm    & 10.7\\
      8& 2MASSJ15572986-2258438 &  15 57 29.86&$-$22 58 43.8&2007-05-25&34000&C269Hr  &Dahm    & 16.4\\
      9& 2MASSJ15581270-2328364 &  15 58 12.70&$-$23 28 36.4&2015-06-01&34000&C247Hr  &Carpenter& 38.4\\
     10& 2MASSJ15582981-2310077 &  15 58 29.81&$-$23 10 07.7&2007-05-24&34000&C269Hr  &Dahm    & 14.8\\
     11& 2MASSJ15583692-2257153 &  15 58 36.92&$-$22 57 15.3&2008-05-23&48000&C199Hb&Herczeg& 57.3\\
     12& 2MASSJ15584772-1757595 &  15 58 47.72&$-$17 57 59.5&2015-06-01&34000&C247Hr  &Carpenter& 38.1\\
     13& 2MASSJ16001330-2418106 &  16 00 13.30&$-$24 18 10.6&2015-06-02&34000&C247Hr  &Carpenter& 30.5\\
     14& 2MASSJ16001730-2236504 &  16 00 17.30&$-$22 36 50.4&2015-06-01&34000&C247Hr  &Carpenter& 20.0\\
     15& 2MASSJ16001844-2230114 &  16 00 18.44&$-$22 30 11.4&2015-06-01&34000&C247Hr  &Carpenter& 10.6\\
     16& 2MASSJ16014086-2258103 &  16 01 40.86&$-$22 58 10.3&2013-06-04&34000&C189Hr  &Carpenter& 12.4\\
     17& 2MASSJ16014157-2111380 &  16 01 41.57&$-$21 11 38.0&2013-06-04&34000&C189Hr  &Carpenter&  2.8\\
     18& 2MASSJ16020039-2221237 &  16 02 00.39&$-$22 21 23.7&2015-06-02&34000&C247Hr  &Carpenter& 13.8\\
     19& 2MASSJ16020287-2236139 &  16 02 02.87&$-$22 36 13.9&2015-06-02&34000&C247Hr  &Carpenter&  4.3\\
     20& 2MASSJ16020757-2257467 &  16 02 07.57&$-$22 57 46.7&2013-06-04&34000&C189Hr  &Carpenter& 12.3\\
     21& 2MASSJ16024152-2138245 &  16 02 41.52&$-$21 38 24.5&2013-06-04&34000&C189Hr  &Carpenter&  5.1\\
     22& 2MASSJ16025123-2401574 &  16 02 51.23&$-$24 01 57.4&2011-04-24&60000&ENG     &Engineering& 51.8\\
     23& 2MASSJ16030161-2207523 &  16 03 01.61&$-$22 07 52.3&2015-06-01&34000&C247Hr  &Carpenter&  2.6\\
     24& 2MASSJ16031329-2112569 &  16 03 13.29&$-$21 12 56.9&2015-06-01&34000&C247Hr  &Carpenter&  5.0\\
     25& 2MASSJ16032225-2413111 &  16 03 22.25&$-$24 13 11.1&2013-06-04&34000&C189Hr  &Carpenter&  8.6\\
     26& 2MASSJ16035767-2031055 &  16 03 57.67&$-$20 31 05.5&2006-06-16&34000&C315Hr  &Carpenter& 42.3\\
     27& 2MASSJ16035793-1942108 &  16 03 57.93&$-$19 42 10.8&2006-06-16&34000&C315Hr  &Carpenter& 18.9\\
     28& 2MASSJ16041740-1942287 &  16 04 17.40&$-$19 42 28.7&2013-06-04&34000&C189Hr  &Carpenter&  9.5\\
     29& 2MASSJ16042165-2130284 &  16 04 21.65&$-$21 30 28.4&2006-06-16&34000&C315Hr  &Carpenter& 38.7\\
     30& 2MASSJ16043916-1942459 &  16 04 39.16&$-$19 42 45.9&2015-06-01&34000&C247Hr  &Carpenter&  3.0\\
     31& 2MASSJ16050231-1941554 &  16 05 02.31&$-$19 41 55.4&2015-06-01&34000&C247Hr  &Carpenter&  1.2\\
     32& 2MASSJ16052459-1954419 &  16 05 24.59&$-$19 54 41.9&2015-06-01&34000&C247Hr  &Carpenter&  4.3\\
     33& 2MASSJ16052556-2035397 &  16 05 25.56&$-$20 35 39.7&2007-05-25&34000&C269Hr  &Dahm    & 12.2\\
     34& 2MASSJ16052661-1957050 &  16 05 26.61&$-$19 57 05.0&2013-06-04&34000&C189Hr  &Carpenter&  8.1\\
     35& 2MASSJ16053215-1933159 &  16 05 32.15&$-$19 33 15.9&2007-05-25&34000&C269Hr  &Dahm    & 11.7\\
     36& 2MASSJ16055863-1949029 &  16 05 58.63&$-$19 49 02.9&2015-06-01&34000&C247Hr  &Carpenter& 12.0\\
     37& 2MASSJ16060061-1957114 &  16 06 00.61&$-$19 57 11.4&2007-05-24&34000&C269Hr  &Dahm    & 18.0\\
     38& 2MASSJ16061330-2212537 &  16 06 13.30&$-$22 12 53.7&2015-06-01&34000&C247Hr  &Carpenter&  5.8\\
     39& 2MASSJ16062196-1928445 &  16 06 21.96&$-$19 28 44.5&2006-06-16&34000&C315Hr  &Carpenter& 30.2\\
     40& 2MASSJ16062277-2011243 &  16 06 22.77&$-$20 11 24.3&2007-05-24&34000&C269Hr  &Dahm    & 15.2\\
     41& 2MASSJ16063539-2516510 &  16 06 35.39&$-$25 16 51.0&2013-06-04&34000&C189Hr  &Carpenter&  4.4\\
     42& 2MASSJ16064102-2455489 &  16 06 41.02&$-$24 55 48.9&2015-06-01&34000&C247Hr  &Carpenter&  4.6\\
     43& 2MASSJ16064115-2517044 &  16 06 41.15&$-$25 17 04.4&2013-06-04&34000&C189Hr  &Carpenter&  7.1\\
     44& 2MASSJ16064385-1908056 &  16 06 43.85&$-$19 08 05.6&2015-06-02&34000&C247Hr  &Carpenter& 25.3\\
     45& 2MASSJ16070014-2033092 &  16 07 00.14&$-$20 33 09.2&2013-06-04&34000&C189Hr  &Carpenter&  8.9\\
     46& 2MASSJ16070211-2019387 &  16 07 02.11&$-$20 19 38.7&2007-05-24&34000&C269Hr  &Dahm    &  6.4\\
     47& 2MASSJ16070873-1927341 &  16 07 08.73&$-$19 27 34.1&2015-06-01&34000&C247Hr  &Carpenter&  2.0\\
     48& 2MASSJ16071971-2020555 &  16 07 19.71&$-$20 20 55.5&2015-06-01&34000&C247Hr  &Carpenter&  2.7\\
     49& 2MASSJ16072625-2432079 &  16 07 26.25&$-$24 32 07.9&2013-06-04&34000&C189Hr  &Carpenter&  7.2\\
     50& 2MASSJ16072747-2059442 &  16 07 27.47&$-$20 59 44.2&2013-06-04&34000&C189Hr  &Carpenter&  7.2\\
     51& 2MASSJ16073939-1917472 &  16 07 39.39&$-$19 17 47.2&2015-06-01&34000&C247Hr  &Carpenter&  6.2\\
     52& 2MASSJ16075796-2040087 &  16 07 57.96&$-$20 40 08.7&2015-06-01&34000&C247Hr  &Carpenter& 20.3\\
     53& 2MASSJ16080555-2218070 &  16 08 05.55&$-$22 18 07.0&2015-06-01&34000&C247Hr  &Carpenter&  9.0\\
     54& 2MASSJ16081566-2222199 &  16 08 15.66&$-$22 22 19.9&2013-06-04&34000&C189Hr  &Carpenter&  8.1\\
     55& 2MASSJ16082751-1949047 &  16 08 27.51&$-$19 49 04.7&2007-05-24&34000&C269Hr  &Dahm    & 13.1\\
     56& 2MASSJ16083455-2211559 &  16 08 34.55&$-$22 11 55.9&2013-06-04&34000&C189Hr  &Carpenter&  3.4\\
     57& 2MASSJ16084894-2400045 &  16 08 48.94&$-$24 00 04.5&2013-06-04&34000&C189Hr  &Carpenter&  7.2\\
     58& 2MASSJ16090002-1908368 &  16 09 00.02&$-$19 08 36.8&2007-05-24&34000&C269Hr  &Dahm    & 13.2\\
     59& 2MASSJ16090075-1908526 &  16 09 00.75&$-$19 08 52.6&2006-06-16&34000&C315Hr  &Carpenter& 31.0\\
     60& 2MASSJ16093558-1828232 &  16 09 35.58&$-$18 28 23.2&2013-06-04&34000&C189Hr  &Carpenter&  5.8\\
     61& 2MASSJ16094098-2217594 &  16 09 40.98&$-$22 17 59.4&2015-06-01&34000&C247Hr  &Carpenter&  9.0\\
     62& 2MASSJ16095361-1754474 &  16 09 53.61&$-$17 54 47.4&2007-05-24&34000&C269Hr  &Dahm    &  9.4\\
     63& 2MASSJ16095441-1906551 &  16 09 54.41&$-$19 06 55.1&2015-06-01&34000&C247Hr  &Carpenter&  9.0\\
     64& 2MASSJ16101473-1919095 &  16 10 14.73&$-$19 19 09.5&2015-06-01&34000&C247Hr  &Carpenter&  5.7\\
     65& 2MASSJ16102819-1910444 &  16 10 28.19&$-$19 10 44.4&2015-06-01&34000&C247Hr  &Carpenter&  2.3\\
     66& 2MASSJ16103956-1916524 &  16 10 39.56&$-$19 16 52.4&2015-06-01&34000&C247Hr  &Carpenter&  4.5\\
     67& 2MASSJ16104202-2101319 &  16 10 42.02&$-$21 01 31.9&2015-06-01&34000&C247Hr  &Carpenter& 13.7\\
     68& 2MASSJ16104636-1840598 &  16 10 46.36&$-$18 40 59.8&2015-06-01&34000&C247Hr  &Carpenter&  3.8\\
     69& 2MASSJ16111330-2019029 &  16 11 13.30&$-$20 19 02.9&2013-06-04&34000&C189Hr  &Carpenter&  8.1\\
     70& 2MASSJ16111534-1757214 &  16 11 15.34&$-$17 57 21.4&2006-06-16&34000&C315Hr  &Carpenter& 28.4\\
     71& 2MASSJ16112057-1820549 &  16 11 20.57&$-$18 20 54.9&2011-04-24&60000&ENG     &Engineering& 49.7\\
     72& 2MASSJ16113134-1838259 &  16 11 31.34&$-$18 38 25.9&2011-04-24&-9999&ENG     &Engineering& 58.0\\
     73& 2MASSJ16115091-2012098 &  16 11 50.91&$-$20 12 09.8&2008-05-23&48000&C199Hb&Herczeg&  7.7\\
     74& 2MASSJ16122737-2009596 &  16 12 27.37&$-$20 09 59.6&2015-06-01&34000&C247Hr  &Carpenter&  2.7\\
     75& 2MASSJ16123916-1859284 &  16 12 39.16&$-$18 59 28.4&2013-06-04&34000&C189Hr  &Carpenter& 15.7\\
     76& 2MASSJ16124893-1800525 &  16 12 48.93&$-$18 00 52.5&2015-06-01&34000&C247Hr  &Carpenter&  5.2\\
     77& 2MASSJ16125533-2319456 &  16 12 55.33&$-$23 19 45.6&2015-06-01&34000&C247Hr  &Carpenter& 26.6\\
     78& 2MASSJ16130996-1904269 &  16 13 09.96&$-$19 04 26.9&2013-06-04&34000&C189Hr  &Carpenter&  4.6\\
     79& 2MASSJ16133650-2503473 &  16 13 36.50&$-$25 03 47.3&2013-06-04&34000&C189Hr  &Carpenter& 10.2\\
     80& 2MASSJ16135434-2320342 &  16 13 54.34&$-$23 20 34.2&2013-06-04&34000&C189Hr  &Carpenter& 10.1\\
     81& 2MASSJ16141107-2305362 &  16 14 11.07&$-$23 05 36.2&2008-06-25&45000&ENG     &hireseng& 82.0\\
     82& 2MASSJ16142029-1906481 &  16 14 20.29&$-$19 06 48.1&2006-06-16&34000&C315Hr  &Carpenter& 15.6\\
     83& 2MASSJ16142893-1857224 &  16 14 28.93&$-$18 57 22.4&2015-06-01&34000&C247Hr  &Carpenter&  9.4\\
     84& 2MASSJ16143367-1900133 &  16 14 33.67&$-$19 00 13.3&2013-06-04&34000&C189Hr  &Carpenter& 12.9\\
     85& 2MASSJ16145918-2750230 &  16 14 59.18&$-$27 50 23.0&2012-06-20&45000&N190Hr  &Soderblom& 77.9\\
     86& 2MASSJ16145928-2459308 &  16 14 59.28&$-$24 59 30.8&2015-06-01&34000&C247Hr  &Carpenter&  8.6\\
     87& 2MASSJ16151239-2420091 &  16 15 12.39&$-$24 20 09.1&2013-06-04&34000&C189Hr  &Carpenter&  2.5\\
     88& 2MASSJ16153456-2242421 &  16 15 34.56&$-$22 42 42.1&2013-06-04&34000&C189Hr  &Carpenter& 13.4\\
     89& 2MASSJ16154416-1921171 &  16 15 44.16&$-$19 21 17.1&2013-06-04&34000&C189Hr  &Carpenter& 15.8\\
     90& 2MASSJ16163345-2521505 &  16 16 33.45&$-$25 21 50.5&2013-06-04&34000&C189Hr  &Carpenter& 15.1\\
     91& 2MASSJ16181904-2028479 &  16 18 19.04&$-$20 28 47.9&2015-06-01&34000&C247Hr  &Carpenter&  4.2\\
     92& 2MASSJ16215466-2043091 &  16 21 54.66&$-$20 43 09.1&2011-06-28&45000&C196Hr  &Hillenbrand& 27.2\\
     93& 2MASSJ16220961-1953005 &  16 22 09.61&$-$19 53 00.5&2015-06-01&34000&C247Hr  &Carpenter& 16.0\\
     94& 2MASSJ16235385-2946401 &  16 23 53.85&$-$29 46 40.1&2015-06-01&34000&C247Hr  &Carpenter& 24.9\\
     95& 2MASSJ16270942-2148457 &  16 27 09.42&$-$21 48 45.7&2015-06-01&34000&C247Hr  &Carpenter&  3.3\\
     96& 2MASSJ15564244-2039339 &  15 56 42.44&$-$20 39 33.9&2015-06-02&34000&C247Hr  &Carpenter& 12.0\\
     97& 2MASSJ15583620-1946135 &  15 58 36.20&$-$19 46 13.5&2015-06-02&34000&C247Hr  &Carpenter&  7.8\\
     98& 2MASSJ15594426-2029232 &  15 59 44.26&$-$20 29 23.2&2015-06-02&34000&C247Hr  &Carpenter&  6.6\\
     99& 2MASSJ16011398-2516281 &  16 01 13.98&$-$25 16 28.1&2015-06-02&34000&C247Hr  &Carpenter& 11.1\\
    100& 2MASSJ16012902-2509069 &  16 01 29.02&$-$25 09 06.9&2015-06-02&34000&C247Hr  &Carpenter& 13.5\\
    101& 2MASSJ16023587-2320170 &  16 02 35.87&$-$23 20 17.0&2015-06-02&34000&C247Hr  &Carpenter& 12.1\\
    102& 2MASSJ16041893-2430392 &  16 04 18.93&$-$24 30 39.2&2015-06-02&34000&C247Hr  &Carpenter& 28.4\\
    103& 2MASSJ16052157-1821412 &  16 05 21.57&$-$18 21 41.2&2015-06-02&34000&C247Hr  &Carpenter& 34.6\\
    104& 2MASSJ16064794-1841437 &  16 06 47.94&$-$18 41 43.7&2015-06-02&34000&C247Hr  &Carpenter& 25.9\\
    105& 2MASSJ16093164-2229224 &  16 09 31.64&$-$22 29 22.4&2015-06-02&34000&C247Hr  &Carpenter& 15.1\\
    106& 2MASSJ16100501-2132318 &  16 10 05.01&$-$21 32 31.8&2015-06-02&34000&C247Hr  &Carpenter& 24.3\\
    107& 2MASSJ16112601-2631558 &  16 11 26.01&$-$26 31 55.8&2015-06-02&34000&C247Hr  &Carpenter& 20.5\\
    108& 2MASSJ16120505-2043404 &  16 12 05.05&$-$20 43 40.4&2015-06-02&34000&C247Hr  &Carpenter& 17.4\\
    109& 2MASSJ16120668-3010270 &  16 12 06.68&$-$30 10 27.0&2015-06-02&34000&C247Hr  &Carpenter& 21.5\\
    110& 2MASSJ16132190-2136136 &  16 13 21.90&$-$21 36 13.6&2015-06-02&34000&C247Hr  &Carpenter&  6.3\\
    111& 2MASSJ16145244-2513523 &  16 14 52.44&$-$25 13 52.3&2015-06-02&34000&C247Hr  &Carpenter&  5.8\\
    112& 2MASSJ16153220-2010236 &  16 15 32.20&$-$20 10 23.6&2015-06-02&34000&C247Hr  &Carpenter& 10.7\\
    113& 2MASSJ16194711-2203112 &  16 19 47.11&$-$22 03 11.2&2015-06-02&34000&C247Hr  &Carpenter&  3.2\\
    114& 2MASSJ16200616-2212385 &  16 20 06.16&$-$22 12 38.5&2015-06-02&34000&C247Hr  &Carpenter&  6.0\\
    115& 2MASSJ16252883-2607538 &  16 25 28.83&$-$26 07 53.8&2015-06-02&34000&C247Hr  &Carpenter& 17.1\\
\enddata
\end{deluxetable*}

\clearpage
\setcounter{table}{3}
\startlongtable
\begin{deluxetable*}{rccrrcccccrccccccc}
\tablecaption{Parameters for the \OIa\ line profile decomposition  \label{tabe_lineprofile}}
\tablewidth{700pt}

\tablehead{\colhead{ID}&\colhead{Name} &\colhead{$FWHM$} &\colhead{$v_{\rm c}$} &\colhead{$EW$}  &\colhead{$L_{\rm 6300}$} &Class \\
           \colhead{  }&\colhead{    } &\colhead{(\kms)}    &\colhead{(\kms)}   &\colhead{(\AA)}&\colhead{($L_{\odot}$)}  &      
}
\startdata
1 & 2MASSJ15514032-2146103&   72.9&$   -6.5$&   0.64&$  -6.50$&SC\\
\hline
2 & 2MASSJ15521088-2125372&   56.6&$   -6.4$&  36.32&$  -5.88$&SC\\
\hline
3 & 2MASSJ15530132-2114135&   67.0&$   -1.3$&   0.68&$  -6.47$&SC\\
\hline
4 & 2MASSJ15534211-2049282&   81.7&$   -7.6$&   4.14&$  -5.30$&BC\\
  &     &   21.6&$   13.7$&   0.47&$  -6.24$&NC\\
\hline
6 & 2MASSJ15562477-2225552&   24.2&$   -1.4$&   0.32&$  -6.56$&SC\\
\hline
7 & 2MASSJ15570641-2206060&   27.3&$  -33.4$&   0.51&$  -6.66$&HVC-B\\
  &     &   62.4&$   11.1$&   1.34&$  -6.25$&SCJ\\
\hline
10 & 2MASSJ15582981-2310077&   56.2&$    3.4$&   0.53&$  -6.38$&BC\\
  &     &   18.8&$   -1.5$&   0.21&$  -6.78$&NC\\
\hline
11 & 2MASSJ15583692-2257153&   66.3&$    2.8$&   0.09&$  -4.44$&BC\\
  &     &   19.1&$    2.1$&   0.02&$  -5.01$&NC\\
\hline
15 & 2MASSJ16001844-2230114&   21.5&$    0.8$&   0.66&$  -5.86$&SC\\
\hline
16 & 2MASSJ16014086-2258103&   49.7&$   -0.8$&   1.20&$  -5.51$&SC\\
\hline
17 & 2MASSJ16014157-2111380&   78.2&$    1.4$&   1.30&$  -6.36$&SC\\
\hline
20 & 2MASSJ16020757-2257467&   52.7&$    0.0$&   0.35&$  -5.75$&SC\\
\hline
25 & 2MASSJ16032225-2413111&   61.1&$    2.9$&   0.97&$  -5.64$&SC\\
\hline
26 & 2MASSJ16035767-2031055&   42.3&$    0.0$&   0.22&$  -4.91$&SC\\
\hline
28 & 2MASSJ16041740-1942287&   68.4&$   -1.8$&   0.42&$  -6.13$&SC\\
\hline
29 & 2MASSJ16042165-2130284&   18.1&$   -0.4$&   0.18&$  -4.70$&SC\\
\hline
34 & 2MASSJ16052661-1957050&   55.8&$    8.1$&   0.29&$  -6.82$&SC\\
\hline
35 & 2MASSJ16053215-1933159&   69.5&$  -10.1$&   0.79&$  -6.51$&SC\\
\hline
39 & 2MASSJ16062196-1928445&   41.5&$   -5.9$&   0.23&$  -5.33$&BC\\
  &     &   13.4&$   -2.7$&   0.17&$  -5.47$&NC\\
\hline
40 & 2MASSJ16062277-2011243&   50.0&$   -6.8$&   0.26&$  -6.65$&SC\\
\hline
42 & 2MASSJ16064102-2455489&   85.5&$   -7.6$&   1.81&$  -6.08$&SC\\
\hline
45 & 2MASSJ16070014-2033092&   24.0&$    0.7$&   0.21&$  -6.07$&SC\\
\hline
49 & 2MASSJ16072625-2432079&   70.5&$    0.4$&   0.70&$  -5.79$&SC\\
\hline
52 & 2MASSJ16075796-2040087&   61.0&$  -76.6$&  13.58&$  -3.87$&HVC-B\\
  &     &  202.5&$   15.1$&   4.82&$  -4.32$&BC\\
  &     &   44.6&$  -14.8$&   5.92&$  -4.23$&NC\\
\hline
59 & 2MASSJ16090075-1908526&   75.0&$   -5.6$&   0.48&$  -5.04$&SC\\
\hline
60 & 2MASSJ16093558-1828232&   48.9&$    2.6$&   0.49&$  -6.14$&SC\\
\hline
62 & 2MASSJ16095361-1754474&   76.8&$    4.8$&   0.43&$  -6.86$&SC\\
\hline
69 & 2MASSJ16111330-2019029&   43.7&$   -3.0$&   0.54&$  -5.67$&SC\\
\hline
70 & 2MASSJ16111534-1757214&   40.3&$   -2.8$&   0.12&$  -5.75$&SC\\
\hline
72 & 2MASSJ16113134-1838259&   56.3&$ -221.9$&   0.11&$  -4.65$&HVC-B\\
  &     &  225.5&$  -85.4$&   0.28&$  -4.23$&HVC-B\\
  &     &   54.5&$  -13.5$&   0.16&$  -4.46$&BC\\
  &     &   14.5&$   -0.7$&   0.11&$  -4.62$&NC\\
\hline
73 & 2MASSJ16115091-2012098&   69.1&$    2.3$&   1.04&$  -5.83$&SC\\
\hline
75 & 2MASSJ16123916-1859284&   95.8&$  -25.8$&   0.23&$  -5.49$&SC\\
\hline
78 & 2MASSJ16130996-1904269&   48.2&$   -5.7$&   2.19&$  -5.55$&SC\\
\hline
79 & 2MASSJ16133650-2503473&   77.8&$    2.5$&   0.98&$  -5.27$&SC\\
\hline
82 & 2MASSJ16142029-1906481&  101.9&$  -32.2$&   4.08&$  -4.19$&HVC-B\\
  &     &   31.6&$   -1.7$&   0.53&$  -5.08$&SCJ\\
\hline
84 & 2MASSJ16143367-1900133&   73.4&$  -20.0$&   1.27&$  -4.93$&SC\\
\hline
86 & 2MASSJ16145928-2459308&   55.1&$    0.4$&   0.38&$  -6.62$&SC\\
\hline
87 & 2MASSJ16151239-2420091&   73.8&$   -0.2$&   1.31&$  -6.47$&SC\\
\hline
88 & 2MASSJ16153456-2242421&   53.8&$   -9.0$&   0.28&$  -5.14$&SC\\
\hline
89 & 2MASSJ16154416-1921171&   92.3&$ -120.7$&   0.51&$  -4.75$&HVC-B\\
  &     &   57.5&$  108.2$&   0.25&$  -5.06$&HVC-R\\
  &     &   80.7&$   -9.8$&   0.78&$  -4.57$&BC\\
  &     &   22.7&$   -3.7$&   0.41&$  -4.85$&NC\\
\hline
100 & 2MASSJ16012902-2509069&   92.0&$    6.1$&   0.60&$  -5.97$&SC\\
\hline
102 & 2MASSJ16041893-2430392&   74.5&$   -3.9$&   0.19&$  -5.72$&SC\\
\hline
103 & 2MASSJ16052157-1821412&   45.1&$    1.6$&   0.19&$  -4.78$&SC\\
\hline
106 & 2MASSJ16100501-2132318&   34.4&$   -1.5$&   0.13&$  -5.45$&SC\\
\hline
109 & 2MASSJ16120668-3010270&   23.8&$    0.3$&   0.28&$  -5.35$&SC\\
\hline
\hline
\enddata
\end{deluxetable*}
\clearpage

\setcounter{table}{8}
\startlongtable
\begin{deluxetable*}{ccccccc}
\tablecaption{Sources used in this work to estimate $r_{\rm base}$ and wind mass-loss rates \label{tabe_Mwind}}
\tabletypesize{\scriptsize}
\tablehead{
\colhead{Name} & \colhead{SpT}   &\colhead{Log~$\dot{M}_{\rm acc}$}  &\colhead{Log~$\dot{M}_{\rm loss}$}  & \colhead{$r_{\rm base}$}  &\colhead{Disk inclination}& \colhead{Ref$^{\alpha}$} \\
\colhead{}     & \colhead{}      & \colhead{($M$\accunit)}           & \colhead{($M$\accunit)}            & \colhead{(au)}            &\colhead{(degree)}               & \colhead{} }
\startdata
\multicolumn{7}{c}{Upper Sco}\\
\hline
2MASSJ15514032-2146103&M4.8&$<-10.77$&-10.30&  0.076&  84.0& 1\\
2MASS J15521088-2125372&M4.9&$-11.30$&-10.60&   0.62&  24.0& 1\\
2MASS J15530132-2114135&M4.7&$-10.31$&-10.63&   0.18&  47.0&1\\
2MASS J15562477-2225552&M4.2&$<-10.67$&-11.53&    1.1&  85.0&1\\
2MASS J16001844-2230114&M5.2&$ -8.22$&-11.58&    6.2&  24.0&1\\
2MASS J16014086-2258103&M4.0&$ -8.46$& -9.89&   0.29&  74.0&1\\
2MASS J16014157-2111380&M5.0&$ -9.85$&-10.00&  0.053&  80.0&1\\
2MASS J16020757-2257467&M2.5&$<-10.28$&-10.18&   0.59&  57.0&1\\
2MASS J16032225-2413111&M3.9&$ -9.06$& -9.92&   0.23&  64.0&1\\
2MASS J16035767-2031055&K5.1&$ -9.29$& -9.86&    1.9&  69.0&1\\
2MASS J16041740-1942287&M3.6&$<-10.46$&-10.29&   0.18&  80.0&1\\
2MASS J16062277-2011243&M4.1&$<-10.93$&-10.96&   0.25&  85.0&1\\
2MASS J16064102-2455489&M4.8&$ -9.64$&-10.05&   0.11&  40.0&1\\
2MASS J16072625-2432079&M3.9&$ -9.61$&-10.18&   0.29&  43.0&1\\
2MASS J16090075-1908526&M0.6&$ -8.92$& -9.34&   0.43&  56.0&1\\
2MASS J16093558-1828232&M3.7&$<-10.27$&-10.57&   0.32&  83.0&1\\
2MASS J16095361-1754474&M5.0&$-10.26$&-10.56&  0.060&  86.0&1\\
2MASS J16111330-2019029&M4.3&$ -8.48$&-11.14&    3.6&  17.0&1\\
2MASS J16115091-2012098&M3.8&$<-10.54$& -9.93&   0.15&  86.0&1\\
2MASS J16123916-1859284&M2.0&$ -8.69$& -9.46&   0.20&  53.0&1\\
2MASS J16133650-2503473&M3.8&$ -8.33$& -9.26&   0.12&  86.0&1\\
2MASS J16143367-1900133&M3.3&$ -9.29$& -9.10&   0.18&  69.0&1\\
2MASS J16153456-2242421&M0.2&$ -8.68$& -9.87&    1.2&  46.0&1\\
\hline
\multicolumn{7}{c}{NGC~2264 sample}\\
\hline
CSIMon-000021&K5.0&$ -9.02$&-10.72&   11.9&  53.0&2\\
CSIMon-000131&K7.0&$ -7.96$& -8.55&   0.76&  90.0&2\\
CSIMon-000137&M2.0&$ -8.54$& -8.67&   0.11&  40.0&2\\
CSIMon-000168&K7.5&$ -7.92$& -9.02&   0.67&  70.0&2\\
CSIMon-000273&M1.0&$ -7.68$& -8.88&   0.34&  90.0&2\\
CSIMon-000296&K2.0&$ -8.77$& -9.16&    2.8&  76.0&2\\
CSIMon-000314&M3.0&$ -8.02$& -8.58&  0.064&  73.0&2\\
CSIMon-000326&M0.0&$ -8.77$&-10.29&    1.5&  44.0&2\\
CSIMon-000328&M1.0&$ -8.62$& -9.48&   0.29&  86.0&2\\
CSIMon-000335&K4.0&$ -8.64$& -9.41&    4.7&  46.0&2\\
CSIMon-000341&M0.5&$ -7.21$& -9.17&   0.47&  90.0&2\\
CSIMon-000412&M1.0&$ -7.00$& -7.87&  0.081&  56.0&2\\
CSIMon-000434&M2.5&$ -8.32$& -8.40&  0.049&  54.0&2\\
CSIMon-000456&K4.0&$ -8.29$& -9.00&    1.2&  60.0&2\\
CSIMon-000462&G0.0&$ -7.49$&-10.27&   57.3&  28.0&2\\
CSIMon-000558&K4.0&$ -7.21$& -8.92&    3.5&  58.0&2\\
CSIMon-000598&M1.0&$ -7.38$& -8.04&   0.25&  90.0&2\\
CSIMon-000619&K8.5&$ -7.40$& -8.30&   0.38&  57.0&2\\
CSIMon-000637&M1.0&$ -8.49$& -9.18&   0.40&  90.0&2\\
CSIMon-000667&K3.0&$ -8.74$& -9.20&    3.2&  66.0&2\\
CSIMon-000717&M0.5&$ -7.55$& -7.71&  0.047&  77.0&2\\
CSIMon-000765&K1.0&$ -8.19$&-10.24&   12.3&  36.0&2\\
CSIMon-000804&K5.5&$ -8.17$& -9.69&    3.0&  45.0&2\\
CSIMon-000811&K6.0&$ -8.17$& -8.65&   0.70&  65.0&2\\
CSIMon-000879&M1.0&$ -7.92$& -9.00&   0.28&  90.0&2\\
CSIMon-000904&K7.0&$-10.13$& -8.71&    1.3&  37.0&2\\
CSIMon-000928&M0.0&$ -8.25$& -9.34&   0.57&  90.0&2\\
CSIMon-001033&K7.0&$ -7.70$& -8.27&   0.20&  90.0&2\\
CSIMon-001038&M0.0&$ -8.82$& -8.72&   0.26&  66.0&2\\
CSIMon-001054&M2.0&$ -6.68$& -7.51&  0.042&  66.0&2\\
CSIMon-001064&M1.0&$ -7.82$& -8.22&   0.38&  35.0&2\\
CSIMon-001094&K5.0&$ -7.80$& -9.92&    4.7&  66.0&2\\
CSIMon-001114&M1.5&$ -7.37$& -9.88&    2.1&  44.0&2\\
CSIMon-001131&M2.0&$ -7.68$& -9.06&   0.82&  38.0&2\\
CSIMon-001140&K4.0&$ -8.11$& -9.56&    3.3&  59.0&2\\
CSIMon-001157&K6.0&$ -8.40$& -9.25&    1.4&  46.0&2\\
CSIMon-001174&M2.0&$ -7.21$& -9.00&   0.47&  73.0&2\\
CSIMon-001234&K6.0&$ -8.82$& -8.52&   0.43&  60.0&2\\
CSIMon-001279&K6.0&$ -9.34$& -8.92&    1.0&  34.0&2\\
\hline
\multicolumn{7}{c}{SFB sample}\\
\hline
CX Tau&M2.5&$ -9.37$&-11.16&    2.2&  61.0&3\\
FP Tau&M2.6&$ -9.17$& -9.63&   0.35&  66.0&3\\
UX Tau&K0.0&$ -8.81$&-10.15&   10.8&  39.0&4\\
DM Tau&M3.0&$ -8.79$&-10.85&    5.5&  34.0&4\\
LKCa 15&K5.5&$ -8.75$& -9.83&    2.3&  51.0&4\\
DS Tau&M0.4&$ -8.42$& -8.50&   0.11&  71.0&3\\
Sz 65A&K6.0&$ -8.90$& -8.57&   0.26&  61.0&5\\
GW Lup&M2.3&$ -8.71$& -9.65&   0.67&  40.0&6\\
GQ Lup&K5.0&$ -7.39$& -8.24&   0.44&  60.0&7\\
IM Lup&K6.0&$ -8.68$& -8.70&   0.31&  48.0&5\\
RY Lup&K2.0&$ -8.59$& -9.82&    5.8&  68.0&8\\
Sz 111&M1.2&$ -8.79$&-10.23&    3.7&  53.0&8\\
DoAr 44&K2.0&$ -8.05$&-10.44&   16.2&  16.0&4\\
RNO 90&G8.0&$ -7.26$& -9.29&    5.0&  37.0&9\\
V1121 Oph&K4.0&$ -8.30$&-11.40&   31.7&  31.0&4\\
RX~J1852.3-3700&K4.0&$ -9.03$&-11.71&   87.7&  16.0&10\\
CY Tau&M2.3&$ -8.19$& -9.80&    2.0&  27.0&11\\
GM Aur&K6.0&$ -8.16$& -9.68&    4.1&  55.0&11\\
GO Tau&M2.3&$ -8.93$& -9.65&   0.30&  53.0&11\\
V836 Tau&M0.8&$ -9.56$& -9.26&   0.81&  51.0&11\\
\enddata
\tablerefs{1. \citet{2017ApJ...851...85B}, 2. \cite{2018AA...620A..87M}, 3. \citet{2017ApJ...844..158S},  4. \citet{2017ApJ...845...44T}, 5. \cite{2017AA...606A..88T}, 6 \citet{2016ApJ...828...46A}, 7. \cite{2017ApJ...835...17M}, 8. \citet{2018ApJ...854..177V}, 9 \cite{2011ApJ...733...84P}, 10. \cite{2010AJ....140..887H}, 11. \citet{2017ApJ...845...44T}}
\tablecomments{For the NGC~2264 sample disk inclinations are estimated by assuming that the star and disk have coaxial rotation}
\end{deluxetable*}
\clearpage



\appendix
\section{Observations}\label{Appen:Obs_log}
Table~\ref{Tab:obs_log} provides a summary of the observation log and includes source coordinates, observational dates, nominal spectral resolution, Program ID, and Principal Investigator (PI) for that program.

\section{Flux calibration and spectral classification}\label{Appen:SPT}
We use one or two stars to derive the response curves for HIRES data each night: HD~109011 (K2V) for 2006 June 16; HR~8634 (B8V) for 2006 August 12; HIP~79410 (B9V) for 2007 May 24; HIP~77859 (B2V) for 2007 May 25; HD~191089 (F5V) for 2008 June 25;  HD~100180 (G0V) for 2011 April 24; HD~141943 (G2V) for 2012 June 20; HD~201320 (A0V) and HD~3765 (K2V) for 2013 June 4; HD~88371 (G2V) for 2015 June 1; and HD~201320 (A0V) for 2015 June 2. While early type stars are ideal for this task, they were not available each night, hence the use of later type stars.

Following common practice, we first obtain the BOSZ Kurucz model atmosphere \citep{2012AJ....144..120M} for stars with spectral type earlier than A0 and the PHOENIX model atmospheres \citep{2013A&A...553A...6H} for those later than A0 to derive the response function. Next, we fit their {\it Gaia} and 2MASS broad-band photometry  using the aforementioned models with two free parameters, extinction and stellar angular radius as  in  \cite{2009A&A...504..461F} and \cite{2013ApJS..207....5F}. Then, we shift and rotationally broaden the best-fit model atmosphere and degrade it to the HIRES spectral resolution of individual nights. In each order, the ratio between the observed and model spectrum is fitted with a 5-order polynomial function from which we derive the conversion from counts to absolute flux. Using these conversions, we obtain the flux-calibrated spectra of our targets.  We use optical extinction coefficients  proper for  Mauna Kea \footnote{The coefficients used are listed in the website: www.gemini.edu/sciops/telescopes-and-sites/observing-condition-constraints/extinction.} to correct for atmospheric absorption due to a different airmass between the spectrophotometric standard and the scientific object. We also correct for missing flux due to point spread function differences between the target and the spectrophotometric standard using the FWHMs of individual echelle orders estimated by the MAKEE pipeline. Since there are only one or two spectrophotometric standards per night, this procedure is not achieving an accurate  flux calibration, but does remove spectral features caused by the instrumental response.

The spectral classification is carried out by finding the best match between each target  flux-calibrated spectrum and a set of pre-main sequence spectral templates  obtained with VLT/X-Shooter (see \citealt{2021ApJ...908...49F} for details). We degrade the HIRES spectra to the resolution (R=3,000) of the spectral templates. Then, to avoid uncertainties in the flux calibration, we normalize each target spectrum and the spectral templates using a 5-order polynomial function. To mimic the filling effect on the the photospheric lines from the excess emission from the accretion shocks,  the spectral template is  added an  excess flux. This excess flux is parameterized as $r_{\rm ex,~7500}=\frac{F_{\rm excess,~7500}}{F_{\rm phot,~7500}}$, where $r_{\rm ex,~7500}$ is the veiling at 7500\,\AA, $F_{\rm excess,~7500}$  is the excess flux at \,7500\AA, and $F_{\rm phot,~7500}$ is the photospheric emission at  7500\,\AA. The accretion continuum spectrum  is the same as in \cite{2014ApJ...786...97H} and approximated as a constant. The veiled templates are then normalized to fit the spectra of the targets.  The best-fit template is obtained by minimizing the $\chi^{2}$.
 The target is then assigned the spectral type of the best-fit template spectrum.

Fig.~\ref{Fig:fitspt} summarizes the steps of our spectral classification for source ID~15. The upper panel (a) shows the HIRES spectrum in gray and the 26 median fluxes, within 150\AA\ and uniformly distributed in wavelength (black circles), that are used to fit a 5-order polynomial function (black dashed line). The source spectrum normalized by the polynomial function is shown in gray in the middle panel (b). In the same panel we superimpose in red and blue the best-fit unveiled template and veiled template, respectively.
The extinction ($A_{V}$), as well as the flux  at 7500~\AA\ ($I_{7500}$), are obtained by minimizing the difference between the synthetic and the observed G, BR, and RP {\it Gaia} photometry. $A_{V}$ and $I_{7500}$ are used to 'flux-calibrate' the template spectrum. The lowest panel of Fig.~\ref{Fig:fitspt}(c) compares the flux-calibrated HIRES spectra with the flux-calibrated veiled  and unveiled templates. We note that there are some systematic shift between them. As discussed above, our HIRES spectral calibration is not expected to be very accurate since we have only one or two spectrophotometric standards. Therefore, in this work all the fluxes rely on the flux-calibrated best-fit templates which have been calibrated according to the {\it Gaia} broad-band photometry.

\begin{figure*}
\begin{center}
\includegraphics[width=\columnwidth]{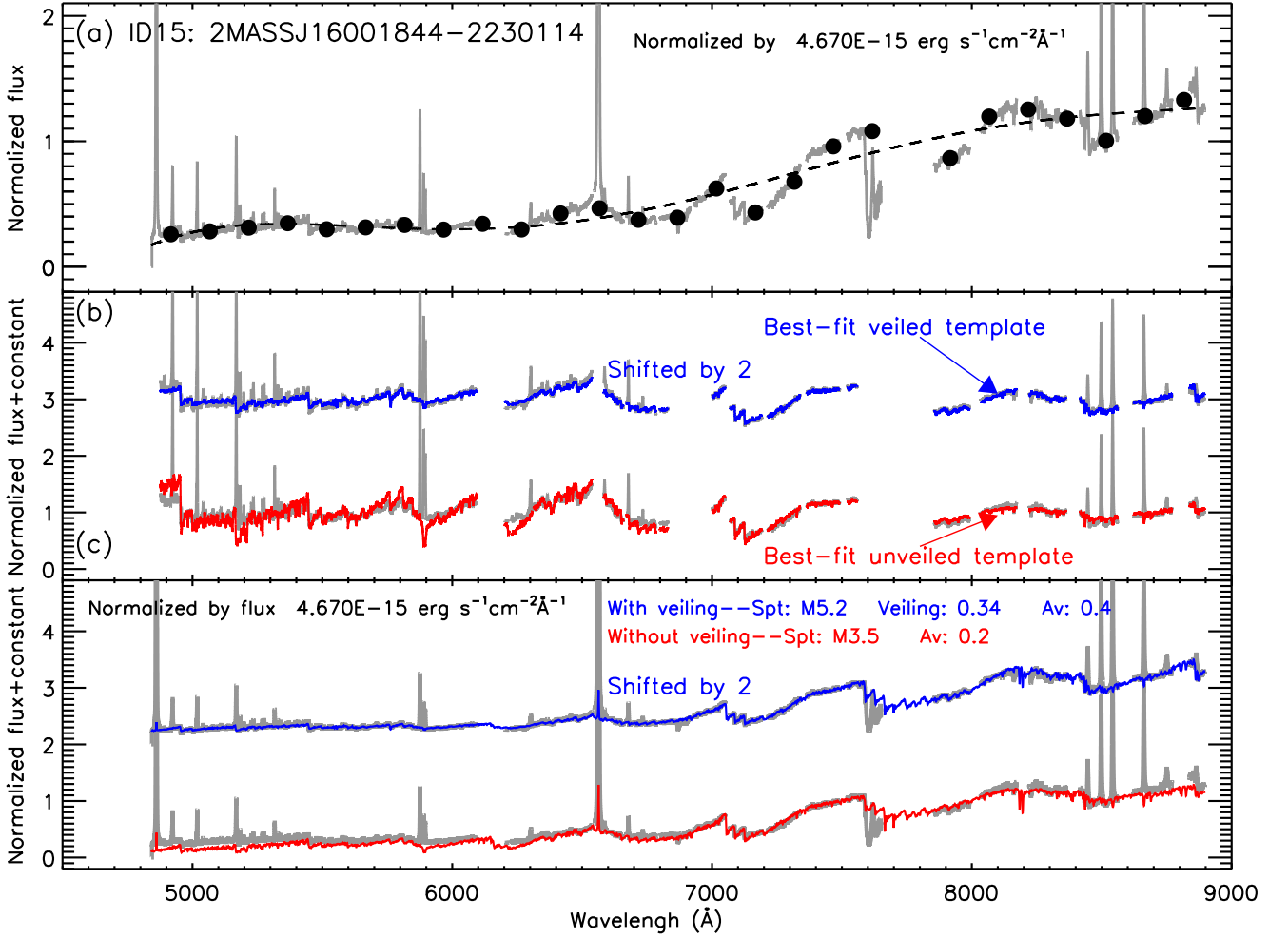}
\caption{(a) The flux-calibrated spectra of the source 15 normalized by 4.67$\times$10$^{-15}$~erg$^{-1}$~s$^{-1}$~cm$^{-2}$~\AA$^{-1}$. The filled circles mark the wavelength bins used to construct the 5-order polynomial function (the dashed line) to normalize the spectra. (b) The spectra (gray lines) of the source 15 normalized by a 5-order polynomial function overlapped with the best-fit unveiled template and veiled template, which are normalized in the same way as the target's spectra. (c) The flux-calibrated spectra of the source 15 normalized by 4.67$\times$10$^{-15}$~erg$^{-1}$~s$^{-1}$~cm$^{-2}$~\AA$^{-1}$\ overlapped with the best-fit unveiled template and veiled template. The best-fit templates have been flux-calibrated with the {\it Gaia} broad-band photometry. 
}\label{Fig:fitspt}
\end{center}
\end{figure*}

\section{Line fluxes and associated uncertainties\label{Appen:UNC}}
We estimate the fluxes of the accretion-related emission lines and the \OIa\ line using the line $EW$s  from fitting the HIRES spectra and the continuum flux near the lines from the flux-calibrated templates. We use two methods to assess the uncertainty in our flux calibration. For the first one, we perform  synthetic photometry in the $r$ band from the panstarr survey and the $J$ band from the 2MASS survey. By comparing the synthetic photometry with the observations, we obtain a mean difference of 0.05~dex in the $r$-band flux and 0.04~dex in the $J$-band flux. 

Our second method uses the overlapping sample of HIRES and  VLT/X-Shooter spectra, a total of 32 sources 31 of which are PDs, and one is a DB. We extract the raw data from ESO archive and obtain the flux-calibrated spectra following the procedure described in \cite{2020A&A...639A..58M}. The uncertainty of the flux-calibration is expected to be about $\pm$10\% \footnote{http://www.eso.org/observing/dfo/quality/PHOENIX/X-Shooter/processing.html}.  Fig.~\ref{Fig:flux_com} compares the 7500\,\AA\ fluxes  from the X-Shooter and our calibrated spectra. From the comparison,  we further estimate the uncertainty on our flux calibration\footnote{Source 11 is not included in the comparison because its HIRES spectra does not cover 7500\,\AA.}.  The mean difference between the two works  is about 0.09\,dex. Sources 21, 69, and 81  show larger differences (0.24-0.52\,dex), which may be attributed to their brightness variability \citep{2018AJ....156...71C}. If excluding the four outliers in Figure~\ref{Fig:flux_com}, the mean difference between two measurements is reduced to $\sim$0.057\,dex. Given that the uncertainty of the flux-calibration on X-Shooter spectra is about $\pm$10\%,  the comparison suggests that the uncertainty of our absolute flux calibration is likely better than 15\%.

\begin{figure*}
\begin{center}
\includegraphics[width=0.6\columnwidth]{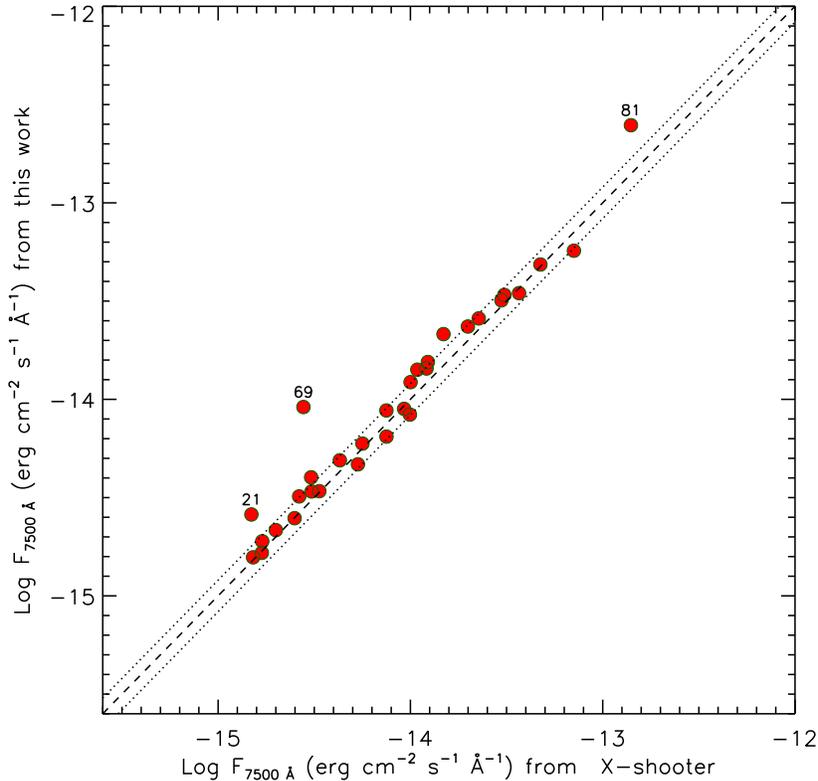}
\caption{Comparison of 7500\,\AA\ fluxes from our work and from  X-Shooter spectra. The dash line shows the 1:1 line while dotted lines mark a $\pm$0.08~dex difference.}\label{Fig:flux_com}
\end{center}
\end{figure*}

\section{Stellar parameters \label{Appen:mass}}
We convert the spectral type to the effective temperature using the conversions in \cite{2017AJ....153..188F}, which are from \cite{2013ApJS..208....9P} for stars earlier than M4 and from \cite{2014ApJ...786...97H} for stars later than M4 as the former does not cover all SpTy in our sample while the latter employs a coarse grid for early SpTy.  With the effective temperatures, we obtain the stellar luminosity using the flux at 7510~\AA\ from flux-calibrated best-fit templates described in Appendix~\ref{Appen:SPT} and the Bolometric Corrections from \cite{2014ApJ...786...97H}. For the sources with accretion, the  flux at 7500~\AA\ only includes the contribution from the stellar photosphere and excludes the excess emission from the accretion shock {using the $r_{\rm ex,~7500}$ derived from the spectral classification (see \S~\ref{Appen:SPT}).}  The {\it Gaia} Data Release 3 geometric distances of individual sources used for computing the stellar luminosity are taken from \citet{2021AJ....161..147B}.

\section{Spectral energy distributions and \OIa\ line profiles} \label{Appen:SED}

Fig.~\ref{Fig:SED1}  show the SEDs of the sources in this work. Their SEDs are constructed using the the Two-Micron All Sky Survey \citep[2MASS, ][]{2006AJ....131.1163S}, the Wide-field Infrared Survey Explorer \citep[WISE,][]{2010AJ....140.1868W}, and the Spitzer photometric data from \cite{2018AJ....156...75E}.  For each source in our sample, we construct a median SED from  CTTs  with similar spectral type in Orion \citep{2009A&A...504..461F,2013ApJS..207....5F,2017AJ....153..188F}. In each panel of Figs~\ref{Fig:SED1}, \ref{Fig:SED2}, and ~\ref{Fig:SED3}, we show in red the upper and lower quartiles of the median SEDs, and in gray the  upper and lower 42.5\% of the median SEDs.

Figures~\ref{Fig:SED_LR1}, \ref{Fig:SED_LR2}, and \ref{Fig:SED_LR3}  show the \OIa\ line profiles for the protoplanetary and debris disks. There are no \OIa\ detections towards the DBs. For the protoplanetary disks, we divide them into two categories, those with and those without \OIa\ detection, and show them in separate figures.

 \begin{figure*}
\begin{center}
\includegraphics[width=\columnwidth]{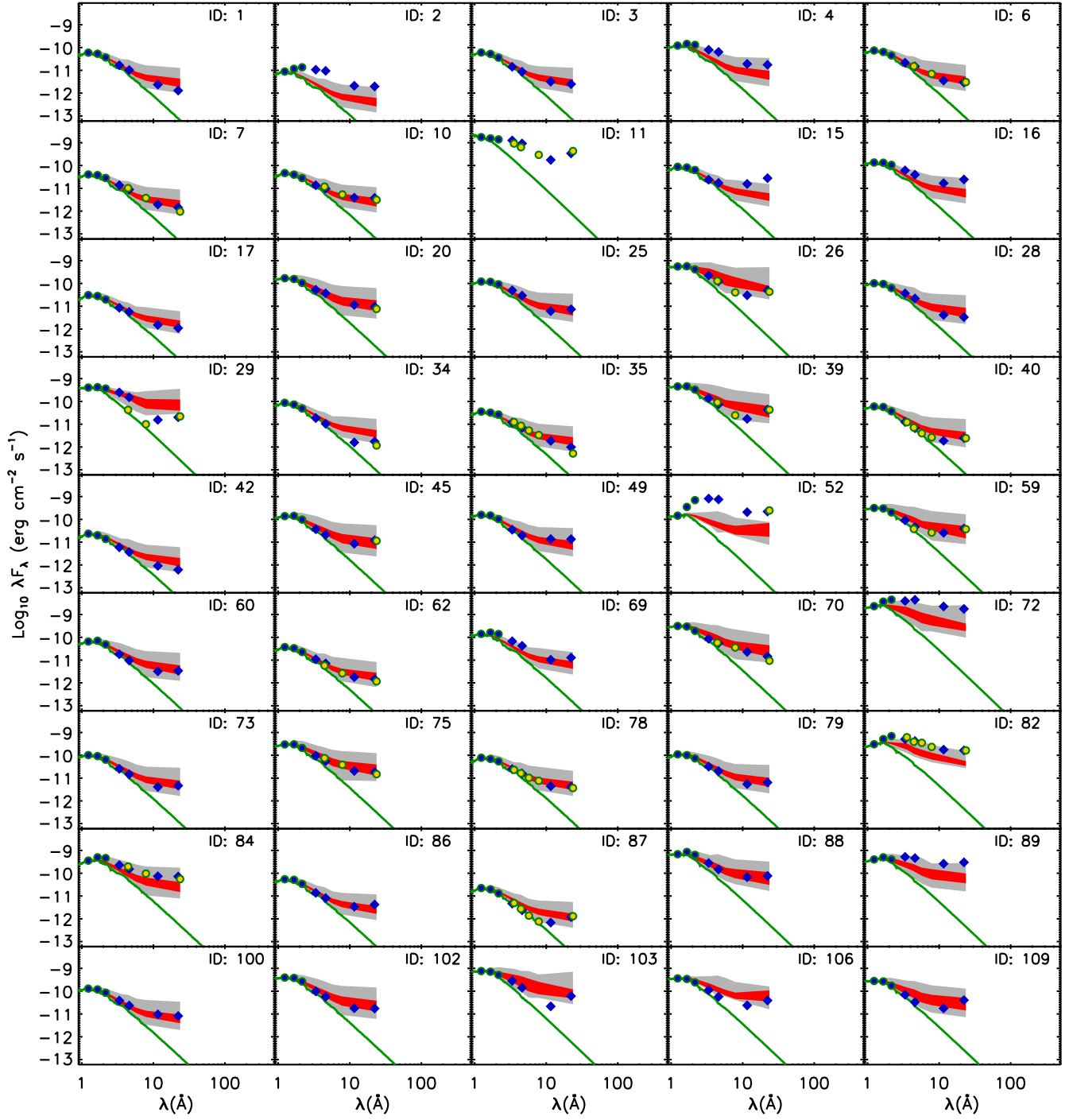}
\caption{SEDs for the protoplanetary disks with \OIa\ detection in  analyzed our sample in this paper. Filled circles are photometric data from 2MASS and WISE (blue), Spitzer (yellow).  The green solid line the stellar photosphere.  The grey region shows the upper and lower quarterlies of the Lynds~1641 classical T~Tauri median SED. The median SED has been reddened  with the extinction of each source, and then normalized to the J-band flux.}\label{Fig:SED1}
\end{center}
\end{figure*}

 \begin{figure*}
\begin{center}
\includegraphics[width=\columnwidth]{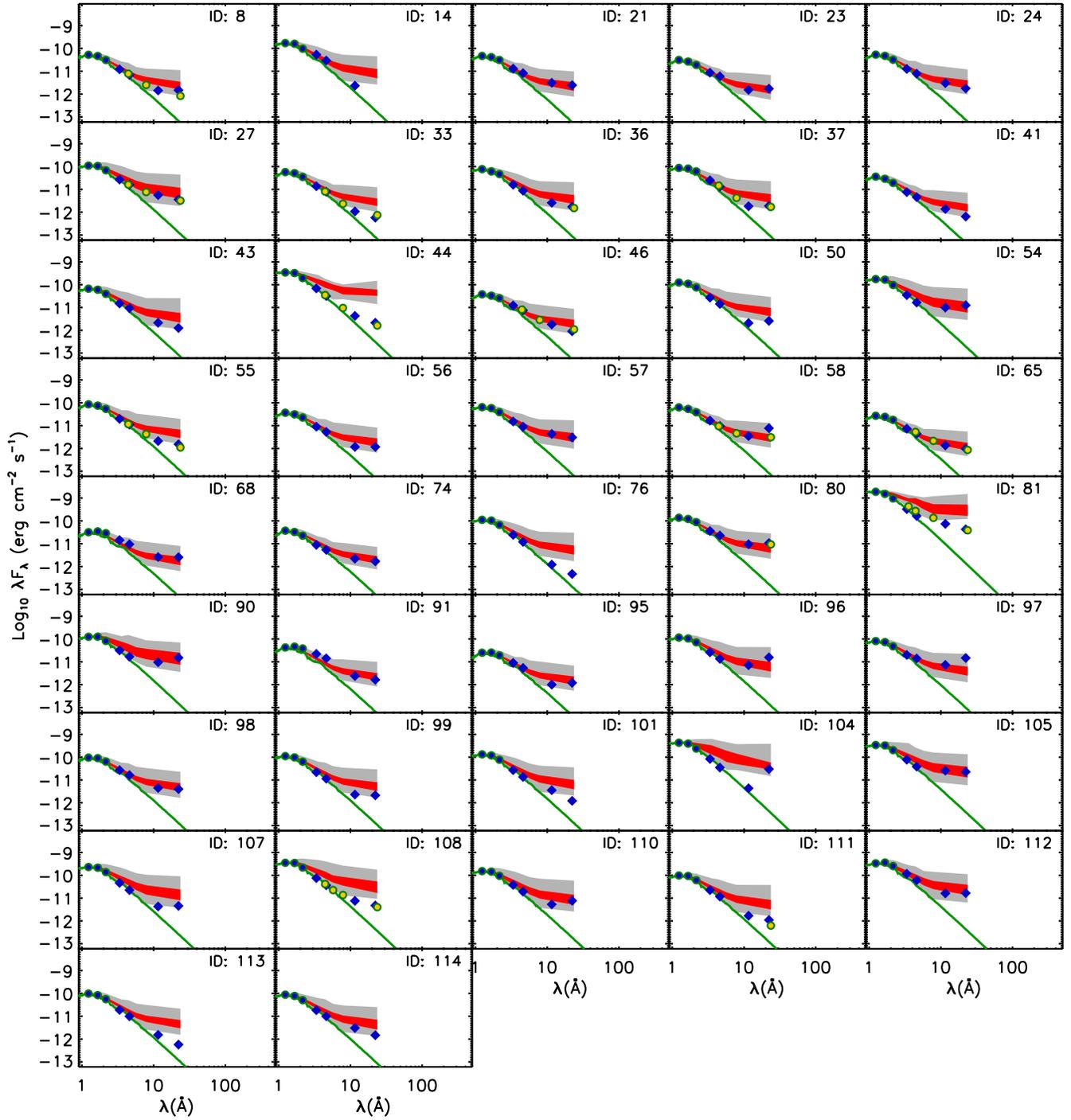}
\caption{Same as Figure~\ref{Fig:SED1} but for the protoplanetary disks without \OIa\ detection. }\label{Fig:SED2}
\end{center}
\end{figure*}

\begin{figure*}
\begin{center}
\includegraphics[width=\columnwidth]{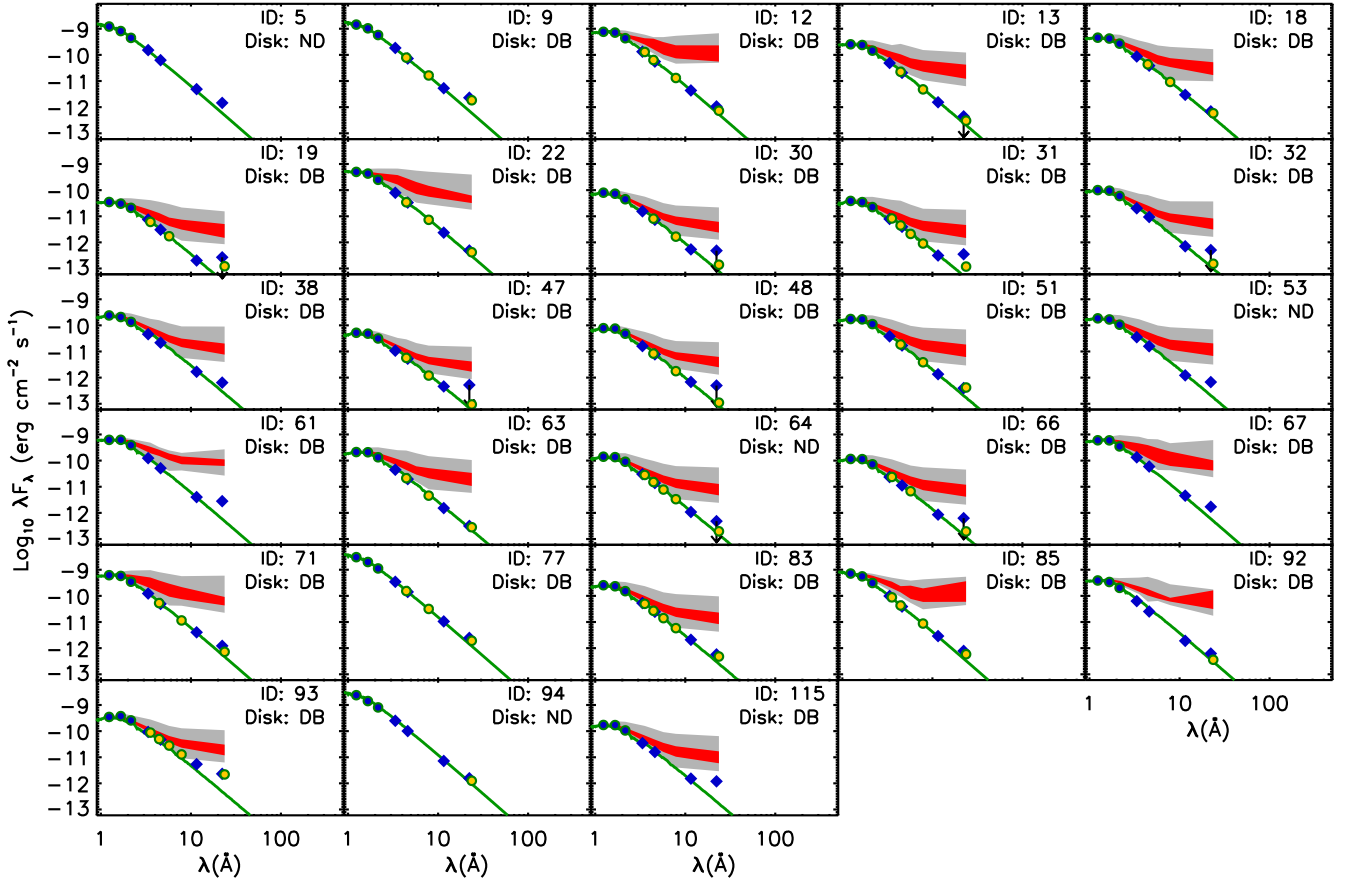}
\caption{Same as Figure~\ref{Fig:SED1} but for sources with debris disks or without disks.}\label{Fig:SED3}
\end{center}
\end{figure*}

\begin{figure*}
\begin{center}
\includegraphics[width=\columnwidth]{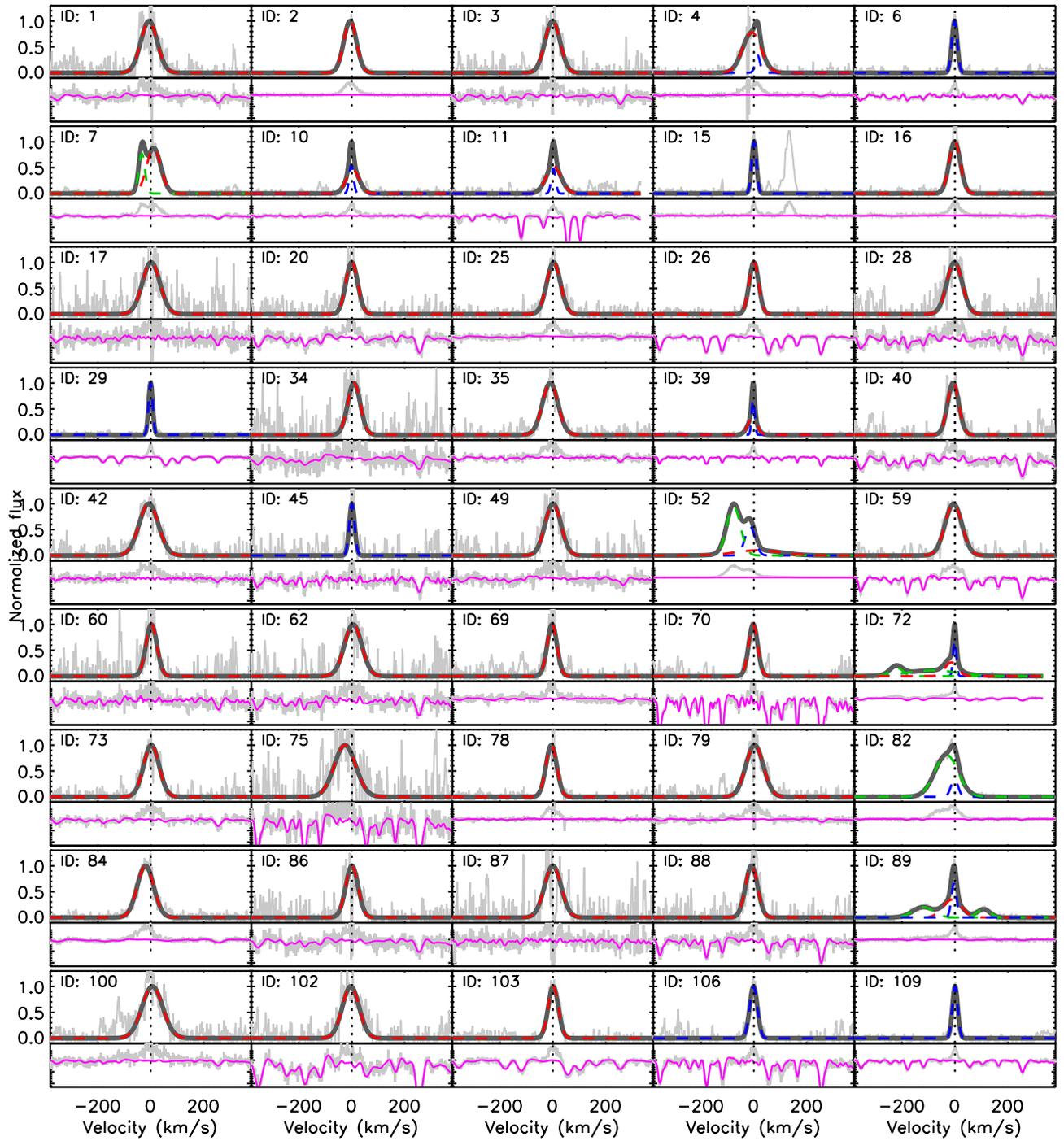}
\caption{Corrected and uncorrected \OIa\ line profiles for the protoplanetary disks with \OIa\ detection in our sample. The figure is same as Fig.~\ref{Fig:example_line}. The SEDs of the corresponding sources are shown in Fig.~\ref{Fig:SED1}.}\label{Fig:SED_LR1}
\end{center}
\end{figure*}

\begin{figure*}
\begin{center}
\includegraphics[width=\columnwidth]{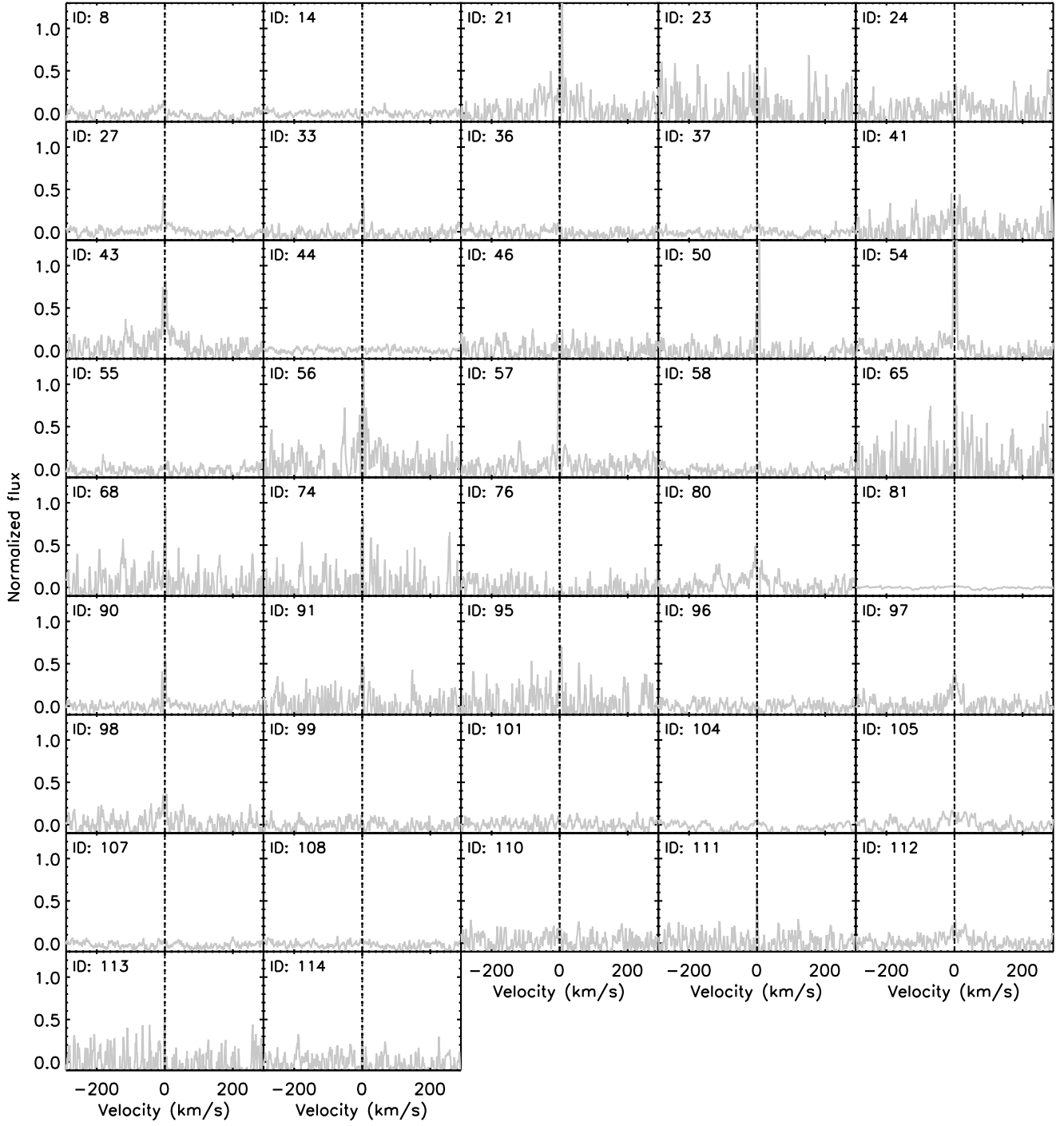}
\caption{\OIa\ line profiles  for the protoplanetary disks without \OIa\ detection.  The SEDs of the corresponding sources are shown in Fig.~\ref{Fig:SED2}.}\label{Fig:SED_LR2}
\end{center}
\end{figure*}

\begin{figure*}
\begin{center}
\includegraphics[width=\columnwidth]{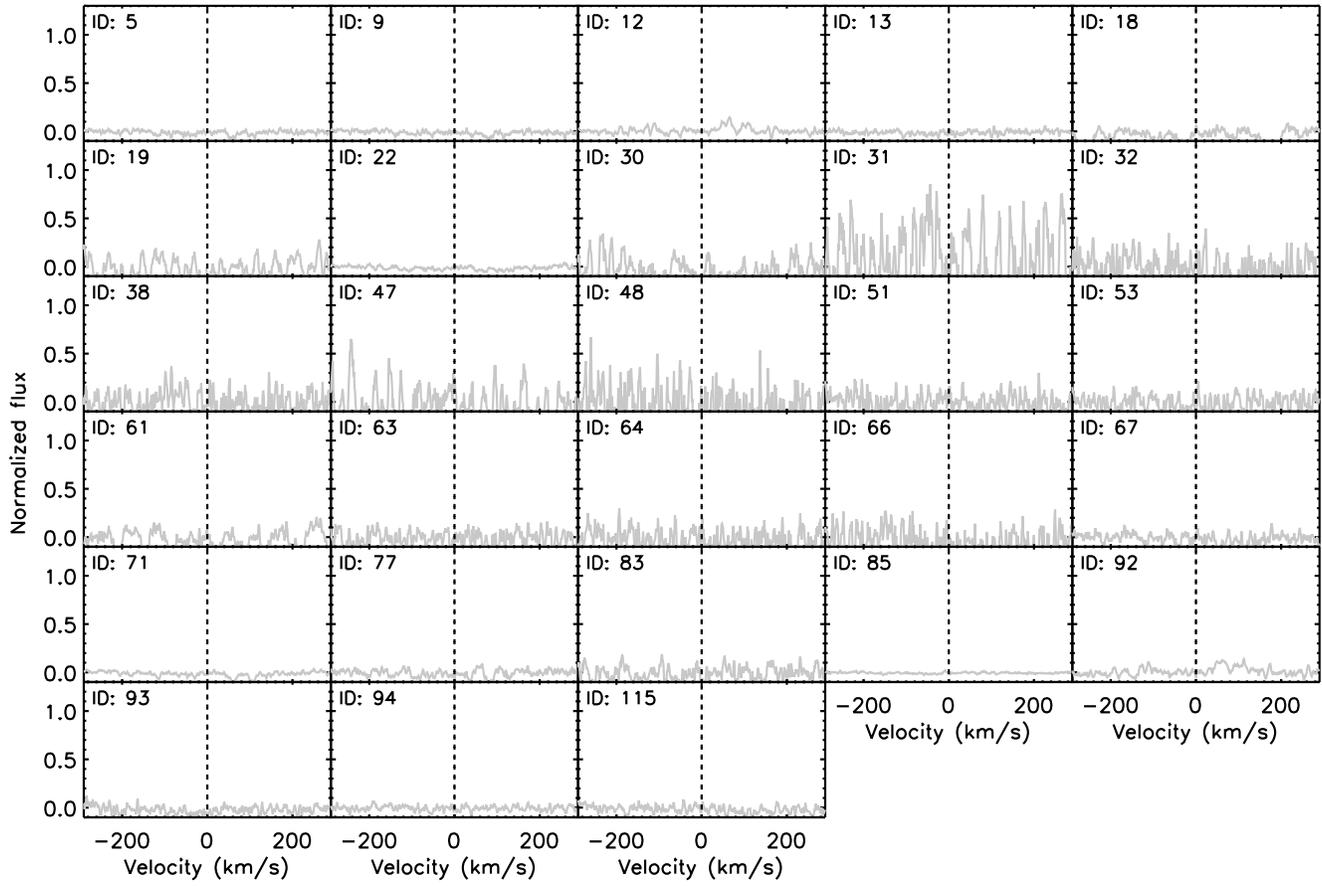}
\caption{\OIa\ line profiles for DBs or NDs. The SEDs of the corresponding sources are shown in Fig.~\ref{Fig:SED3}.}\label{Fig:SED_LR3}
\end{center}
\end{figure*}

\clearpage

\section{The dependence of H$\alpha$ $FW_{\rm H\alpha 10\%}$ on spectral types \label{Appen:FW10}}

We collect a sample of young stars which show accretion activities indicated by the presence of the Balmer continuum excess in their U-band spectra in the literature \citep{2008ApJ...681..594H,2014A&A...561A...2A,2017A&A...600A..20A,2016A&A...585A.136M,2017A&A...604A.127M,2018A&A...609A..70R} and a sample of young stars with debris disks or without disks which show no accretion activities (this work; \citealt{2003ApJ...592..282J};  \citealt{2005ApJ...625..906M}; \citealt{2006ApJ...648.1206J}; \citealt{2012ApJ...745..119N}). 

In Fig.~\ref{Fig:CFW10}, we show the H$\alpha$ $FW_{\rm H\alpha 10\%}$  vs. spectral types for these young stars. Based on it, we propose the thresholds on the H$\alpha$ $FW_{\rm H\alpha 10\%}$ to separate WTTSs from CTTSs within two spectral type ranges. 
We consider a young star to be a CTTS  if H$\alpha$ $FW_{\rm H\alpha 10\%}>250$\,\kms\ for the ones with spectral types earlier than M4 \citep{2013ApJS..207....5F}  and if H$\alpha$ $FW_{\rm H\alpha 10\%}>200$\,\kms\ for later type ones \citep{2003ApJ...592..282J}. With these criteria, we can separate WTTSs from CTTSs well. However, it is also noted that there are some outliers in Fig.~\ref{Fig:CFW10}. The WTTSs with higher H$\alpha$ $FW_{\rm H\alpha 10\%}$ than the thresholds could be fast rotators or spectroscopic binaries. The CTTSs which show lower H$\alpha$ $FW_{\rm H\alpha 10\%}$ than the thresholds are lower accretors and their H$\alpha$ emissions are dominated by chromospheric emission.

 \begin{figure*}
\begin{center}
\includegraphics[width=0.9\columnwidth]{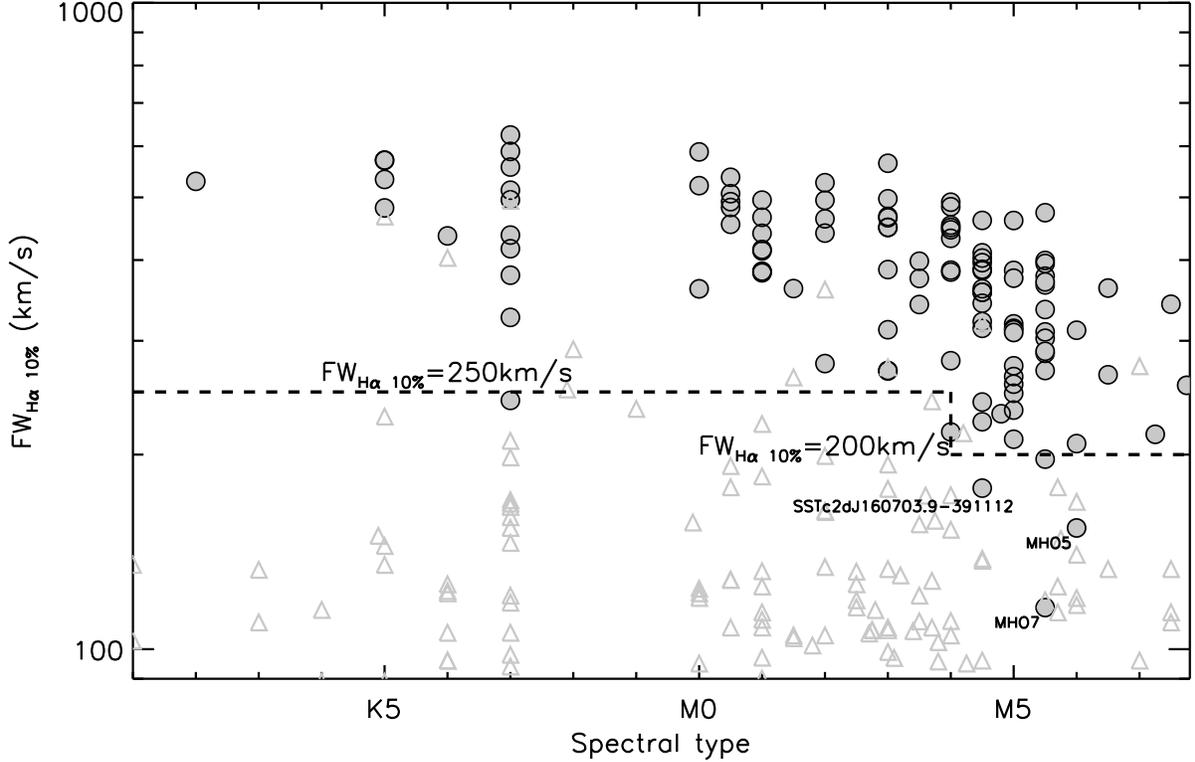}
\caption{H$\alpha$ $FW_{\rm H\alpha 10\%}$  vs. spectral types for accreting young stars (gray filled circles) and nonaccreting young stars (gray open diamonds). The dashed lines mark the thresholds which classify the CTTS/WTTS within two spectral type bins, earlier than M4 and later than M4.  \label{Fig:CFW10}}
\end{center}
\end{figure*}

\begin{figure*}
\begin{center}
\includegraphics[width=\columnwidth]{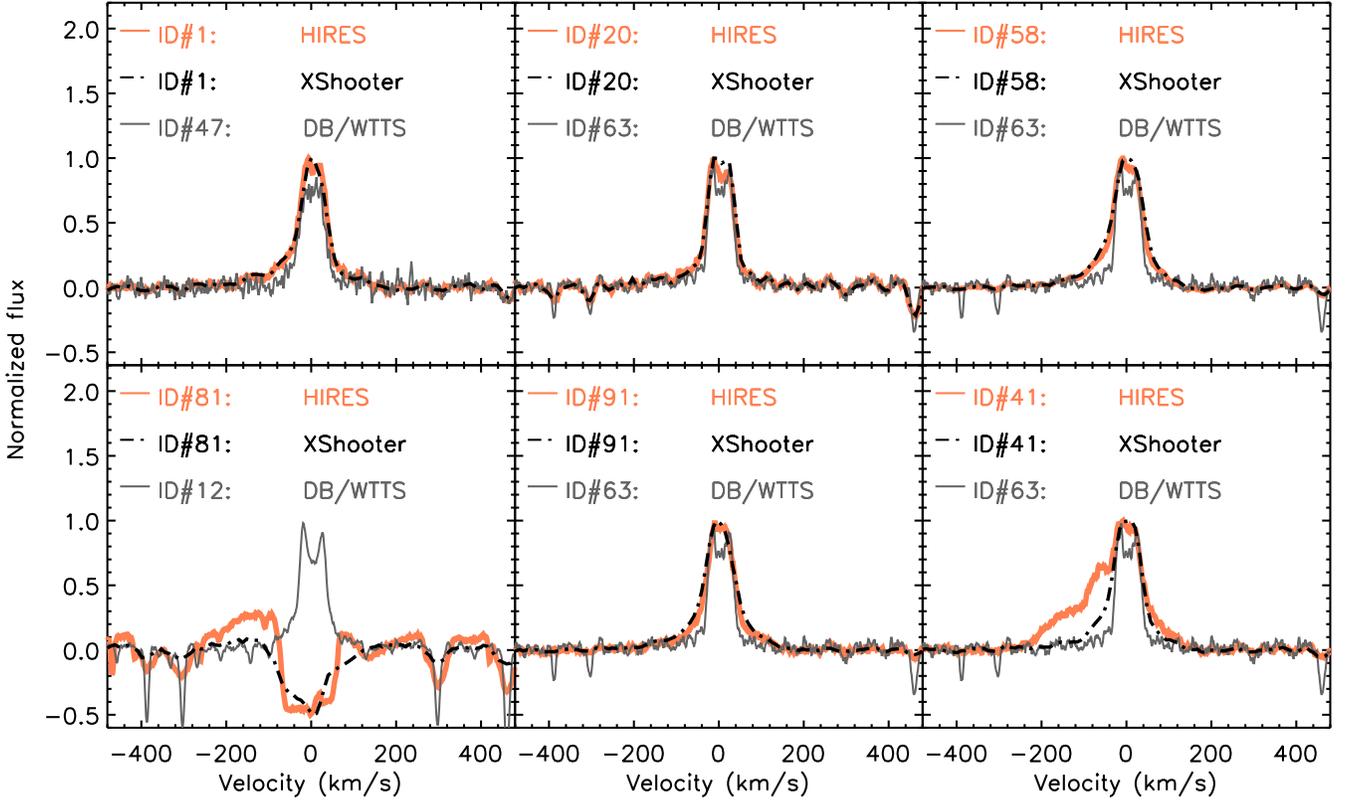}
\caption{HIRES H$\alpha$ line profiles of 5 WTTSs (ID~1, 20, 58, 81, and 91) and one CTTS (ID~41) classified in this work that have different WTTS/CTTS classification in \cite{2020A&A...639A..58M} based on X-Shooter spectra.  As a comparison, for each source we show its  X-Shooter H$\alpha$  profile (black) and one H$\alpha$ profile from a DB WTTS (gray) of similar spectral type.
}\label{Fig:com_Halpha}
\end{center}
\end{figure*}

\begin{figure}
\begin{center}
\includegraphics[width=0.7\columnwidth]{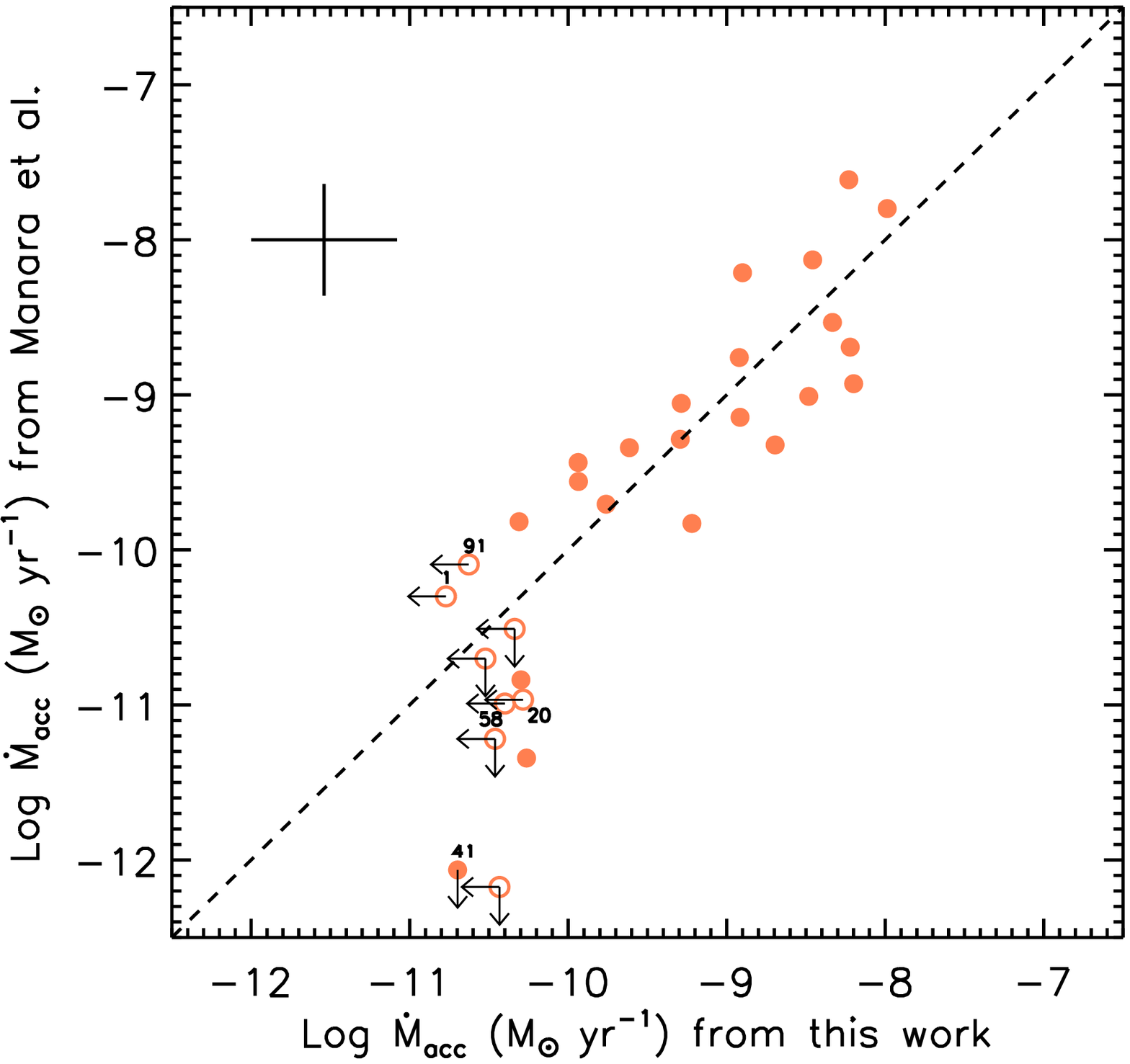}
\caption{Comparison of accretion rates  from this work and from \cite{2020A&A...639A..58M}.  Left pointing arrows indicate sources classified as WTTSs in this work while downward pointing arrows indicate WTTSs in \cite{2020A&A...639A..58M}. The dashed line shows the one-to-one relation. The cross indicates typical uncertainties. The IDs for 5 sources with discrepant WTTS/CTTS classification are marked. ID~81 is not included  because there are no accretion-related emission lines in the HIRES spectra.  Other symbols are  as in Fig.~\ref{Fig:EW_FW10}.
}\label{Fig:com_acc}
\end{center}
\end{figure}

\section{Comparisons of Accretion properties in this work and in the literature}\label{Appen:ACC_com}

 Figure~\ref{Fig:com_Halpha} shows the H$\alpha$ profiles of the 6 sources: the HIRES and X-Shooter spectra as well as one H$\alpha$ profile from the DB WTTs with the closest spectral type. We note that for 4 sources (Source 1, 20, 58, and 91) both HIRES and X-Shooter spectra show identical line profiles  which are similar to the ones of DB WTTSs, hence our classification as WTTSs. These sources are classified as CTTSs in \cite{2020A&A...639A..58M} but have estimated mass accretion rates below $10^{-10}$\,M\accunit\, which may be difficult to detect in the H$\alpha$ profiles. For Source 81, both the HIRES and X-Shooter spectra show H$\alpha$ in absorption. On the contrary, for Source 41, the HIRES spectrum clearly shows a  blue-shifted broader wing than the X-Shooter one and thus is classified as a CTTS in this work.  For this source the discrepancy on the WTTS/CTTS classification might be due to  accretion variability.

Figure~\ref{Fig:com_acc} compares our mass accretion rate estimates with those obtained by 
\cite{2020A&A...639A..58M} by modeling the Balmer continuum excess for the 32 sources in common with our sample. Note that Source~81 (2MASS~J16141107-2305362), which is classified as a CTTS in \cite{2020A&A...639A..58M}, does not clearly show any accretion-related emission lines in our HIRES spectra (see Figure~\ref{Fig:com_Halpha}), hence it is not included in the plot. 
\cite{2020A&A...639A..58M} report typical uncertainties of 0.36~dex in mass accretion rates which compare well with our overall uncertainty of 0.42~dex \footnote{This includes typical uncertainties of 0.36~dex and a flux calibration uncertainty of 15\% in our Keck data}.
Figure~\ref{Fig:com_acc} shows an overall good agreement between the two works and the mean difference in the accretion rates of the 21 CTTSs among the two works is 0.38~dex, within the quoted uncertainties.

\cite{2022AJ....163...74T} employ a new scheme to classify low accretors based the He~{\scriptsize I}~$\lambda$10830 line profile. We collect the X-Shooter spectra for the 32  common young stars from the ESO Phase 3 spectral data archive, and obtain their line profiles of He {\scriptsize I} $\lambda$10830, see Fig.~\ref{Fig:HeI}. \cite{2022AJ....163...74T} classify the He {\scriptsize I} $\lambda$10830 line profiles of young stars into six types including Types r, b, br, c, e and f. Type r line profiles show redshifted absorption below the continuum  covering velocities comparable to the free-fall velocities, Type b show detectable blueshifted absorption below the continuum, Type bf  consist of both  blueshifted and redshifted absorption features, Type c  show  detectable absorption at or near the line center, Type e  only show   emission features, and Type f  are featureless. \cite{2022AJ....163...74T}  classify the accretion level of young stars using the $\lambda$10830 line profiles: stars are considered to be accreting  if their line profiles are types r, b or br, and non accreting if their line profiles are types c, e, or f. 

\cite{2022AJ....163...74T}  have observed a sample of sources in Upper~Sco using the FIRE spectrograph at the Magellan Baade Telescope, and among them 22  (see Table~\ref{Tab:CTTS_UpperSco}) are  in common  with our sample, including 4 (IDs~20, 28, 29, and 49) that are among the above 33 sources. For the 4 sources which have both X-Shooter and FIRE spectra, three  (IDs~20, 29, and 49) have the same WTTS/CTTS classification based on the two line profiles of He~{\scriptsize I}~$\lambda$10830, and one (ID~28) has different classification which might be due to the accretion variability. If including all the 51 sources with line profiles of He~{\scriptsize I}~$\lambda$10830  from either X-Shooter spectra or FIRE spectra, 76\% (39/51) of them have same classification  in \cite{2022AJ....163...74T} and our work. Among the 27 WTTSs classified in this work, 8 sources are classified as CTTSs based on the line profiles of He~{\scriptsize I}~$\lambda$10830. In our sample, the fraction (30\%, 8/27) of low accretors among WTTSs classifed based on H$\alpha$ line is comparable to the one (27\%) for the sources with ages lager than 5\,Myr  in \cite{2022AJ....163...74T}. We also note that among the 24 CTTSs classified in this work  5 sources are classified as WTTSs  based on the line profiles of He~{\scriptsize I}~$\lambda$10830. This might be due these sources having variable accretion.
 In Table~\ref{Tab:CTTS_UpperSco}, we compare the  CTTS/WTTS classification of the Upper~Sco sources using different schemes. 
This comparison further underlines the difficulty in properly classifying low accretors using emission lines. }

\begin{figure*}
\begin{center}
\includegraphics[width=\columnwidth]{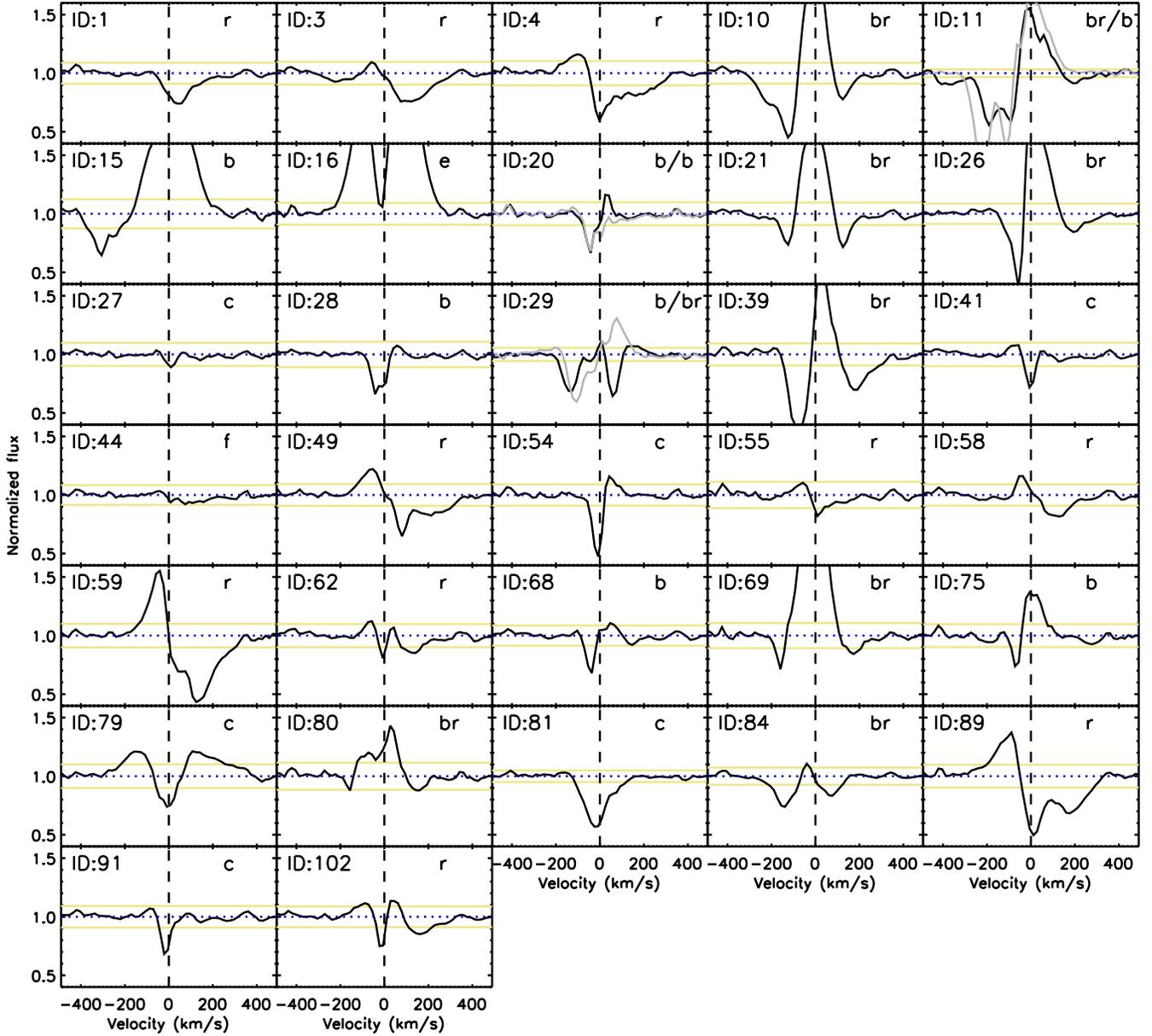}
\caption{Line profiles of He {\scriptsize I}  $\lambda$10830 for 32 sources with  X-Shooter data. The yellow solid lines mark the flux at 3$\sigma$ from the continuum. The features where their extrema above 3$\sigma$  away from the continuum are considered as detections. The type of each profile is shown in each panel. Three sources (IDs~11, 20, and 29) have two X-Shooter spectra. Two of them are shown as comparison.}\label{Fig:HeI}
\end{center}
\end{figure*}

\section{Outliers: WTTS with \OIa\ detections and CTTS without}\label{App:outliers}

\begin{figure*}
\begin{center}
\includegraphics[width=\columnwidth]{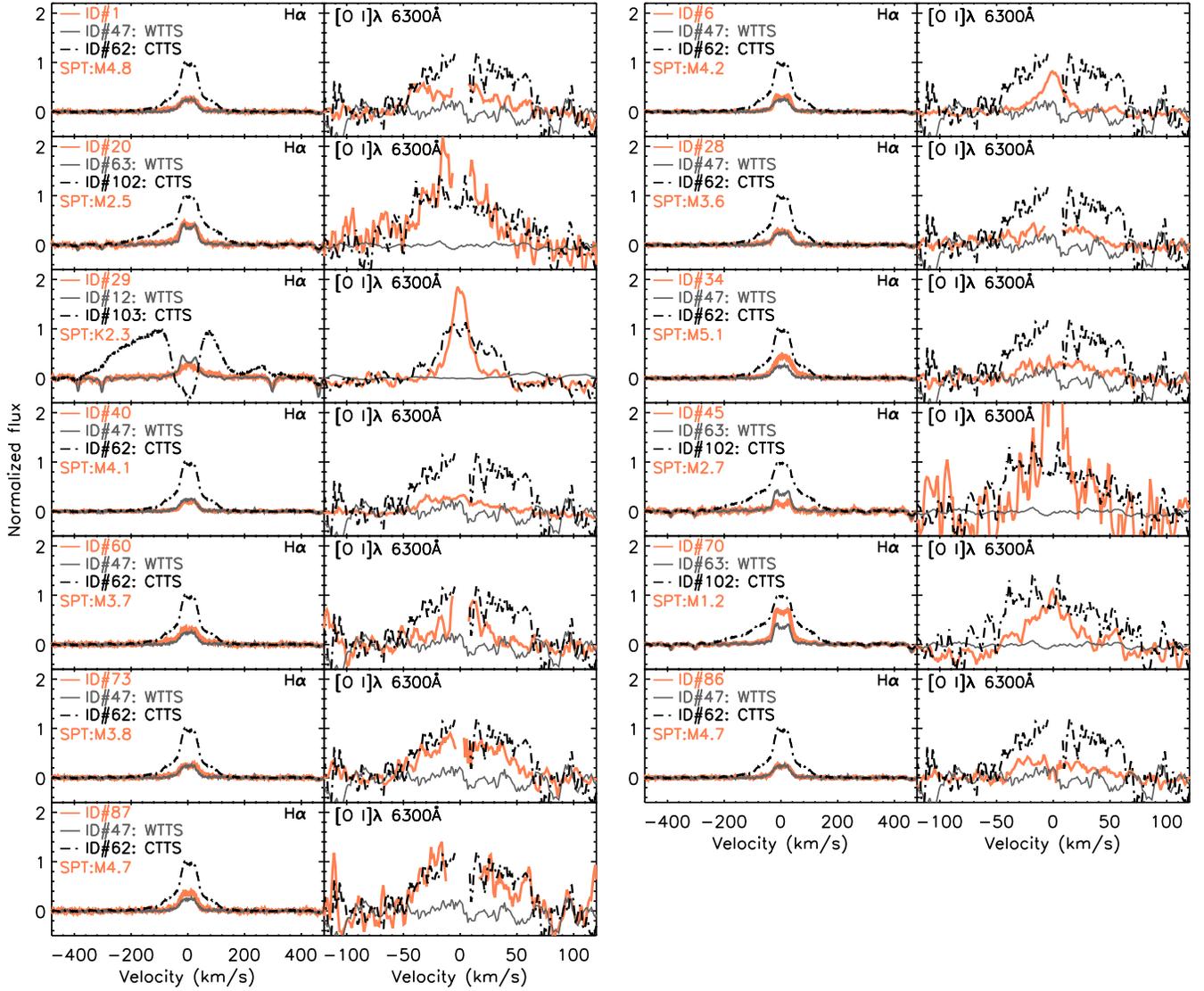}
\caption{Thirteen WTTSs (orange solid line) with \OIa\ detection in Upper Sco. For each source, its H$\alpha$ and \OIa\  line profiles are shown, as well the ones of a CTTS (black dash-dotted line) and WTTS (gray solid line) with the similar spectral type to that source. The line profiles in each panel are normalized by the peak flux of the CTTS line profile. \label{Fig:WTTS_Halpha_OI}}
\end{center}
\end{figure*}

\begin{figure*}
\begin{center}
\includegraphics[width=\columnwidth]{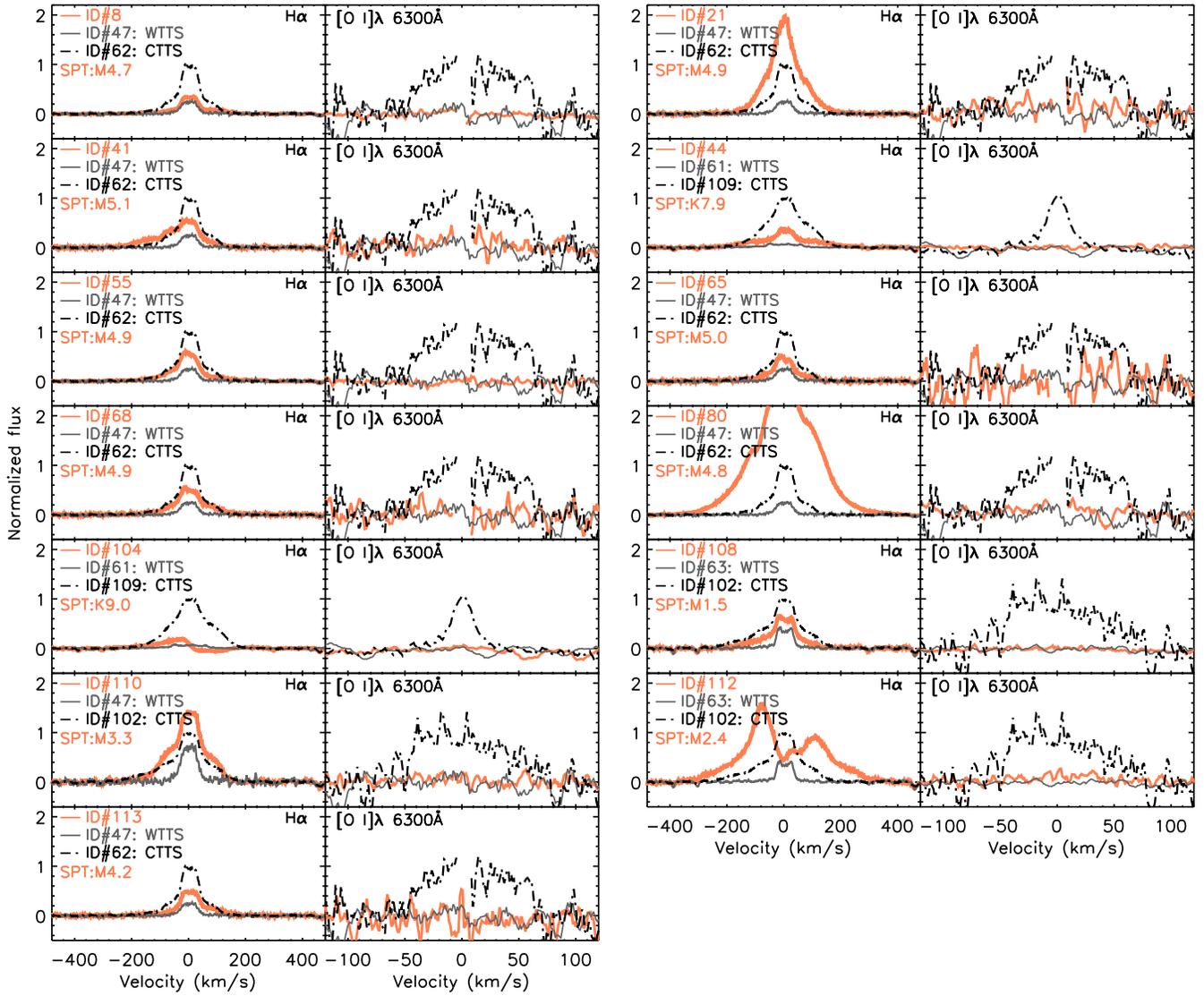}
\caption{Thirteen CTTSs (orange) without \OIa\ detection in Upper~Sco. In each panel, the lines are same as Fig.~\ref{Fig:WTTS_Halpha_OI}, beside the orange line is for the CTTS without \OIa\ detection. \label{Fig:CTTS_Halpha_OI}}
\end{center}
\end{figure*}

Our Upper~Sco PD sample includes   13 WTTSs with \OIa\ detection  and 13 CTTSs without \OIa\ detection. The H$\alpha$ line profiles and \OIa\ lines of these sources are shown in Figs.~\ref{Fig:WTTS_Halpha_OI} and ~\ref{Fig:CTTS_Halpha_OI}. As a comparison, for each source the H$\alpha$ line profiles and \OIa\ lines of a CTTS and WTTS with similar spectral type are shown.

 The \OIa\ detections in the 13 WTTSs suggest that these sources are still surrounded by a gaseous disk but the H$\alpha$ lines, which are similar to the DB/WTTS, are not sensitive enough to detect small mass accretion rates, see Fig.~\ref{Fig:WTTS_Halpha_OI}. We note that the \OIa\ detection rate (33\%) among the Upper~Sco WTTSs is comparable to the fraction (27\%) of low accretors among the sample of the old ($>5$\,Myr) WTTSs based on the line profiles of He~{\scriptsize I} $\lambda$10830 (this work and \citealt{2022AJ....163...74T}).  This suggests that the \OIa\ line could be another sensitive tracer of low accretion in addition to the He~{\scriptsize I} $\lambda$10830 line. Indeed, among the 13 WTTSs with the  \OIa\ detections, 9  (Sources~1, 20, 28, 29, 34, 45, 71, 74, and 87) have observed He~{\scriptsize I} $\lambda$10830 line (see Fig.~\ref{Fig:HeI} and \citealt{2022AJ....163...74T})  and 5 of them  (Sources~1, 20, 28, 29, and 74) have been also classified as low accretors based on the line profiles of He~{\scriptsize I} $\lambda$10830. For the other 4 WTTSs (Sources~34, 45, 71 and 87) which are also classified as  nonaccretors based on the line profiles of He~{\scriptsize I} $\lambda$10830 in \cite{2022AJ....163...74T}, more observations are required to distinguish if they are true nonaccretors or low variable accretors.

 The 13 PDs classified as CTTSs but  without \OIa\ detection are shown in Figure~\ref{Fig:CTTS_Halpha_OI}. These sources have H$\alpha$  profiles with strengths and widths comparable to other CTTSs with spectral types similar to them, but do not shown clear \OIa\ line.  Among them, 6 sources (IDs~21, 41, 44, 56, 69, and 81) have been observed with the He~{\scriptsize I} $\lambda$10830 line, and 4 of them are also classified as accretors based on the line profiles of He~{\scriptsize I} $\lambda$10830. Sources 41 and 44 are classified as a nonaccretor based on the He~{\scriptsize I} $\lambda$10830 line. However, as shown in Fig.~\ref{Fig:com_Halpha} for Source 41 this is probably due accretion variability. Source 44 has been classified as a CTTS both in this work and in \cite{2020A&A...639A..58M}. As shown in Fig.~\ref{Fig:LVC_acc},  5 sources (IDs~8, 21, 41, 56 and 81) have upper limit \footnote{For sources without \OIa\ detection, the upper limits are calculated assuming a Gaussian profile with the FWHM of an unresolved line ($\sim$8.8\,\kms), and a peak of 3$\times rms$}  on their \OIa\ line luminosity which are  one order of magnitude below the expected ones from their accretion luminosity.  \cite{2019ApJ...870...76B} presented two other young CTTSs, DN~Tau and CI~Tau, without an \OIa\ line detection. A comparison with the data in the literature show that for DN~Tau its \OIa\ line was present in 1987 \citep{1995ApJ...452..736H} but disappeared in 2006 \citep{2016ApJ...831..169S}.

\clearpage

\startlongtable
\begin{deluxetable*}{rccccccccccccccccc}
\tablecaption{A comparison of CTTS/WTTS classification in three schemes \label{Tab:CTTS_UpperSco}}
\tablewidth{700pt}
\tabletypesize{\scriptsize}
\tablehead{\colhead{ID}&\colhead{Name} & \colhead{RA} & \colhead{DEC} & \multicolumn{4}{c}{CTTS} &\colhead{[O {\tiny I] }}\\
\cline{5-8}
\colhead{}&\colhead{} & \colhead{} & \colhead{} & \colhead{Manara+20}& \colhead{Xshooter He{\tiny I}} &\colhead{Thanathibodee+22} & \colhead{This work} &\colhead{det}}
\startdata
       1& 2MASSJ15514032-2146103 &  15 51 40.32&$-$21 46 10.3&Y&Y&\nodata&N&Y\\
      3& 2MASSJ15530132-2114135 &  15 53 01.32&$-$21 14 13.5&Y&Y&\nodata&Y&Y\\
      4& 2MASSJ15534211-2049282 &  15 53 42.11&$-$20 49 28.2&Y&Y&\nodata&Y&Y\\
     10& 2MASSJ15582981-2310077 &  15 58 29.81&$-$23 10 07.7&Y&Y&\nodata&Y&Y\\
     11& 2MASSJ15583692-2257153 &  15 58 36.92&$-$22 57 15.3&Y&Y&\nodata&Y&Y\\
     15& 2MASSJ16001844-2230114 &  16 00 18.44&$-$22 30 11.4&Y&Y&\nodata&Y&Y\\
     16& 2MASSJ16014086-2258103 &  16 01 40.86&$-$22 58 10.3&Y&N&\nodata&Y&Y\\
     20& 2MASSJ16020757-2257467 &  16 02 07.57&$-$22 57 46.7&Y&Y&Y&N&Y\\
     21& 2MASSJ16024152-2138245 &  16 02 41.52&$-$21 38 24.5&Y&Y&\nodata&Y&N\\
     26& 2MASSJ16035767-2031055 &  16 03 57.67&$-$20 31 05.5&Y&Y&\nodata&Y&Y\\
     27& 2MASSJ16035793-1942108 &  16 03 57.93&$-$19 42 10.8&N&N&\nodata&N&N\\
     28& 2MASSJ16041740-1942287 &  16 04 17.40&$-$19 42 28.7&N&Y&N&N&Y\\
     29& 2MASSJ16042165-2130284 &  16 04 21.65&$-$21 30 28.4&N&Y&Y&N&Y\\
     39& 2MASSJ16062196-1928445 &  16 06 21.96&$-$19 28 44.5&Y&Y&\nodata&Y&Y\\
     41& 2MASSJ16063539-2516510 &  16 06 35.39&$-$25 16 51.0&N&N&\nodata&Y&N\\
     44& 2MASSJ16064385-1908056 &  16 06 43.85&$-$19 08 05.6&Y&N&\nodata&Y&N\\
     49& 2MASSJ16072625-2432079 &  16 07 26.25&$-$24 32 07.9&Y&Y&Y&Y&Y\\
     54& 2MASSJ16081566-2222199 &  16 08 15.66&$-$22 22 19.9&N&N&\nodata&N&N\\
     55& 2MASSJ16082751-1949047 &  16 08 27.51&$-$19 49 04.7&Y&Y&\nodata&Y&N\\
     58& 2MASSJ16090002-1908368 &  16 09 00.02&$-$19 08 36.8&Y&Y&\nodata&N&N\\
     59& 2MASSJ16090075-1908526 &  16 09 00.75&$-$19 08 52.6&Y&Y&\nodata&Y&Y\\
     62& 2MASSJ16095361-1754474 &  16 09 53.61&$-$17 54 47.4&Y&Y&\nodata&Y&Y\\
     68& 2MASSJ16104636-1840598 &  16 10 46.36&$-$18 40 59.8&Y&Y&\nodata&Y&N\\
     69& 2MASSJ16111330-2019029 &  16 11 13.30&$-$20 19 02.9&Y&Y&\nodata&Y&Y\\
     75& 2MASSJ16123916-1859284 &  16 12 39.16&$-$18 59 28.4&Y&Y&\nodata&Y&Y\\
     79& 2MASSJ16133650-2503473 &  16 13 36.50&$-$25 03 47.3&Y&N&\nodata&Y&Y\\
     80& 2MASSJ16135434-2320342 &  16 13 54.34&$-$23 20 34.2&Y&Y&\nodata&Y&N\\
     81& 2MASSJ16141107-2305362 &  16 14 11.07&$-$23 05 36.2&Y&N&\nodata&N&N\\
     84& 2MASSJ16143367-1900133 &  16 14 33.67&$-$19 00 13.3&Y&Y&\nodata&Y&Y\\
     89& 2MASSJ16154416-1921171 &  16 15 44.16&$-$19 21 17.1&Y&Y&\nodata&Y&Y\\
     91& 2MASSJ16181904-2028479 &  16 18 19.04&$-$20 28 47.9&Y&N&\nodata&N&N\\
    102& 2MASSJ16041893-2430392 &  16 04 18.93&$-$24 30 39.2&Y&Y&\nodata&Y&Y\\
     23& 2MASSJ16030161-2207523 &  16 03 01.61&$-$22 07 52.3&\nodata&\nodata&Y&N&N\\
     24& 2MASSJ16031329-2112569 &  16 03 13.29&$-$21 12 56.9&\nodata&\nodata&N&N&N\\
     34& 2MASSJ16052661-1957050 &  16 05 26.61&$-$19 57 05.0&\nodata&\nodata&N&N&Y\\
     36& 2MASSJ16055863-1949029 &  16 05 58.63&$-$19 49 02.9&\nodata&\nodata&N&N&N\\
     37& 2MASSJ16060061-1957114 &  16 06 00.61&$-$19 57 11.4&\nodata&\nodata&N&N&N\\
     42& 2MASSJ16064102-2455489 &  16 06 41.02&$-$24 55 48.9&\nodata&\nodata&N&Y&Y\\
     43& 2MASSJ16064115-2517044 &  16 06 41.15&$-$25 17 04.4&\nodata&\nodata&N&N&N\\
     45& 2MASSJ16070014-2033092 &  16 07 00.14&$-$20 33 09.2&\nodata&\nodata&N&N&Y\\
     50& 2MASSJ16072747-2059442 &  16 07 27.47&$-$20 59 44.2&\nodata&\nodata&N&N&N\\
     57& 2MASSJ16084894-2400045 &  16 08 48.94&$-$24 00 04.5&\nodata&\nodata&N&N&N\\
     70& 2MASSJ16111534-1757214 &  16 11 15.34&$-$17 57 21.4&\nodata&\nodata&N&N&Y\\
     73& 2MASSJ16115091-2012098 &  16 11 50.91&$-$20 12 09.8&\nodata&\nodata&Y&N&Y\\
     74& 2MASSJ16122737-2009596 &  16 12 27.37&$-$20 09 59.6&\nodata&\nodata&Y&N&N\\
     86& 2MASSJ16145928-2459308 &  16 14 59.28&$-$24 59 30.8&\nodata&\nodata&N&N&Y\\
     90& 2MASSJ16163345-2521505 &  16 16 33.45&$-$25 21 50.5&\nodata&\nodata&N&N&N\\
    101& 2MASSJ16023587-2320170 &  16 02 35.87&$-$23 20 17.0&\nodata&\nodata&N&N&N\\
    105& 2MASSJ16093164-2229224 &  16 09 31.64&$-$22 29 22.4&\nodata&\nodata&N&N&N\\
    111& 2MASSJ16145244-2513523 &  16 14 52.44&$-$25 13 52.3&\nodata&\nodata&N&N&N\\
\enddata
\end{deluxetable*}

\end{document}